\documentclass[preprint,showpacs,preprintnumbers,amsmath,amssymb,nofootinbib]{revtex4}
\usepackage{graphicx}
\usepackage{dcolumn}
\usepackage{bm}
\usepackage{amssymb}
\usepackage{epsfig}

\newcommand\plotwidth{5in}

\newcommand{\calA}{{\cal A}}
\newcommand{\calB}{{\cal B}}
\newcommand{\calV}{{\cal V}}
\newcommand{\calN}{{\cal N}}
\newcommand{\calO}{{\cal O}}
\newcommand{\calG}{{\cal G}}
\newcommand{\calL}{{\cal L}}
\newcommand{\calC}{{\cal C}}
\newcommand{\calD}{{\cal D}}

\newcommand{\bfx}{{\bf x}}

\newcommand{\Transpose}{T}
\newcommand{\sgn}{\rm{sgn}}
\newcommand{\dof}{\rm{d.o.f.}}
\renewcommand{\Re}{{\rm Re\,}}

\newcommand\beq{\begin{eqnarray}}
\newcommand\eeq{\end{eqnarray}}
\newcommand\Table[1]{Table~\ref{tab:#1}}
\newcommand\Fig[1]{Fig.~\ref{fig:#1}}
\newcommand\Figure[1]{Figure~\ref{fig:#1}}
\newcommand\Eq[1]{Eq.~\ref{eq:#1}} 
\newcommand\Sec[1]{Sec.~\ref{sec:#1}} 
\newcommand\Appendix[1]{Appendix~\ref{app:#1}} 
\newcommand\bal{\begin{align}}
\newcommand\eal{\end{align} }

\newcommand{\Pf}{\mathop{\rm Pf}}
\newcommand{\Tr}{{\rm Tr\,}}
\newcommand{\Dslash}{\ensuremath \raisebox{0.025cm}{\slash}\hspace{-0.27cm} D}
\newcommand{\mybar}[1]{\kern 0.6pt\overline{\kern -0.6pt#1\kern -0.6pt}\kern 0.6pt}

\begin{document}

\preprint{CU-TP-1188}

\title{Dynamical simulation of $\calN=1$ supersymmetric Yang-Mills theory with domain wall fermions}
\author{Michael G. Endres}
\email{mge2112@columbia.edu}
\affiliation{Physics Department, Columbia University , New York, NY 10027, USA}

\begin{abstract}
We present results from a numerical study of $\calN=1$ supersymmetric Yang-Mills theory using domain wall fermions.
In this particular lattice formulation of the theory, supersymmetry is expected to emerge accidentally in the continuum and chiral limits without any fine-tuning of operators.
Dynamical simulations were performed for the gauge group $SU(2)$ on $8^3\times 8$ and  $16^3\times 32$ lattice space-time volumes and at three different values of the coupling: $\beta = 2.3$, $2.35\mybar3$ and $2.4$.
Results from this study include measurements of the static potential, residual mass, and a chirally extrapolated value for the gluino condensate at $\beta=2.3$.
In addition to these, we study the low lying eigenvalues and eigenvectors of the five dimensional Hermitian domain-wall fermion Dirac operator and present evidence that, for the choice of parameters under investigation, features of the spectrum appear qualitatively consistent with strong coupling and the presence of a large residual mass.
From the five dimensional eigenvalues we explore the possibility of using the Banks-Casher relation to determine an independent value for the gluino condensate in the chiral limit.

\end{abstract}
\date{\today}

\pacs{11.15.Ha, 11.30.Pb} 

\maketitle

\section{Introduction}
\label{sec:0}

In recent years, substantial effort has been devoted toward formulating supersymmetric (SUSY) gauge theories on the lattice.\footnote{For recent reviews, see e.g., \cite{Feo:2002yi,Giedt:2007hz}.}
These efforts have been partially motivated by the fascinating theoretical and technical difficulties associated with the problem, as well as the obvious potential role of SUSY in beyond the standard model physics.
Of crucial importance pertaining to the latter point is an understanding of a variety of nonperturbative phenomena, including--but not limited to--dynamical SUSY breaking.
An understanding of nonperturbative aspects such as these may in principle be achieved with numerical simulations, provided an appropriate lattice discretization of the theory may be found.

Since na\'ive lattice discretizations typically break SUSY explicitly, simulations of such theories will generally require a substantial degree of fine-tuning in order to cancel any undesirable SUSY breaking operators which may arise through radiative corrections.
Such a task is exceedingly difficult, even when the number of parameters which require fine-tuning are relatively small.
However, it has been realized for some time that one of the simplest SUSY theories, $\calN=1$ supersymmetric Yang-Mills (SYM), may be simulated using conventional lattice discretizations and yet require only a minimal degree of fine-tuning.

The field content of $\calN=1$ SYM consists of a vector field and a single Majorana fermion which transforms as an adjoint under the gauge group, which for our purposes will be taken to be $SU(N_c)$.
The theory possesses an anomalous $U(1)_A$ axial symmetry, however, a discrete $Z_{2N_c}$ subgroup of $U(1)_A$ survives at the quantum level.
It is believed that $\calN=1$ SYM shares a variety of features in common with Quantum Chromodynamics (QCD), such as confinement and chiral symmetry breaking \cite{Witten:1997ep}.
But in contrast to QCD, $\calN=1$ SYM is believed to possess discrete chiral symmetry breaking from $N_{2N_c}\to Z_2$ which results in the possible formation of domain walls but no Goldstone bosons.
Although this theory does not possess dynamical SUSY breaking, as implied by a nonvanishing Witten index \cite{Witten:1982df}, it is still believed to exhibit a variety of other features which may be of interest to explore nonperturbatively.
Of perhaps greatest interest is whether or not a gluino condensate forms, the spectrum of the theory, and the relationship between the domain wall tension and the gluino condensate.

The low energy spectrum of $\calN=1$ SYM is believed to consist of massive supermultiplets which may involve glue-glue, glue-gluino as well as gluino-gluino bound states.
Although the states within a given supermultiplet are degenerate, at finite gluino mass one expects mass splittings which may be calculated using a variety of effective theories \cite{Veneziano:1982ah,Farrar:1997fn}.
Lattice simulations of $\calN=1$ SYM may be used to directly test the predictions of these effective theories, as well as to determine the effects of soft SUSY breaking on observables other than the spectrum.

$\calN=1$ SYM theory provides an ideal starting point for studying aspects of SUSY numerically while still allowing one to utilize familiar and well understood fermion discretizations.
Because the field content of this theory consists only of a vector field and Majorana fermion (and in particular no scalar fields), with conventional lattice discretizations of this theory, the only relevant SUSY violating operator which may arise radiatively is a gluino mass term.
In the chiral and continuum limits, SUSY is therefore expected to emerge {\it accidentally} at infinite volume.
For fermion discretizations that explicitly break chiral symmetry, taking the chiral limit requires tuning the input mass parameter to some critical value, such that the bare mass vanishes.

With the use of a Ginsparg-Wilson \cite{Ginsparg:1981bj} fermion discretization such as overlap fermions \cite{Neuberger:1997fp}, fine-tuning of the gluino mass term may be entirely avoided.
This study utilizes the closely related domain wall fermion (DWF) formalism \cite{Kaplan:1992bt,Narayanan:1992wx,Shamir:1993zy}, which realizes chiral symmetry in the limit of infinite fifth dimension extent and vanishing input gluino mass.
In a typical DWF simulation where the fifth dimension is finite, however, the action receives residual chiral symmetry violating radiative corrections, including an additive mass renormalization.
The important advantage that DWF fermions have over other discretizations (which lack chiral symmetry) is that these renormalizations may be removed by taking a controlled and unambiguously determined limit.

In the past, a variety of numerical studies have employed Wilson fermions in order to simulate $\calN=1$ SYM \cite{Montvay:1995ea,Donini:1996nr,Koutsoumbas:1996kz,Montvay:1996pz,Donini:1997hh,Koutsoumbas:1997de,Montvay:1997ak,Kirchner:1998mp,Campos:1999du,Montvay:1999gn,Feo:1999hw,Feo:1999hx,Farchioni:2001wx,Montvay:2001aj,Peetz:2002sr,Farchioni:2004ej,Farchioni:2004fy,Demmouche:2008ms}.
This discretization requires fine-tuning and formally possesses a sign problem, since the Pfaffian obtained from “integrating out” the fermion degrees of freedom is generally not positive definite.\footnote{While one may prove that the fermion Pfaffian for this theory is positive definite in the continuum, positivity is not necessarily guaranteed on the lattice.}
Studies, however, have found that the Pfaffian phase becomes innocuous in the parameter regime of physical interest and may be effectively handled by phase reweighting techniques.
It was observed in \cite{Kaplan:1999jn} that DWFs have an additional advantage over Wilson fermions in that the fermion Pfaffian is positive definite.
With the use of DWFs, one may therefore avoid the task of phase reweighting altogether.

The first and until recently the only study to use DWFs to investigate $\calN=1$ SYM focused on the chiral limit of the gluino condensate \cite{Fleming:2000fa} at a single lattice spacing.
This study was performed by using an inexact, hybrid molecular dynamics R (HMDR) algorithm \cite{Gottlieb:1987mq} which, for the theory under consideration, is prone to finite integration step size errors.
Our study offers an improvement over \cite{Fleming:2000fa}, by utilizing the rational hybrid Monte Carlo (RHMC) algorithm \cite{Clark:2003na,Clark:2004cp,Clark:2005sq}, which is an exact algorithm.
Furthermore, due to algorithmic improvements and faster computers, we are able to explore the theory on larger lattices and smaller couplings as well.
Recently an independent study of $\calN=1$ SYM using DWFs was reported by Giedt, et al. in \cite{Giedt:2008cd,Giedt:2008xm}.

The primary purpose of our study is to establish a set of sensible simulation parameters, perform basic measurements and provide the necessary ground work for more detailed studies of the theory in the future.
As such, we expand the work of \cite{Fleming:2000fa} in several important respects: we
1) establish the lattice scale by measuring the static potential and provide evidence for confinement which is consistent with expectations,
2) determine the size of the residual mass in order to ascertain the proximity to the SUSY point,
3) extrapolate the chiral condensate to the chiral limit using a recent, theoretically motivated fit formula for its $L_s$ dependence and which differs from the formula used in \cite{Fleming:2000fa},
4) study the structure of the low lying eigenvectors and eigenvalues of the Hermitian DWF Dirac operator and attempt to extract an independent value for the chiral condensate using the Banks-Casher relation \cite{Banks:1979yr}.
With exception to the third and final points, questions such as these could not be easily addressed in \cite{Fleming:2000fa} due to the limited space-time volumes employed.

Portions of this work have been reported at Lattice 2008 \cite{Endres:2008tz}, including results obtained for points 1-3.
In this paper we go into greater detail on the analysis of these results, and expand this analysis by including the eigenvalue and eigenvector studies outlined in point 4.
We organize this paper as follows.
In \Sec{1} we describe the details of our numerical simulations, including a brief review of the domain wall fermion formalism and conventions, the RHMC algorithm, ensembles and measurement methods.
\Sec{2} is devoted to issues concerning thermalization and autocorrelations, while \Sec{3} pertains to measurement details, data analysis and results.
In \Sec{4}, we summarize our findings.
Finally, in \Appendix{0} we review some basic properties of Marjorana fermions, which may be less familiar to those readers who specialize in lattice QCD, and specifically derive expressions for some of the gluino-dependent observables studied in this paper.

\section{Simulation details}
\label{sec:1}

\subsection{Lattice action and conventions}
\label{sec:1_0}

Dynamical numerical simulations of $\calN=1$ SYM were performed using a Wilson gauge action and domain wall fermions.
The full form of the partition function in Euclidean space-time is given by:
\beq
Z = \int [dU] [d\Psi] [d\Phi] \, e^{-S_G[U] - S_F[\Psi,U] - S_{PV}[\Phi,U]}\ ,
\label{eq:partition_function}
\eeq
where $S_G$ represents the gauge action and $S_F$ represents the action for domain wall fermions.
$S_{PV}$ represents the action for the  Pauli-Villars (PV) fields which are required in order to cancel off the UV contributions associated with the bulk fermions in the fifth dimension.
Simulations are performed on a $V\times L_s$ lattice, where $V = L^3\times T$ represents the lattice space-time volume of the four physical dimensions with spatial extent $L$ and temporal extent $T$, and $L_s$ represents the size of the unphysical fifth dimension.

The Wilson gauge action is given by:
\beq
S_G[U] =  \beta \sum_{x,\mu\nu} ( 1 -\frac{1}{2} \Re \Tr  P_{x,\mu\nu} )\ ,
\eeq
where $P_{x,\mu\nu} = \Tr ( U_{x,\mu} U_{x+\mu,\nu} U^\dagger_{x+\nu,\mu} U^\dagger_{x,\nu})$ represents a $1\times1$ plaquette and $U_{x,\mu}$ are link variables which belong to fundamental representation of $SU(N_c)$ and are associated with the four dimensional space-time coordinate ($x$) and orientation ($\mu$).
As is usual with the domain wall fermion discretization, the gauge field is taken to be independent of the fifth dimension coordinate ($s$).
The bare coupling is given by $\beta = 4/g^2$.

The five dimensional DWF Dirac operator $D$ is defined by:
\beq
D_{x,s;x^\prime,s^\prime}(M_5,m_f) = \delta_{s,s^\prime} D^\parallel_{x,x^\prime}(M_5) + \delta_{x,x^\prime} D^{\perp}_{s,s^\prime}(m_f)\ ,
\label{eq:dwf_dirac_op}
\eeq
where
\beq
D^\parallel_{x,x^\prime}(M_5) = \sum_{\mu=1}^4 \left[P^-_\mu V_{x,\mu} \delta_{x+\mu,x^\prime} + P^+_\mu V_{x^\prime,\mu}^\Transpose \delta_{x-\mu,x^\prime}
                              + (M_5 - 4) \delta_{x,x^\prime} \right]\ ,
\eeq
\beq
D^{\perp}_{s,s^\prime}(m_f) = \left\{\begin{array}{ll}
                              - m_f P_- \ ,&\qquad \textrm{if $s^\prime = 0$ and $s = L_s-1$} \\
                              - m_f P_+ \ ,&\qquad \textrm{if $s = 0$ and $s^\prime = L_s-1$} \\
                              P_- \delta_{s+1,s^\prime} + P_+ \delta_{s-1,s^\prime} - 2 \delta_{s,s^\prime} \ ,&\qquad \textrm{otherwise}\ ,
                              \end{array} \right.
\eeq
and the spin projectors $P_\pm$\footnote{$P_\pm = P_{R/L}$ in the notation of, for example, reference \cite{Furman:1994ky}.} and $P^\pm_\mu$ are given by:
\beq
P_\pm = \frac{1}{2} (1 \pm \gamma_5)\ , \qquad P^\pm_\mu = \frac{1}{2} (1 \pm \gamma_\mu)\ .
\eeq
The gamma matrices $\gamma_\mu$ in Euclidean space are taken to be Hermitian and satisfy the properties: $\Tr \left(  \gamma_\mu \gamma_\nu \right) = 4 \delta_{\mu\nu}$ and $(\gamma_\mu)^2 = 1$ (no sum on $\mu$); the four dimensional chirality operator is given by $\gamma_5 = \gamma_0 \gamma_1 \gamma_2 \gamma_3$.
The real link variables $V_{x,\mu}$ belong to the adjoint representation of the gauge group.
These may be expressed in terms of fundamental representation link variables $U_{x,\mu}$ which appear in the gauge action via the identity:
\beq
V^{ab} = 2  \Tr \left(T^a U^\dagger T^b U \right)\ ,
\eeq
where $T^a$ are generators of the fundamental representation of the gauge group which satisfy $\Tr \left (T^a T^b \right) = \delta^{ab}/2$, with $a=1,\ldots,N_c^2-1$.
Finally, $M_5$ and $m_f$ appearing in \Eq{dwf_dirac_op} represent the domain wall height and input gluino mass, respectively.

The DWF Dirac operator satisfies the following properties:
\beq
\Gamma_5^\dagger D \Gamma_5  &=& D^\dagger \ , \cr
\calC^\dagger D \calC &=& D^* \ ,
\label{eq:dwf_dirac_op_properties}
\eeq
where $\Gamma_5 = R_5 \gamma_5$, $R_5$ is the reflection operator about the mid-plane in the fifth direction, $\calC = R_5 C$ is the DWF charge conjugation operator and $C$ is the standard four dimensional charge conjugation operator which satisfies the usual relations:
\beq
C^\dagger \gamma_\mu C &=& -\gamma_\mu^\Transpose\ , \cr
C^\dagger \gamma_5 C &=& \gamma_5^\Transpose\ , \cr
C^\Transpose &=& -C\ , \cr
C^\dagger C &=& 1\ , \cr
C^\dagger T_{adj}^a C &=& -(T_{adj}^a)^\Transpose\ .
\label{eq:charge_conjugation_matrix}
\eeq
Here, $T_{adj}^a$ are the generators of the adjoint representation of the gauge group which satisfy $\Tr T_{adj}^a T_{adj}^b = N_c \delta^{ab}$.
The DWF and PV actions are given by:
\beq
S_F[\Psi,U] = \frac{1}{2} \mybar\Psi D(M_5,m_f) \Psi\ ,\qquad S_{PV}[\Phi,U] = \frac{1}{2} \mybar\Phi D(M_5,1) \Phi\ ,
\eeq
where we have imposed the DWF and PV Majorana conditions: $\mybar\Psi = \Psi^\Transpose \calC$ and $\mybar\Phi = \Phi^\Transpose \calC$.
Note that the PV action used in these simulations is not the one introduced in \cite{Furman:1994ky} but rather a variant of that action \cite{Vranas:1997da}. 

All gluino dependent observables in this study are expressed in terms of the four dimensional gluino interpolating fields $q(x)$, which are defined at the boundaries of the fifth dimension.
In terms of the five dimensional fields $\Psi(x,s)$, the gluino interpolating fields are given by:
\beq
q(x) &=& P_-\Psi(x,0) + P_+ \Psi(x,L_s-1)\ , \cr
\mybar q(x) &=& \mybar\Psi(x,L_s-1) P_- + \Psi(x,0)P_+\ .
\label{eq:wall_gluino}
\eeq
For convenience, we also define here the four-dimensional gluino fields $q_m(x)$ which may be associated with the mid-plane in the fifth dimension:
\beq
q_m(x) &=& P_-\Psi(x,L_s/2) + P_+ \Psi(x,L_s/2-1)\ , \cr
\mybar q_m(x) &=& \mybar\Psi(x,L_s/2-1) P_- + \Psi(x,L_s/2)P_+\ .
\label{eq:midpoint_gluino}
\eeq
Note that each of the interpolating fields in \Eq{wall_gluino} and \Eq{midpoint_gluino} satisfy the appropriate four dimensional Majorana conditions: $\mybar q = q^\Transpose C$ and $\mybar q_m = q_m^\Transpose C$.

Following \cite{Furman:1994ky} we may define four-dimensional vector and axial currents $\calV_\mu(x)$ and $\calA_\mu(x)$ by:
\beq
\calV_\mu(x) &=& \sum_{s=0}^{L_s-1} j_\mu(x,s) \ , \cr
\calA_\mu(x) &=& \sum_{s=0}^{L_s-1} \sgn\left(s-\frac{L_s-1}{2} \right) j_\mu(x,s)\ , 
\eeq
where $j_\mu(x,s)$ are the first four components of the five dimensional conserved current, given by:
\beq
j_\mu(x,s) = \mybar\Psi(x+\mu,s) P_\mu^+ V_{x,\mu}^\Transpose \Psi(x,s) - \mybar\Psi(x,s) P_\mu^- V_{x,\mu} \Psi(x+\mu,s)\ ,
\eeq
and the last component is given by:
\beq
j_5(x,s) = \left\{\begin{array}{ll}
             \mybar\Psi(x,s) P_- \Psi(x,s+1) - \mybar\Psi(x,s+1) P_+ \Psi(x,s)\ , &\qquad 0\leq s < L_s-1 \ , \\
             \mybar\Psi(x,L_s-1) P_- \Psi(x,0) - \mybar\Psi(x,0) P_+ \Psi(x,L_s-1)\ , &\qquad  s = L_s-1 \ . 
             \end{array} \right.
\eeq

The divergence of the four dimensional vector and axial currents satisfy
\beq
\Delta_\mu \calV_\mu(x) &=& 0 \ , \cr
\Delta_\mu \calA_\mu(x) &=& 2 m_f J_5(x) + 2 J_{5q}(x) \ ,
\eeq
where $\Delta_\mu f(x) = f(x+\mu)- f(x)$ is the forward finite difference operator and
\beq
J_5(x) &=& j_5(x,L_s-1) = \mybar q(x) \gamma_5 q(x) \ , \cr
J_{5q}(x) &=& j_5(x,L_s/2-1) = \mybar q_m(x) \gamma_5 q_m(x) \ ,
\eeq
are the pseudo-scalar densities defined on the walls and midpoint of the fifth dimension, respectively.
The axial Takahashi-Ward identity (ATWI) is given by:
\beq
\Delta_\mu \langle \calA_\mu \calO \rangle = 2 m_f \langle J_5 \calO \rangle + 2\langle J_{5q} \calO \rangle + i \langle \delta \calO \rangle\ ,
\eeq
where it may be verified that the term involving the midpoint pseudo-scalar density $J_{5q}(x)$ gives rise to the desired axial anomaly \cite{Shamir:1993yf}.

\subsection{Simulation method}
\label{sec:1_1}
In order to simulate the partition function given by \Eq{partition_function}, it is necessary to first integrate out the DWF and PV degrees of freedom.
After performing this integration, we are left with an effective action which depends only on the gauge fields and is given by:
\beq
e^{-S_{eff}[U]} = e^{S_G[U]}\times \frac{\Pf\left[ \calC D(M_5,m_f) \right]}{\Pf\left[ \calC D(M_5,1) \right]}\ ,
\label{eq:pfaffians}
\eeq
where, using the relation $\det \calC=1$, the Pfaffian may be expressed as
\beq
\Pf \left(\calC D\right) = \sqrt{\det{D}}\ .
\eeq
Numerical simulations of $\calN =1$ SYM requires a positive definite fermion Pfaffian in order to unambiguously define the Pfaffian as the square root of a determinant.
It may be shown that for domain wall fermions, this is indeed the case and demonstrates an advantage of using this formalism over other discretizations, such as Wilson fermions \cite{Kaplan:1999jn}.
By exploiting $\Gamma_5$-Hermiticity described by \Eq{dwf_dirac_op_properties}, we may rewrite the effective action as
\beq
e^{-S_{eff}[U]} = e^{S_G[U]} \times \det\left[\frac{\calD(M_5,m_f)}{\calD(M_5,1)}\right]^{1/4} \ ,
\label{eq:effective_action}
\eeq
where $\calD(M_5,m_f) = D(M_5,m_f)^\dagger D(M_5,m_f)$.
For the purpose of numerical simulation, we introduce a single complex pseudo-fermion field $\chi$ in order to reproduce the effects of the ratio of fermion determinants given by \Eq{effective_action}.\footnote{The simultaneous treatment of normal and PV fermion contributions with a single pseudo-fermion field was an idea of M. Clark, and was first introduced in \cite{Allton:2007hx}.}
The resulting partition function is given by:
\beq
Z = \int [dU] [d\chi^\dagger][ d\chi] \, e^{-S_G[U] - S_{PF}[\chi^\dagger,\chi,U] }\ ,
\label{eq:pf_partition_function}
\eeq
where the pseudo-fermion action $S_{PF}$ is given by:
\beq
S_{PF}[\chi^\dagger,\chi,U] = \chi^\dagger \calD^{1/8}(M_5,1) \calD^{-1/4}(M_5,m_f) \calD^{1/8}(M_5,1) \chi \ .
\eeq

Dynamical numerical simulations of \Eq{pf_partition_function} with $N_c=2$ were performed using a modified version of the Columbia Physics System (CPS)--a software system which is developed and maintained by the RBC collaboration for the purpose of studying lattice Quantum Chromodynamics.
Modifications to the software were specifically made in order to accommodate the adjoint character of the fermions being simulated as well as to reduce the gauge group under consideration from $SU(3)$ down to $SU(2)$.
We perform numerical simulations of the partition function given by \Eq{pf_partition_function} via the exact rational hybrid Monte Carlo (RHMC) algorithm \cite{Clark:2003na,Clark:2004cp,Clark:2005sq}.
Details of the parameters used in the approximation of $\calD^p(M_5,m_f)$ and $\calD^q(M_5,1)$, for the appropriate rational powers $p$ and $q$ (i.e., $p=1/8$ and $q = 1/8$ for the pseudo-fermion refreshment step, and  $p=1/4$ and $q = 1/8$ for the evolution) used in this simulation, may be found in \Table{rhmc_parameters_small} and \Table{rhmc_parameters}.
Parameters attributed to the PV fields in this study are labeled with the superscript PV.
The parameters $\lambda_{min}$ ($\lambda_{min}^{PV}$) and $\lambda_{max}$ ($\lambda_{max}^{PV}$) specify the lower and upper bounds on the eigenvalue range of $\calD(M_5,m_f)$, over which we require the rational approximation to be valid.
The rational approximation is used both in the molecular dynamics (MD) evolution of the gauge field as well as in the accept/reject Monte Carlo (MC) step, which is performed at the end of the trajectory in order to remove any errors associated with the finite step size $\delta\tau$ in the MD evolution.
The parameters $n_{MD}$ ($n_{MD}^{PV}$) and $n_{MC}$ ($n_{MD}^{PV}$) represent the degree and therefore accuracy of the rational approximation in each of these steps.
In our simulations we require a greater accuracy in the accept/reject step compared to the evolution and therefore take $n_{MC} > n_{MD}$.
Similarly, for the MD evolution and MC accept/reject step, the conjugate gradient stopping conditions used in the operator inversions were $1\times10^{-7}$ and $1\times10^{-10}$, respectively.

Finally, the MD evolution was performed using a two level Omelyan integrator \cite{Omelyan:2003aa,Takaishi:2005tz} with the Omelyan integrator parameter set to $\lambda=0.215$.
The length of each trajectory in the evolution is given by $5\times\delta\tau$ MD time units.
At the end of each trajectory, we project the gauge field back onto the gauge group.
This step breaks reversibility, however the errors induced by the projection are negligible.
Parameters and characteristics of the MD evolution and accept/reject MC step are provided in \Table{rhmc_statistics_small} and \Table{rhmc_statistics}.
In each ensemble, the time step $\delta \tau$ was chosen such that the acceptance rate was approximately $70-80\%$, however for some ensembles the acceptance rates were as low as $60\%$ or as high as $90\%$.
One may verify from \Table{rhmc_statistics_small} and \Table{rhmc_statistics} that the equality $\langle e^{-\Delta H} \rangle = 1$ holds within errors for all ensembles, which  indicates that the algorithm is working correctly with our modifications to the code \cite{Creutz:1988wv}.
 
\subsection{Ensembles and simulation parameters}
\label{sec:1_2}

Numerical simulations where performed on two different lattice space-time volumes: $V=8^3\times8$ and $16^3\times32$.
For the gauge fields we impose periodic conditions (BCs) in all space-time directions, whereas for the gluino we impose periodic BCs in the spatial directions and anti-periodic BCs in the temporal direction.
The $8^3\times8$ ensembles were generated using the parameters: $\beta=2.3$, $L_s=12$, $16$, $20$ and $24$, $m_f=0.02$ and $M=1.9$, in order to check our code against the results of \cite{Fleming:2000fa}.
Simulation parameters for these configurations are listed in \Table{rhmc_parameters_small} and \Table{rhmc_statistics_small}.
Estimates of the gluino condensate and average plaquette are given in \Table{meas_small}, and are consistent with \cite{Fleming:2000fa} at the 1-2$\sigma$ level.

Our $16^3\times32$ ensembles were generated at $\beta=2.3$, $L_s=16$, $20$, $24$ and $28$, and input gluino masses: $m_f = 0.01$, $0.02$ and $0.04$.
As with the $8^3\times8$ ensembles, the domain wall height is set to $M=1.9$.
For $\beta=2.3$ and $m_f=0.02$, we extended the range of $L_s$ values to include $L_s = 32$, $40$ and $48$ in order to investigate the dependence of the residual mass and gluino condensate on the size of the fifth dimension.
Several simulations where performed at the weaker couplings $\beta = 2.35\mybar3$ and $\beta=2.4$ as well in order to investigate the coupling dependence of the residual mass and eigenvalues of the five dimensional Hermitian DWF Dirac operator.
\Table{rhmc_parameters} and \Table{rhmc_statistics} provide a complete list of the ensembles associated simulation parameters used.
Gluino condensate and average plaquette results for these ensembles are provided in \Table{meas}.

\subsection{Measurement methods}
\label{sec:1_3}

\subsubsection{Residual mass}
\label{sec:1_3_0}

At long distances, finite lattice spacing effects may be characterized by a continuum Symanzik effective Lagrangian:
\beq
\calL_{Symanzik} = \calL_{SYM} + a^{-1} \calL_{-1} + a \calL_1+ \ldots\ ,
\label{eq:symanzik}
\eeq
where the leading contribution $\calL_{SYM}$ represents the target continuum $\calN=1$ SYM theory.
The contributions $\calL_n$ with $n=-1,1,\ldots$ characterize finite lattice spacing effects at order $\calO(a^n)$, up to possible logarithms.
Specifically, the lowest order contributions to $\calL_{Symanzik}$ may be expressed as:
\beq
a^{-1} \calL_{-1} = \frac{1}{2} (m_f + m_{res}) J_5 \ ,\qquad a \calL_{1} = \frac{1}{2} c_{sw} J_{5f} \ ,
\label{eq:symanzik_ops}
\eeq
where
\beq
J_5 = \mybar\psi\psi \ ,\qquad J_{5f} = f^{abc} \mybar\psi^a \sigma_{\mu\nu} F^b_{\mu\nu} \psi^c \ ,
\eeq
and $\psi$ and $F_{\mu\nu}$ are the four dimensional continuum fermion field and color field strength tensor.
The invariant tensor $f^{abc} = -2i\, \Tr T^a[T^b,T^c]$ represents the structure constants of the gauge group.
The residual mass $m_{res}$ in \Eq{symanzik_ops} characterizes the leading chiral symmetry breaking effects due to the finite extent of $L_s$, and is defined in such a way that the bare gluino mass is given by the simple sum:
\beq
m_g = m_f + m_{res}\ .
\label{eq:gluino_mass}
\eeq
The Sheikholeslami-Wohlert term which is proportional to $c_{sw}$ will in turn depend on both $m_f$ and $L_s$, and is expected to vanish in the $m_f\to0$ and $L_s\to\infty$ limits.

In the continuum effective theory, the ATWI will read:
\beq
\Delta_\mu \langle \calA_\mu(x) \calO(x^\prime) \rangle = 2 (m_f+m_{res}) \langle J_5(x) \calO(x^\prime) \rangle  +  \langle \rho_{top}(x) \calO(x^\prime) \rangle + i \langle \delta\calO(x^\prime)\rangle + \ldots\ ,
\label{eq:atwi}
\eeq
where
\beq
\rho_{top} = \frac{N_c}{32 \pi^2} \Tr F\tilde F\ ,
\eeq
and the contribution proportional to $c_{sw}$ has been omitted as it is higher order in the lattice spacing and is suppressed at large $L_s$.
Comparing the continuum ATWI with that of the lattice expression, we expect at long distances and sufficiently close to the continuum limit the identity:
\beq
J_{5q}(x) \approx m_{res} J_5(x) +  \rho_{top}(x).
\label{eq:midpoint_wall_identity}
\eeq
It should be emphasized that there are no anomalous contributions to the gluino mass (as there may be for the quark mass in one flavor QCD) due to the underlying $Z_{2N_c}$ symmetry which is present in the target $\calN=1$ SYM theory.

Next we describe details on how the residual mass may be extracted from the the low energy identity \Eq{midpoint_wall_identity} obtained above.
The cleanest method for extracting the residual mass in the context of QCD is to study the long time behavior of the flavor non-singlet wall-midpoint pseudo-scalar density correlator divided by the wall-wall pseudo-scalar density correlator.
In the case of $\calN=1$ SYM, however, the method is complicated by the fact that there are no flavor non-singlet pseudo-scalars, and that the flavor singlet pseudo-scalar correlator is contaminated by anomalous contributions.

In order to proceed, we consider separately the connected and disconnected contributions to the wall-midpoint and wall-wall pseudo-scalar density correlators, which may be defined in a standard fashion.
For the wall-midpoint correlator we have:
\beq
\left\langle J_{5q}(x) P(x^\prime) \right\rangle = 2 \left\langle J_{5q}(x) P(x^\prime) \right\rangle_{connected} + \left\langle J_{5q}(x) P(x^\prime) \right\rangle_{disconnected}\ ,
\label{eq:wall_midpoint_correlator}
\eeq
where, if the pseudo-scalar density $P(x)$ is taken to be $J_5(x)$, for example, the connected and disconnected parts of the wall-midpoint correlator are given by:
\beq
\left\langle J_{5q}(x) J_5(x^\prime) \right\rangle_{connected} &=& \left\langle \overbrace {\mybar q_m(x) \gamma_5 \underbrace{ q_m(x)  \mybar q(x^\prime) } \gamma_5 q(x^\prime) } \right\rangle\ , \cr
\left\langle J_{5q}(x) J_5(x^\prime) \right\rangle_{disconnected} &=& \left\langle \underbrace{\mybar q_m(x) \gamma_5 q_m(x)} \underbrace{ \mybar q(x^\prime)\gamma_5 q(x^\prime) } \right\rangle\ ,
\label{eq:pseudoscalar_correlator_parts}
\eeq
respectively.
In each of these equations the braces indicate which gluino fields are contracted.
The factor of two appearing in the connected contribution to \Eq{wall_midpoint_correlator} above accounts for the fact that we are working with Majorana rather than Dirac fermions (i.e., the quark fields $\mybar q$ and $q$ may be contracted with themselves; for more details, see \Appendix{0}).
Similar expressions may be defined for the wall-wall pseudo-scalar correlator.

At low energies and long distances we expect the relation:
\beq
\left\langle J_{5q}(x) P(x^\prime) \right\rangle &\approx& m_{res} \left\langle J_5(x) P(x^\prime) \right\rangle + \left\langle \rho_{top}(x) P(x^\prime) \right\rangle\ ,
\label{eq:mres_relation}
\eeq
which we assume for the moment may be decomposed in terms of connected and disconnected parts as:
\beq
\left\langle J_{5q}(x) P(x^\prime) \right\rangle_{connected} &\approx& m_{res} \left\langle J_5(x) P(x^\prime) \right\rangle_{connected}\ ,\cr
\left\langle J_{5q}(x) P(x^\prime) \right\rangle_{disconnected} &\approx& m_{res} \left\langle J_5(x) P(x^\prime) \right\rangle_{disconnected} + \left\langle \rho_{top}(x) P(x^\prime) \right\rangle\ .
\label{eq:mres_decomposition}
\eeq
If this decomposition is indeed valid, then the residual mass may be extracted from the ratio of long time correlation functions:
\beq
m_{res}^\prime(m_f) = \lim_{t\to\infty} R(t)\ ,\qquad
R(t) = \frac{\left\langle \sum_{\bf x} J_{5q}({\bf x},t) P(0) \right\rangle_{connected} }{\left\langle \sum_{\bf x} J_5({\bf x},t)  P(0) \right\rangle_{connected}}\ .
\label{eq:mres_ratio}
\eeq
Note that there is an $\calO(a)$ ambiguity in this particular expression for the residual mass due to the presence of $m_f$ and $L_s$ dependent higher order contributions to \Eq{atwi} which have been omitted in \Eq{midpoint_wall_identity}.
As such, we adopt the notation $m_{res}^\prime(m_f)$ rather than $m_{res}$ to represent the quantity which is extracted from the ratio $R(t)$.

In the case of two- or two plus one-flavor QCD, the correctness of the decomposition given in \Eq{mres_decomposition} may be proved trivially because there is an underlying flavor symmetry which allows one to relate the connected contributions to these diagrams to their non-anomalous flavor non-singlet counterparts.
In $\calN=1$ SYM, however, there is no such underlying flavor symmetry and therefore \Eq{mres_ratio} remains based upon  an assumption.
One may verify from numerical simulations that the ratio $R(t)$ indeed tends to a constant value, which suggests that the time dependence of the connected part of the wall-midpoint and wall-wall correlators are the same.
However this observation does not preclude the possibility that the disconnected contribution to the wall-midpoint correlator on the left-hand side of \Eq{mres_relation} contributes to the connected part of the wall-wall correlators on the right-hand side at low energies, thus giving an incorrect estimate of the residual mass.
Ultimately, establishing the validity of \Eq{mres_ratio} will require additional theoretical analysis or corroboration from an independent calculation of the residual mass.
Close to the continuum limit one may, for instance, be able to compute the residual mass from the valence mass ($m_v$) dependence of the Hermitian DWF Dirac operator eigenvalues and verify that the value extracted from $R(t)$ measurements, modulo $\calO(a)$ corrections, is consistent.
Details of this approach are explained in greater detail in \Sec{1_3_3}.

\subsubsection{Static potential}
\label{sec:1_3_1}

The potential associated with two static, fundamental representation sources separated by a distance $|\bfx|$ may be obtained from the Wilson loop: $\langle W(\bfx,t) \rangle$.
Wilson loops were measured using Coulomb gauge fixed gauge field configurations with link fields belonging to the fundamental representation of the gauge group.
In order to reduce the statistical errors associated with this observable, Wilson loops were measured with their time axis oriented along each of the four directions of the lattice and the results were then averaged.
The space-time dependence of the Wilson loop is expected to be of the form:
\beq
\langle W(\bfx,t) \rangle = C(\bfx) e^{-V(\bfx) t} + \textrm{excited states}\ ,
\label{eq:wilson_loop}
\eeq
and from this the static potential $V(\bfx)$ may be extracted from:
\beq
V_{eff}(\bfx,t) = \log \frac{\langle W(\bfx,t)\rangle}{\langle W(\bfx,t+1)\rangle}
\label{eq:effective_potential}
\eeq
at asymptotically large times.
The signal to noise ratio associated with \Eq{effective_potential} generally deteriorates in this limit, therefore moderate values of time are instead used to extract the potential.

In this work, we use \Eq{effective_potential} to locate the plateau region of $V_{eff}(\bfx,t)$ for each value of $\bfx$.
We then fit the Wilson loops to the functional form given by \Eq{wilson_loop} in order to extract the potential.
Finally, the extracted values of $V(\bfx)$ may be fit to the Cornell potential which is given by:
\beq
V(\bfx) = V_0 - \frac{\alpha}{|\bfx|} + \sigma |\bfx|\ ,
\label{eq:potential}
\eeq
allowing us to determine the constant term ($V_0$), Coulomb term ($\alpha$), string tension ($\sigma$) and Sommer scale ($r_0$) defined by \cite{Sommer:1993ce}:
\beq
\left. |\bfx|^2 \frac{\partial V(\bfx)}{\partial |\bfx|} \right|_{|\bfx|=r_0} = 1.65\ ,
\label{eq:sommer_scale}
\eeq
at fixed values of the lattice spacing and gluino mass.

\subsubsection{Gluino condensate}
\label{sec:1_3_2}

The gluino condensate
\beq
\langle \mybar q q \rangle = \frac{1}{12 V} \sum_x \langle \mybar q(x) q(x) \rangle
\label{eq:condensate}
\eeq
was measured using a stochastic estimator with a single random Gaussian space-time volume source.
Note that we normalize the gluino condensate following the conventions of \cite{Blum:2000kn} (i.e., we divide by 4 spin $\times$ 3 adjoint color components), such that $\langle \mybar q q \rangle\sim m_f^{-1}$ for large mass.
For these measurements, the conjugate gradient stopping condition used for the inversion of the Dirac operator was set to $1\times10^{-12}$.
Measurements of the condensate were made for several different values of $m_f$ and $L_s$, then a chiral limit extrapolation of the data was performed.
Details of this analysis are provided in \Sec{3_2}\ .

A second, independent measurement of the chiral condensate may be obtained using the Banks-Casher relation \cite{Banks:1979yr}, which in the continuum reads:
\beq
\lim_{m_g\to0} \lim_{V\to\infty} -\langle \mybar \psi \psi \rangle = \frac{\pi}{12} \rho(0)\ ,
\label{eq:banks_casher}
\eeq
where $\psi$ is a continuum Majorana gluino field, and $\rho(\lambda)$ is the density of eigenvalues per unit volume of the four-dimensional continuum Dirac operator $\Dslash$, given by:
\beq
\rho(\lambda) = \lim_{m_g\to0} \lim_{V\to\infty}\frac{1}{V} \left\langle \sum_i \delta(\lambda-\lambda_i) \right\rangle\ .
\label{eq:eig_density}
\eeq
In \Sec{1_3_3} we describe how one may obtain a numerical estimate of $\rho(\lambda)$ by studying the low-lying spectrum of the five dimensional Hermitian DWF Dirac operator.
In this study we are unable to measure $\rho(\lambda)$ per se, but rather a closely related quantity $\rho^\prime(\lambda;m_g)$, which is given by \Eq{eig_density} prior to taking the $V\to\infty$ and $m_g\to0$ limits.
Note that $\rho^\prime(\lambda;m_g)$ implicitly depends on the gluino mass through the distribution of gauge fields which has been sampled at finite $m_f$ and $L_s$.
The gluino condensate at a finite mass may be extracted from $\rho^\prime(0;m_g)$ in the infinite volume limit, in a way very much analogous to \Eq{banks_casher}, by taking the valence mass goes to zero limit.
Unlike the condensate obtained from \Eq{condensate}, however, the condensate extracted by this method is free of UV divergent contributions and, as such, should only depend on $m_f$ and $L_s$ in the particular combination: $m_g = m_f + m_{res}$, provided higher order terms in the Symanzik action may be neglected.
Hence, by measuring this quantity as opposed to \Eq{condensate}, we eliminate having to perform two independent extrapolations (i.e., $m_f\to0$ and $L_s\to\infty$) in favor of a single extrapolation of $m_g\to 0$.

\subsubsection{Eigenvalues of the Hermitian DWF Dirac operator}
\label{sec:1_3_3}

Here we describe some properties of the Hermitian DWF Dirac operator and its eigenvalues.
Before proceeding, however, we first review some of the expected features of the continuum, four dimensional Dirac operator $D^4 = \Dslash + m_g$ and the four dimensional Hermitian Dirac operator $D^4_H = \gamma_5 D^4$.
Using the anti-Hermiticity of $\Dslash = \gamma_\mu D_\mu$, one may show that analogous to \Eq{dwf_dirac_op_properties}, $D^4$ satisfies the following relations:
\beq
\gamma_5 D^4 \gamma_5 = (D^4)^\dagger \ , \cr
C^\dagger D^4 C = (D^4)^* \ .
\eeq
From these properties it is easy to show that the eigenvalues of $D^4$ come in complex conjugate pairs given by $\pm i \lambda + m_g$, where $\pm i \lambda$ are eigenvalues of $\Dslash$ and $\lambda \in {\mathbb R}$.
Furthermore, the eigenvalues of $D^4$ are two-fold degenerate.
The eigenvalues of $D^4_H$, which we shall denote $\lambda_H$, are real and given by $\pm \sqrt{\lambda^2+m_g^2}$.
Similarly, the eigenvalues of $D^4_H$ also have a two-fold degeneracy which follows from the property
\beq
(\gamma_5 C)^\dagger D^4_H (\gamma_5 C) = (D^4_H)^* \ .
\label{eq:hermitian_dirac_op_properties}
\eeq
Specifically, if $D^4_H \psi_{\lambda_H} = \lambda_H \psi_{\lambda_H}$, then it follows that $D^4_H \psi^c_{\lambda_H} = \lambda_H \psi^c_{\lambda_H}$, where $\psi^c_{\lambda_H} = \gamma_5 C \psi^*_{\lambda_H}$.
Orthogonality of $\psi_{\lambda_H}$ and $\psi^c_{\lambda_H}$ may be established from the antisymmetry of $\gamma_5 C$ and implies that $\psi_{\lambda_H}$ and $\psi^c_{\lambda_H}$ are linearly  independent eigenvectors of $D^4_H$.

The eigenvectors $\psi_{\pm\lambda_H}$ of $D^4_H$ may be expressed as linear combinations of $\psi_{\pm\lambda}$, with coefficients that depend on $\lambda$ and $m_g$.
When $\lambda=0$, the eigenvectors of $D_H$ are simply given by $\psi_{\lambda_H} = \psi_\lambda$.
However when $\lambda\neq0$, they and are given by:
\beq
\psi_{\pm\lambda_H} = \frac{1}{\sqrt{2|\lambda_H|}} \left[ \sqrt{-i\lambda+m_g} \,\psi_{\lambda} \pm \sqrt{i\lambda+m_g} \,\psi_{-\lambda}  \right]\ .
\eeq
In this basis, the matrix elements of $\gamma_5$ are given by:
\beq
\langle \lambda_H^\prime|\gamma_5| \lambda_H \rangle = \frac{1}{ \lambda_H} \left[ m_g \delta_{\lambda_H^\prime,\lambda_H} + i |\lambda| \delta_{-\lambda_H^\prime,\lambda_H} \right]\ .
\label{eq:gamma5_matrix_elements}
\eeq
Note in particular, that in the limit $\lambda\ll m_g$, the eigenstates of $D_H$ are approximately given by  $\psi_{\pm\lambda_H} \approx \sqrt{2} P_\pm \psi_\lambda$ and thus they become near eigenstates of $\gamma_5$ with eigenvalue $\pm1$.
On the other hand, in the limit $\lambda\gg m_g$ the chirality operator is off-diagonal in this basis for modes which are not zero modes.

The Hermitian Dirac operator also satisfies the commutator relation:
\beq
[D^4_H, \gamma_5] \psi_0 = 0\ ,
\label{eq:gamma5_commutator}
\eeq
where $\psi_0$ represents one of possibly many zero modes of $D^4_H$.
\Eq{gamma5_commutator} implies that $\gamma_5$ may be diagonalized within the subspace of zero modes and has eigenvalues given by $\pm1$.
Hence, the zero modes of $D^4$ and $D^4_H$ are also chiral modes.
On a given background gauge field configuration the four dimensional Dirac operator $D^4$ has an index given by $2 N_c \nu = n_+ - n_-$, where $n_+$ and $n_-$ are the number of left- and right-handed chiral zero modes of $D^4$, and the winding number $\nu$ according to the Atiyah-Singer index theorem is given by:
\beq
\nu = \frac{1}{32 \pi^2} \int dx^4 F_{\mu\nu}^a \tilde F_{\mu\nu}^a \ ,
\label{eq:winding_number}
\eeq
where $\tilde F_{\mu\nu} = \epsilon_{\mu\nu\sigma\rho} F_{\sigma\rho}$.
Note that since $n_+-n_-$ is an even integer, $\nu$ may be a rational number of the form: $k/N_c$, with $k\in \mathbb{Z}$.
If periodic BCs are used, fractional values for the winding number are permitted for this theory \cite{Leutwyler:1992yt}.

The five dimensional Hermitian DWF Dirac operator is given by $D_H = \Gamma_5 D$, where $D$ is the DWF Dirac operator defined in \Eq{dwf_dirac_op}.
$D_H$ satisfies a modified form of \Eq{hermitian_dirac_op_properties}, namely, 
\beq
(\Gamma_5 \calC)^\dagger D_H (\Gamma_5 \calC) = D_H^*\ .
\eeq
Hence, with eigenvalues and eigenvectors of $D_H$ given by:
\beq
D_H \Psi_{\Lambda_H} = \Lambda_H \Psi_{\Lambda_H}\ ,
\eeq
it similarly follows that $\Psi^c_{\Lambda_H} = \Gamma_5 \calC \Psi_{\Lambda_H}^*$ are eigenvectors of $D_H$ with eigenvalues $\Lambda_H$.
Due to the antisymmetry of $\calC \Gamma_5$, one may also verify that $\Psi_{\Lambda_H}$ and $\Psi^c_{\Lambda_H}$ are linearly independent and orthogonal vectors.

In this paper we study several quantities which may be extracted from the five dimensional eigenvectors of $D_H$.
We may define a four dimensional norm:
\beq
\calN_{\Lambda_H}(s) = \sum_x  \Psi_{\Lambda_H}^\dagger(x,s)  \Psi_{\Lambda_H}(x,s) \ ,\qquad \sum_s \calN_{\Lambda_H}(s) = 1 \ ,
\label{eq:four_dim_norm}
\eeq
which characterizes the profile of each eigenstate of $D_H$ in the fifth dimension.
The low energy modes of $D_H$ which describe the four dimensional effective theory will appear as bound states, whereas at high energies, the modes will appear as propagating waves in the fifth dimension.
We may also define a physical chirality operator $\Gamma_s$ whose matrix elements are given by:
\beq
\left\langle \Lambda_H^\prime|\Gamma_s|\Lambda_H \right\rangle = \sum_{x,s} \sgn\left( \frac{L_s-1}{2}-s \right) \Psi_{\Lambda_H^\prime}^\dagger(x,s)  \Psi_{\Lambda_H}(x,s)\ .
\label{eq:phys_gamma5}
\eeq
Near the continuum limit, matrix elements of $\Gamma_s$ which involve the low lying modes should exhibit all of the properties of $\gamma_5$ in the continuum four dimensional theory.

In order to extract the residual mass and four dimensional eigenvalue density $\rho(\lambda)$, we consider the valence mass ($m_v$) dependence of the eigenvalues of $D_H^2$ on background gauge field configurations, which have been generated at the input gluino mass value $m_f$.
For small $m_v$, we may parameterize the $m_v$ dependence of $\Lambda_H^2$ using the reparameterized Taylor expansion \cite{Blum:2000kn}:
\beq
\Lambda_{H,i}^2(m_v) = n_{5,i}^2 \left(\lambda_i^2+(m_v+\delta m_i)^2 \right) + \calO(m_v^3)\ ,
\label{eq:eig2_parameterization}
\eeq
and determine eigenvalue by eigenvalue the best-fit values of $n_{5,i}$, $\lambda_i$ and $\delta m_i$, where $i$ labels $i$-th eigenvalue of $D_H$ .
One may show from \Eq{condensate}, that with this parameterization of $\Lambda_H^2$ evaluated at $m_v =m_f$, the gluino condensate may be expressed at order $\calO(m_f^3)$ as \cite{Blum:2000kn}:
\beq
\langle \mybar q q \rangle = \frac{1}{12V} \sum_i \left\langle \frac{m_f+\delta m_i}{\lambda_i^2 + (m_f+\delta m_i)^2} \right\rangle \ .
\label{eq:dwf_condensate}
\eeq
Note that the normalization factor $n_{5,i}$ drops out of this relation.

We may then compare the result of \Eq{dwf_condensate} to the analogous continuum expression obtained from the operator:
\beq
\Dslash + m_f + m_{res} + c_{sw} T^a_{adj} \sigma_{\mu\nu} F^a_{\mu\nu} + \ldots\ ,
\label{eq:symanzik_dirac_op}
\eeq
which is extracted from the Symanzik effective Lagrangian defined in \Eq{symanzik}, and where we have used the identity $(T^a_{adj})^{bc} = f^{abc}$.
Assuming irrelevant operators appearing in \Eq{symanzik} and consequently \Eq{symanzik_dirac_op} are negligible, the form of \Eq{dwf_condensate} allows us to interpret the parameters $\lambda_i$ as the four dimensional eigenvalues of $i \Dslash$, and $\delta m_i$ as an effective residual mass attributed to the $i$-th eigenvalue on a given background gauge field configuration.
We may furthermore estimate the eigenvalue density $\rho^\prime(\lambda;m_g)$ for the ensemble via the formula:
\beq
\rho^\prime(\lambda;m_g) \approx \frac{ \delta N(\lambda) }{ V \times \delta\lambda \times N_{conf}  }\ ,
\eeq
where $\delta N(\lambda)$ is the number of eigenvalues within the interval $\lambda$ and $\lambda+\delta\lambda$ obtained from measurements on $N_{conf}$ gauge field configurations.

Close to the continuum and chiral limits, we expect for the lowest lying modes that the distribution of $\delta m_i$ values will be localized around the value for $m_{res}$, and that the fluctuations in $\delta m_i$ around $m_{res}$ are controlled by the size of finite lattice spacing artifacts.
If on the other hand irrelevant operators in \Eq{symanzik} are non-negligible, the interpretation of the fit parameters $n_{5,i}$, $\lambda_i$ and $\delta m_i$ becomes less clear.
The valence $L_s$ dependence will no longer reveal itself only in the parameter $\delta m_i$, but also in the parameters $\lambda_i$ and $n_{5,i}$, due to the presence of the chiral symmetry breaking dimension five operators whose coefficients are controlled by the size of the fifth dimension.
Hence, $\lambda_i$ obtained from \Eq{eig2_parameterization} may no longer be identified as an eigenvalue $i \Dslash$, but rather of some more complicated operator involving higher order terms in the lattice spacing.

\section{Thermalization and autocorrelations}
\label{sec:2}

\begin{figure}[htbp]
\centering
\includegraphics[width=\plotwidth,angle=-90]{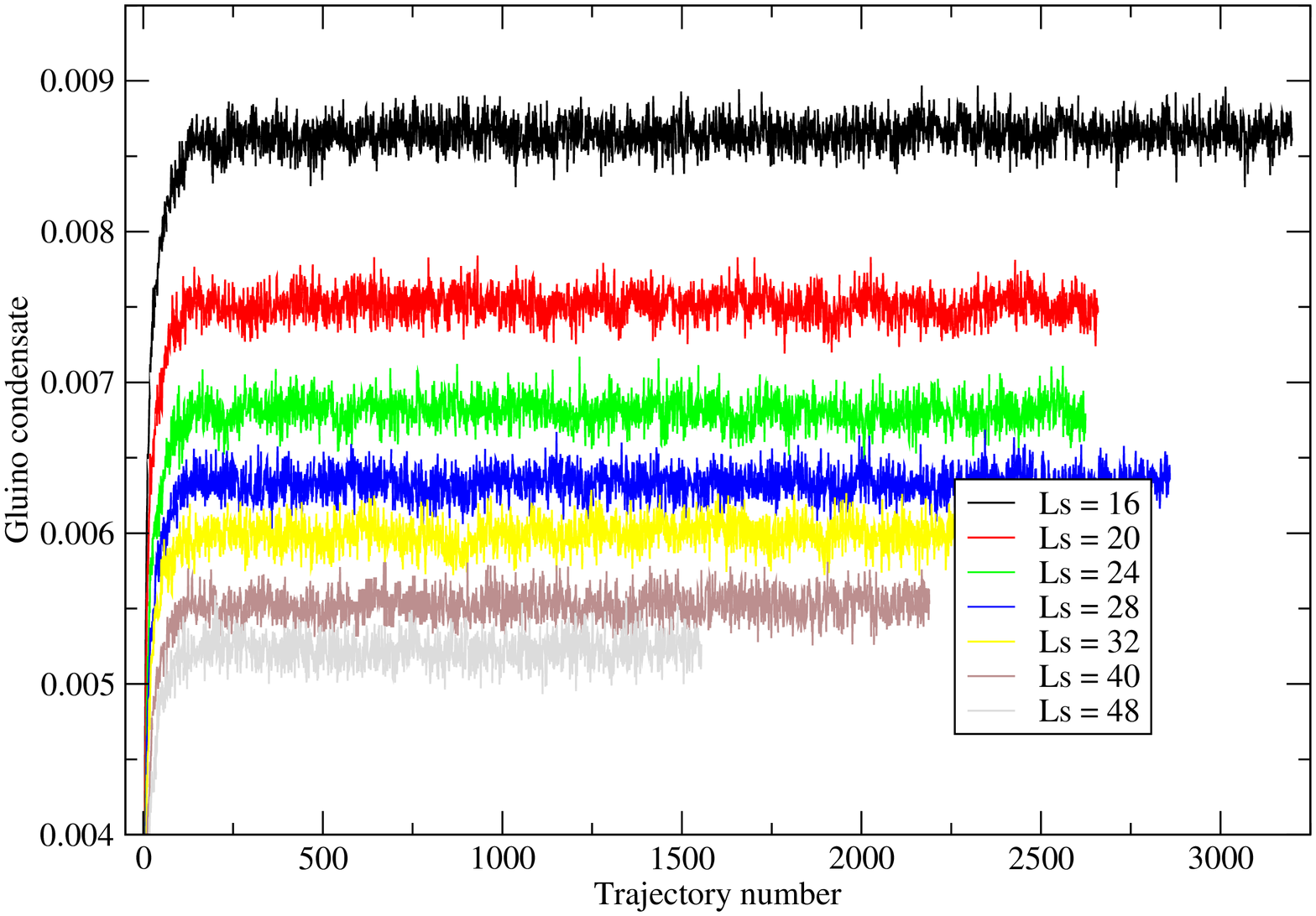}
\caption{%
Gluino condensate as a function of trajectory number for $16^3\times32$ lattices with $\beta=2.3$ and $m_f=0.02$.
}
\label{fig:thermalization}
\end{figure}

\begin{figure}[htbp]
\centering
\includegraphics[width=\plotwidth,angle=-90]{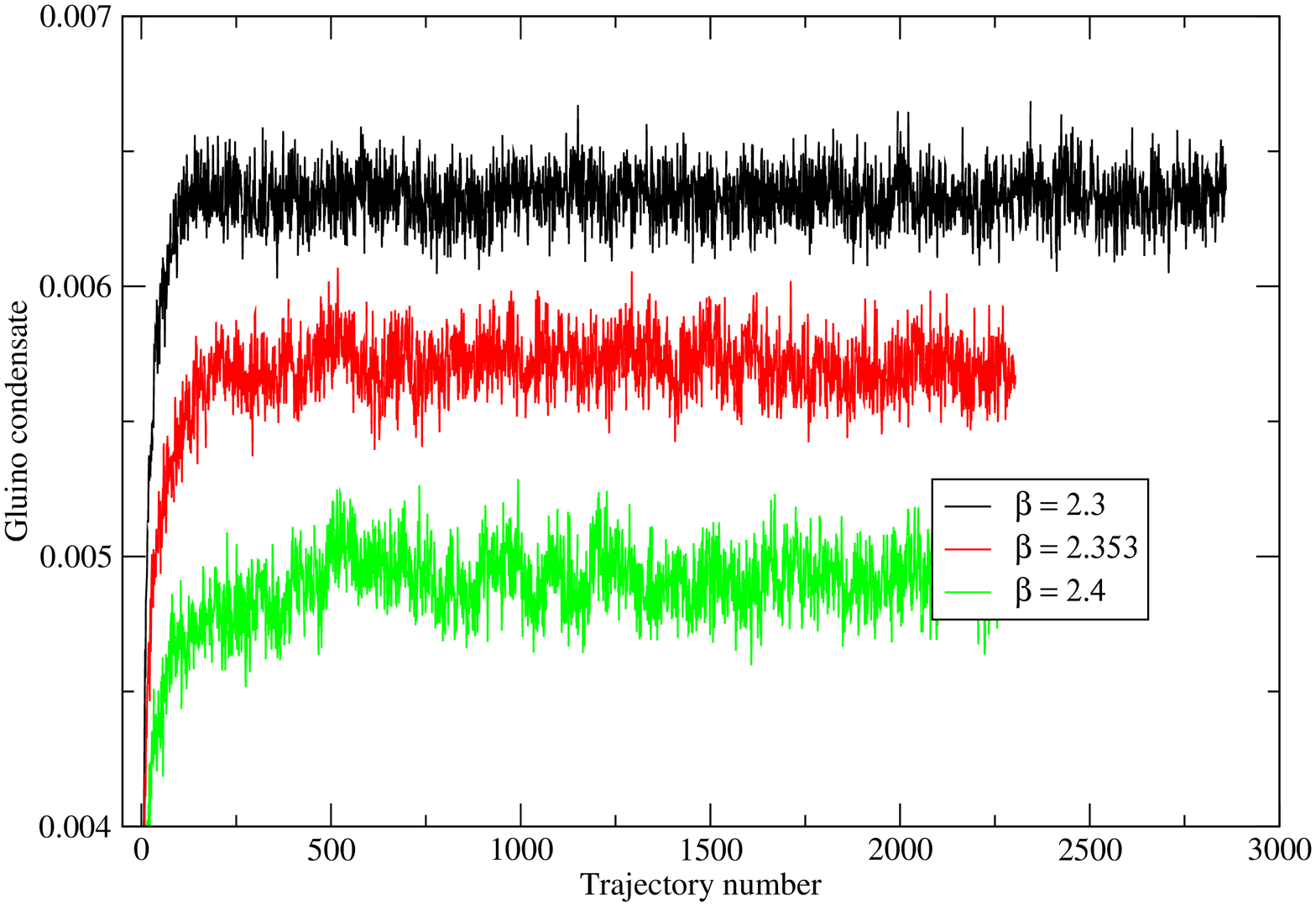}
\caption{%
Gluino condensate as a function of trajectory number for $16^3\times32$ lattices with $L_s=28$ and $m_f=0.02$.
}
\label{fig:thermalization2}
\end{figure}

For each $8^3\times8$ ensemble, a total of 700 to 1000 trajectories were generated starting from an ordered configuration, where all gauge link fields where set to unity.
For these ensembles, equilibrium was achieved within the first $250$ trajectories.
Approximately 1500 to 3000 trajectories where generated for the $16^3\times32$ ensembles from an ordered start and thermalization was achieved within the first 500 to 700 trajectories.
Specific thermalization times for these ensembles are indicated in \Table{rhmc_statistics_small} and \Table{rhmc_statistics}.
Measurements of observables were made using uncorrelated configurations which were generated thereafter.
A plot of the gluino condensate time history is shown for a variety of couplings and $L_s$ values in \Fig{thermalization} and \Fig{thermalization2} on $16^3\times32$ lattices.

\begin{figure}[htbp]
\centering
\includegraphics[width=\plotwidth,angle=-90]{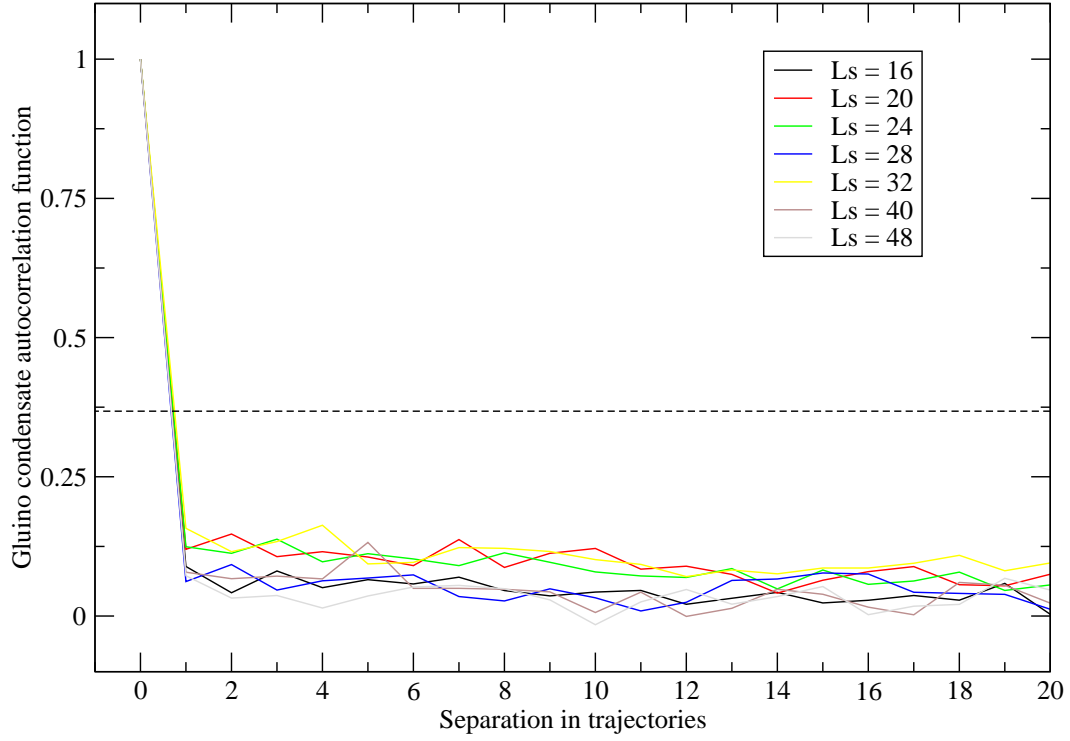}
\caption{%
Autocorrelation function associated with the gluino condensate for $16^3\times32$ lattices with $\beta=2.3$ and $m_f=0.02$.
The dashed line indicates the location of $e^{-1}$.
}
\label{fig:pbp_autocorr}
\end{figure}

\begin{figure}[htbp]
\centering
\includegraphics[width=\plotwidth,angle=-90]{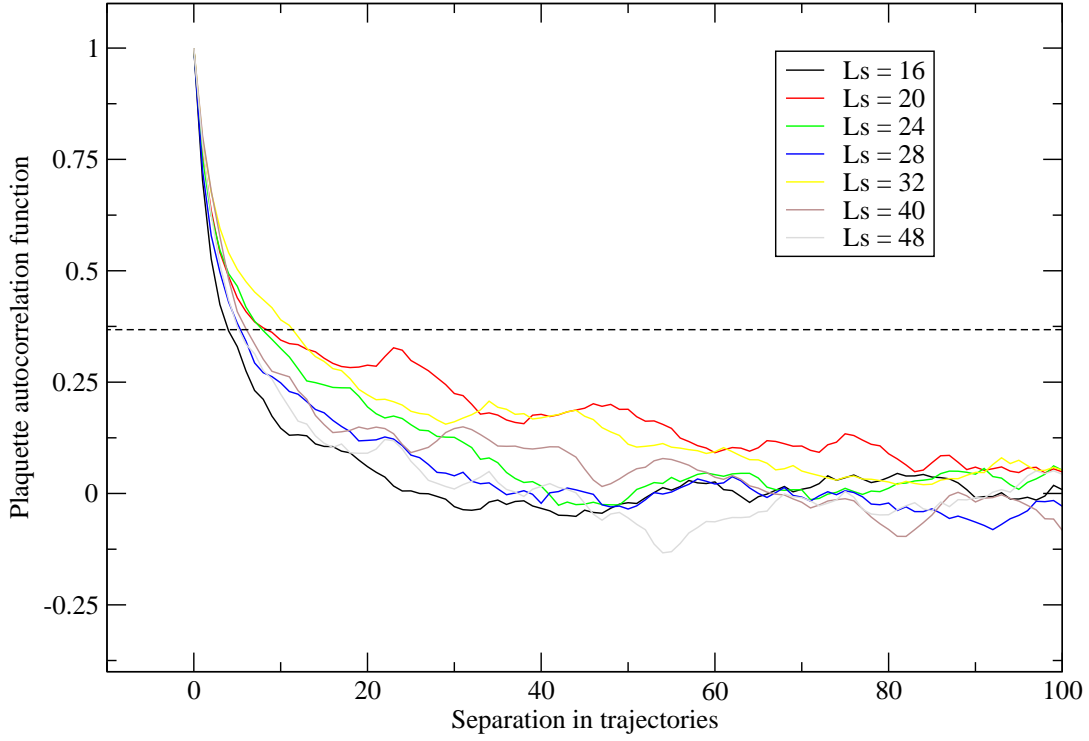}
\caption{%
Autocorrelation function associated with the average plaquette for $16^3\times32$ lattices with $\beta=2.3$ and $m_f=0.02$.
The dashed line indicates the location of $e^{-1}$.
}
\label{fig:plaq_autocorr}
\end{figure}

\begin{figure}[htbp]
\centering
\includegraphics[width=\plotwidth,angle=-90]{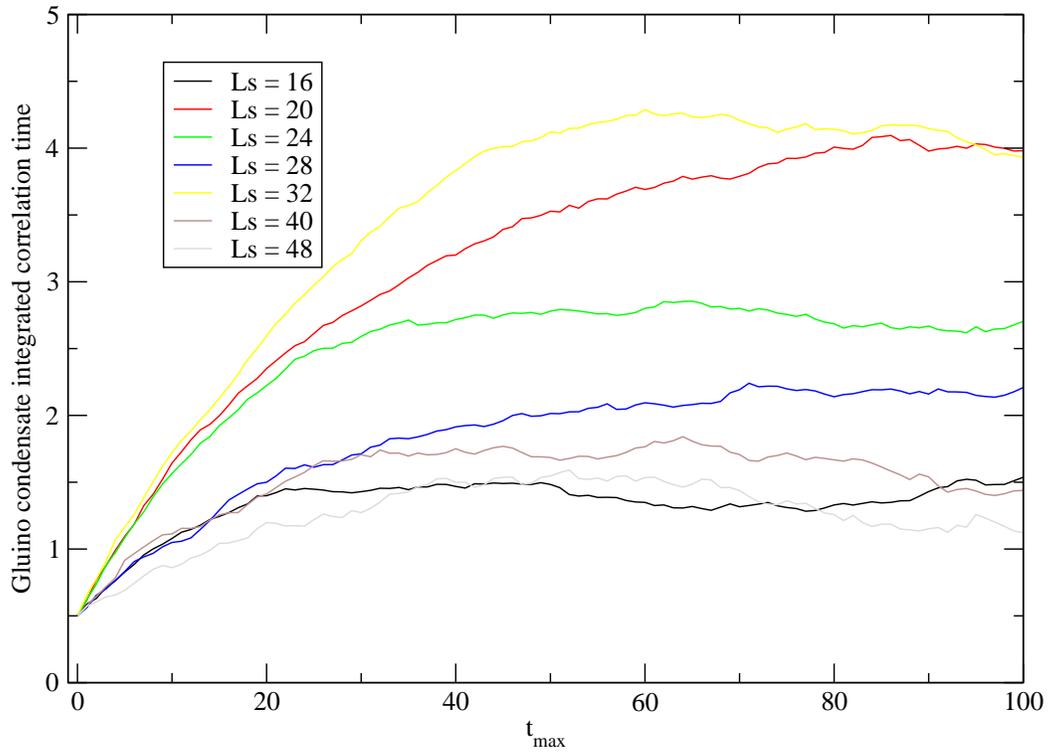}
\caption{%
Integrated correlation time associated with the gluino condensate for $16^3\times32$ lattices with $\beta=2.3$ and $m_f=0.02$.
}
\label{fig:pbp_intcorr}
\end{figure}

\begin{figure}[htbp]
\centering
\includegraphics[width=\plotwidth,angle=-90]{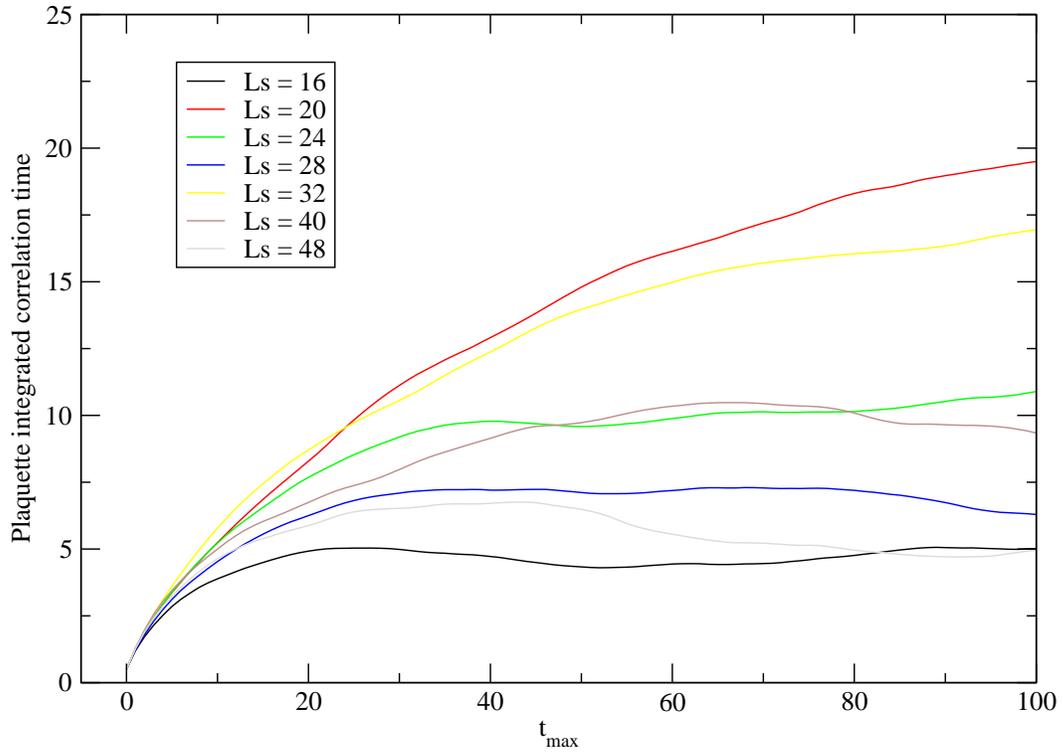}
\caption{%
Integrated correlation time associated with the average plaquette for $16^3\times32$ lattices with $\beta=2.3$ and $m_f=0.02$.
}
\label{fig:plaq_intcorr}
\end{figure}

In order to correctly assess the statistical errors associated with various observable measurements, we first determine the integrated correlation time $\tau_{int}$ associated with the observable $\calO$.
This may be obtained from the normalized auto-correlation function $\rho(\tau)$, defined by:
\beq
\rho(\Delta\tau) = \frac{1}{N-\Delta\tau} \sum_{\tau=1}^{N-\Delta\tau} \frac{\left(\calO_\tau-\mybar \calO\right)\left(\calO_{\tau+\Delta\tau}-\mybar \calO\right)}{\sigma^2_\calO}
\eeq
where
\beq
\mybar \calO = \frac{1}{N} \sum_{\tau=1}^N \calO_\tau\ ,\qquad \sigma^2_\calO = \frac{1}{N} \sum_{\tau=1}^{N} \left(\calO_\tau-\mybar \calO\right)^2
\eeq
The integrated correlation time associated with the observable is computed using:
\beq
\tau^{int}(\tau_{max}) = \frac{1}{2} + \sum_{\tau=1}^{\tau_{max}} \rho(\tau)\ .
\eeq
\Figure{pbp_autocorr} and \Fig{plaq_autocorr} show the dependence of the auto-correlation function associated with the gluino condensate and average plaquette respectively.
The exponential correlation time can be estimated from the value of $\Delta \tau$ at which $\rho(\Delta\tau) \sim e^{-1}$.
The integrated correlation time as a function of $\tau_{max}$ is plotted for these same observables in \Fig{pbp_intcorr} and \Fig{plaq_intcorr}.
Because the gluino condensate was measured using a stochastic estimator, one might expect decorrelation of this observable to occur relatively quickly in comparison to the average plaquette.
However, the $t_{max}$ dependence of the integrated correlation time suggest that longer correlations are merely obscured by the presence of random noise.

\begin{figure}[htbp]
\centering
\includegraphics[width=\plotwidth,angle=-90]{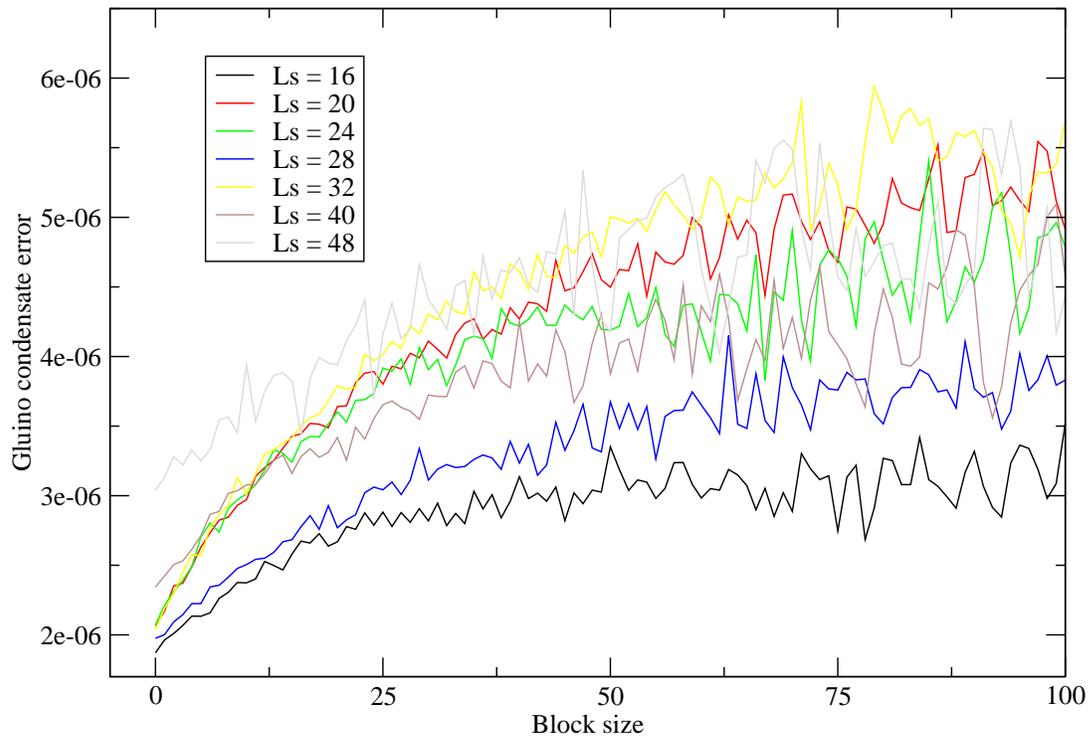}
\caption{%
Gluino condensate error as a function of block size for $16^3\times32$ lattices with $\beta=2.3$ and $m_f=0.02$.
}
\label{fig:pbp_err}
\end{figure}

\begin{figure}[htbp]
\centering
\includegraphics[width=\plotwidth,angle=-90]{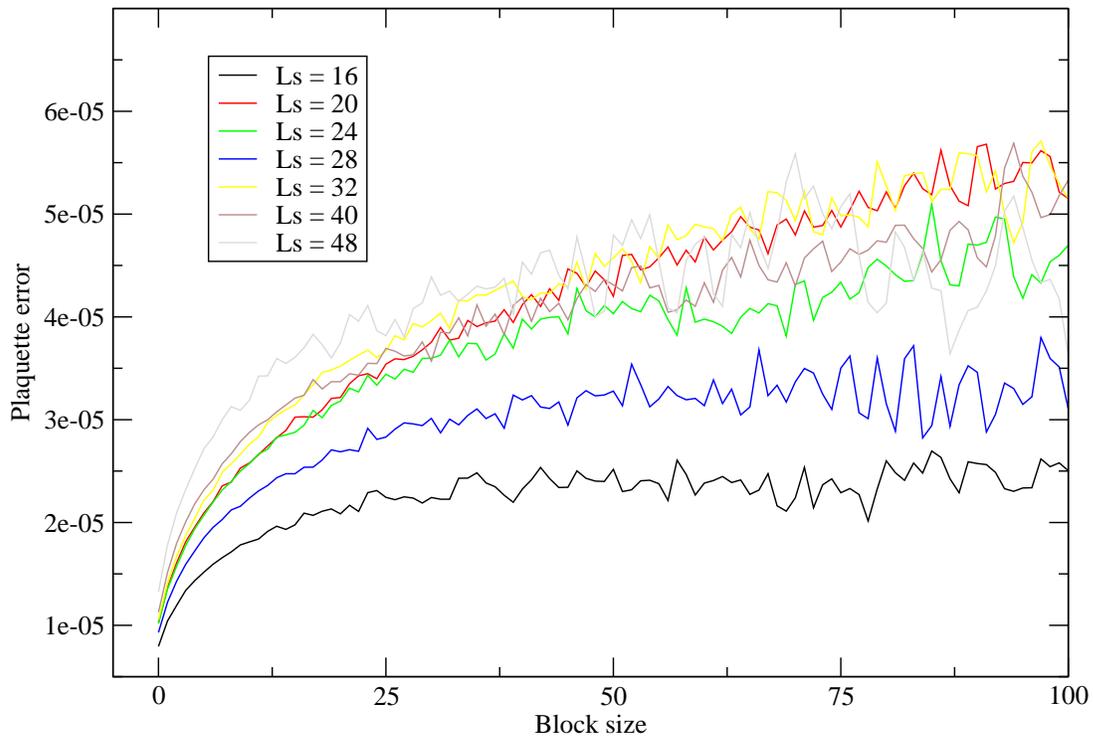}
\caption{%
Average plaquette error as a function of block size for $16^3\times32$ lattices with $\beta=2.3$ and $m_f=0.02$.
}
\label{fig:plaq_err}
\end{figure}

To better understand the correlations, we block our data with block size $N_\calB$ and consider the $N_\calB$ dependence of the errors on each observable.
In \Fig{pbp_err} and \Fig{plaq_err} we plot the error as a function of block size for the gluino condensate and average plaquette.
We find that for the gluino condensate, the decorrelation time is indeed much longer than what is implied by \Fig{pbp_autocorr}.
Based on these considerations, we determine the gluino condensate and plaquette using a block size $N_\calB = 50$, where measurements have been made on every trajectory.
For measurements of the static quark potential, we use a block size $N_\calB=4$, where Wilson loops have been measured on every fifth trajectory.
Unless otherwise noted, all other ensemble averages were performed using measurements made on every five trajectories.

\section{Measurements, Analysis and Results}
\label{sec:3}

\subsection{Residual mass}

\begin{figure}[htbp]
\centering
\includegraphics[width=\plotwidth,angle=-90]{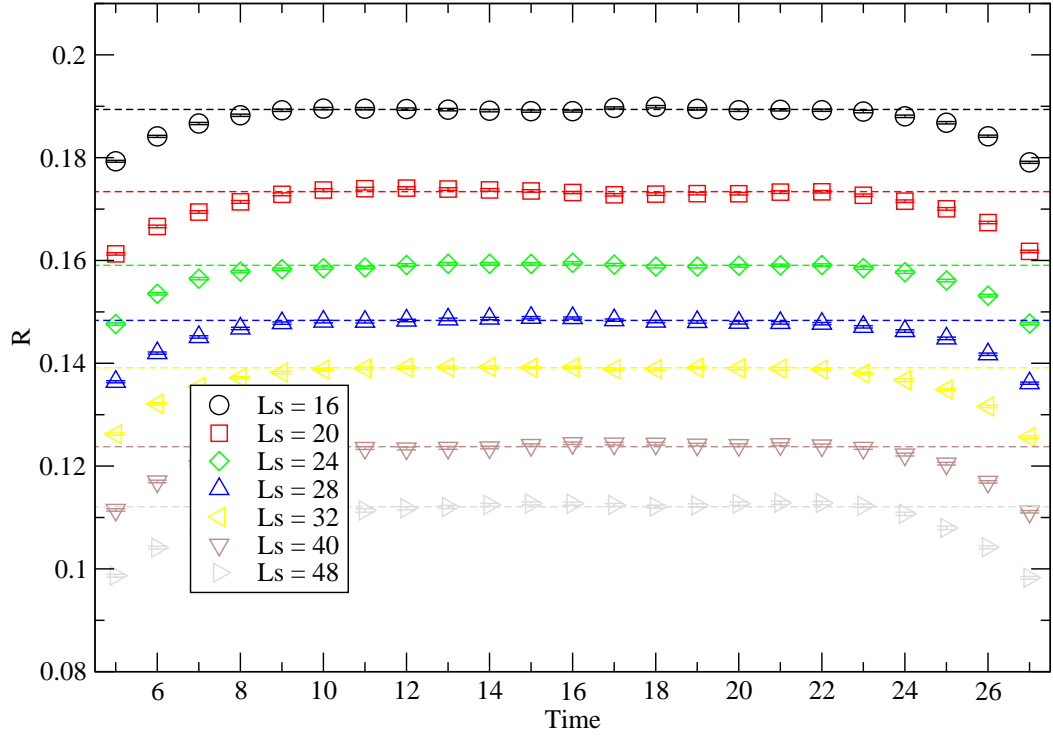}
\caption{%
R as a function of time for $16^3\times32$ lattices with $\beta=2.3$ and $m_f=0.02$.
Solid lines indicate the value of $m_{res}^\prime(m_f)$ obtained from a constant fit to data, as given in \Table{mres_fits}.
}
\label{fig:R}
\end{figure}

\begin{figure}[htbp]
\centering
\includegraphics[width=\plotwidth,angle=-90]{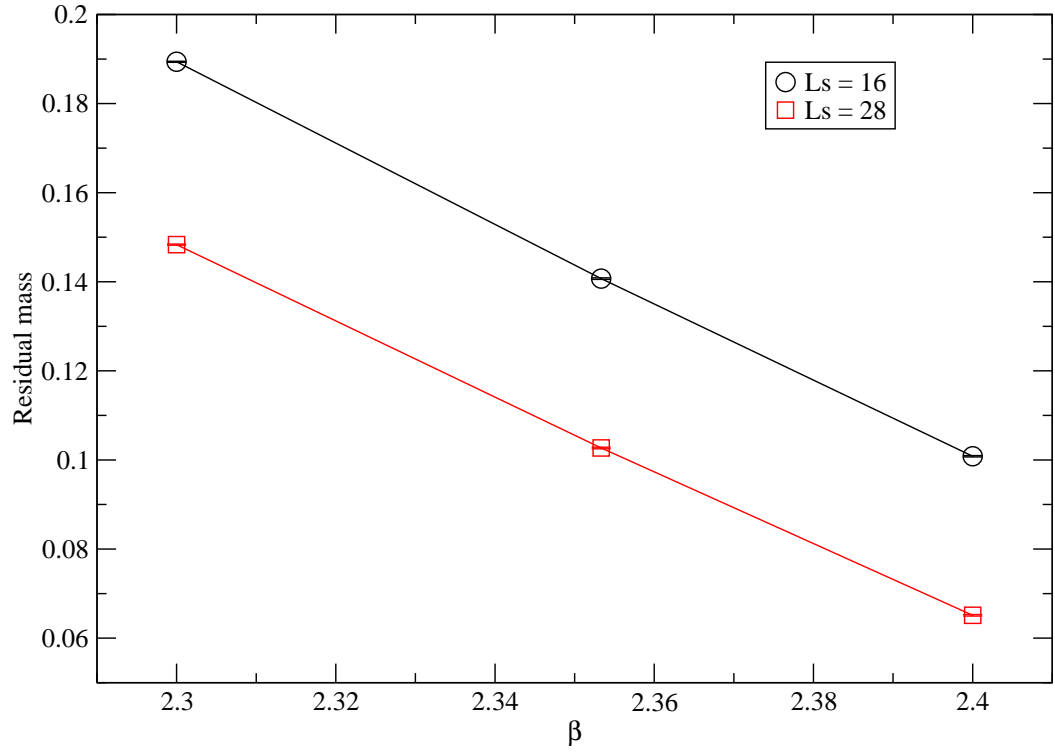}
\caption{%
Residual mass $m_{res}^\prime(m_f)$ as a function of $\beta$ for $16^3\times32$ lattices with $m_f=0.02$.
}
\label{fig:mres_vs_beta}
\end{figure}

Measurements of the residual mass where obtained from $R(t)$ using \Eq{mres_ratio} with wall gluino sources.
The residual mass was computed from a constant fit of $R(t)$ over the plateau region: 10-21.
Fit results may be found in \Table{mres_fits} and in \Fig{R}.
In the large $L_s$ limit, the residual mass is expected to behave as
\beq
m_{res} = a_0 \frac{e^{- a_1 L_s}}{L_s} + \frac{a_2}{L_s}\ ,
\label{eq:mres}
\eeq
where the exponential term is a perturbative contribution which comes from extended modes above the mobility edge \cite{Antonio:2008zz}.
The $L_s^{-1}$ contribution to $m_{res}$ may be attributed to lattice dislocations and is proportional to the density of near unit modes of the five dimensional transfer matrix.
For these simulations, the residual mass is roughly 5-20 times that of the input gluino mass.
\Figure{mres_vs_beta} shows a plot of the residual mass as a function of the coupling.
The strong coupling dependence of $m_{res}$ suggests that the dislocation contribution to the residual mass dominates over that of the perturbative part.

\begin{figure}[htbp]
\centering
\includegraphics[width=\plotwidth,angle=-90]{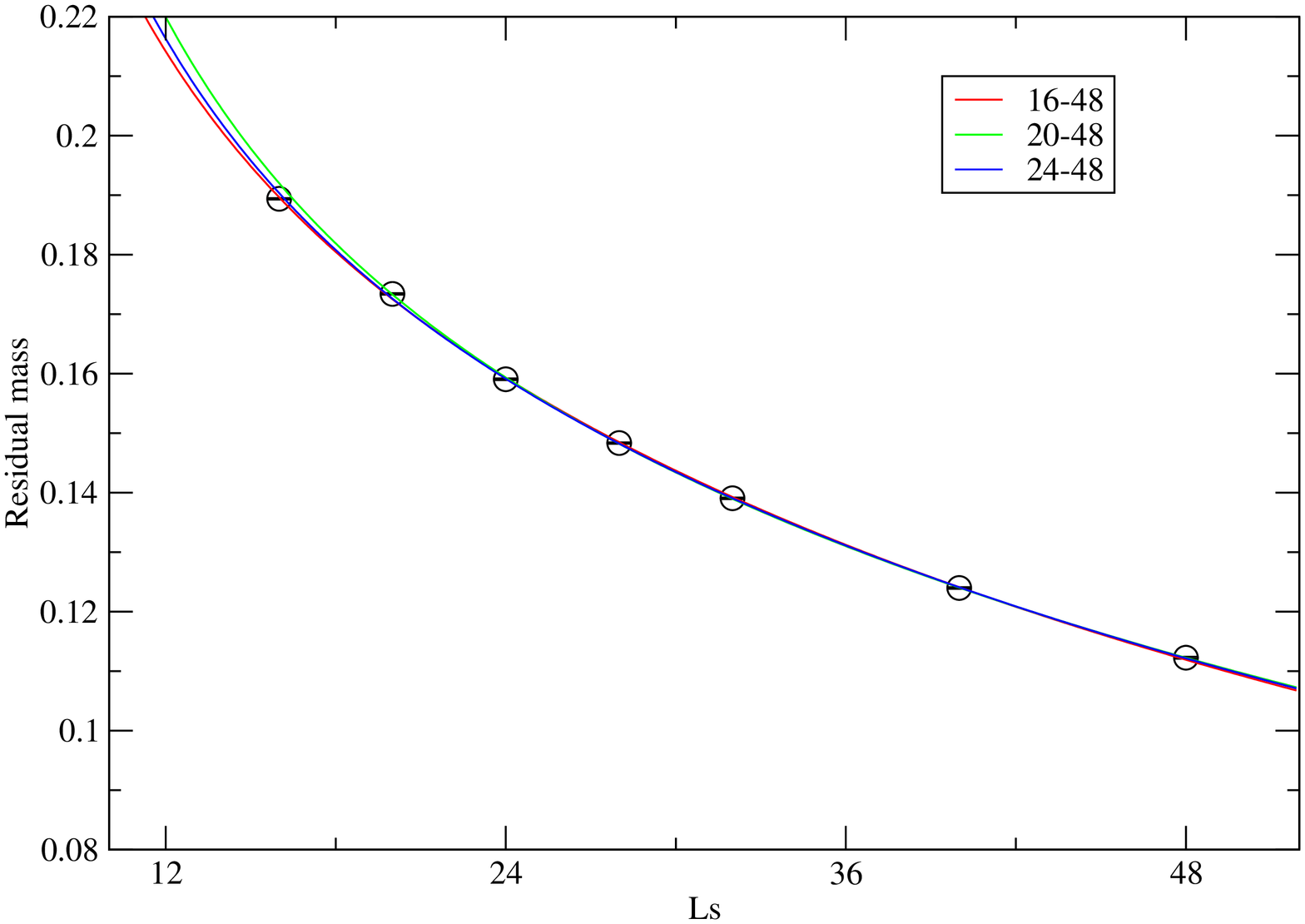}
\caption{%
Residual mass as a function of $L_s$ for $16^3\times32$ lattices with $\beta=2.3$ and $m_f=0.02$.
Solid curves represent fit results obtained for a variety of different $L_s$ ranges.
Details of the fits may be found in \Table{mres_ls_fits}.
}
\label{fig:mres_vs_ls}
\end{figure}

\Figure{mres_vs_ls} shows the dependence of the residual mass on $L_s$ at fixed coupling and $m_f$.
For $\beta=2.3$ and $m_f=0.02$, we fit the residual mass as a function of $L_s$ using the functional form of \Eq{mres}.
Fits were performed for a variety of different fit ranges in order to estimate the systematic errors associated with the fit.
Results may be found in \Table{mres_ls_fits}.
These fits yield a negative value for the coefficient $a_0$, which is inconsistent with the na\'ive expectation that $a_0$ should be positive.
The na\'ive expectation, however, assumes that we are simulating the theory at couplings which are sufficiently close to the continuum limit so that the low energy identity \Eq{midpoint_wall_identity} holds.
Provided the underlying assumptions made in the calculation of $m_{res}^\prime(m_f)$ are correct, one possible explanation for a negative value for $a_0$ is that higher order contributions to \Eq{symanzik} have contaminated our estimate of the residual mass obtained from \Eq{mres_ratio}.

\subsection{Static potential}

\begin{figure}[htbp]
\centering
\includegraphics[width=\plotwidth,angle=-90]{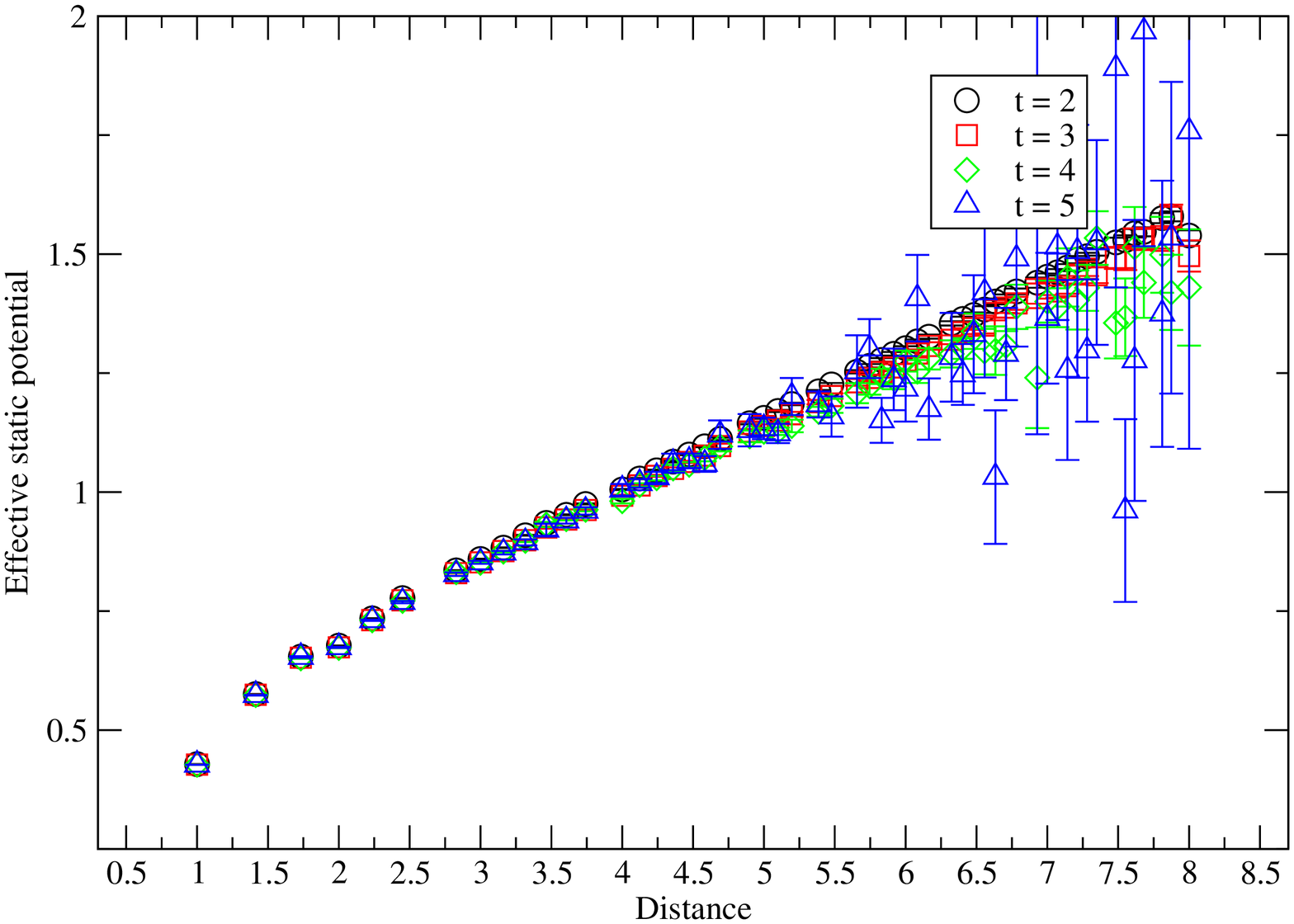}
\caption{%
Effective static potential $V_{eff}(\bfx,t)$ as a function of distance $|\bfx|$ at fixed time $t$ for a $16^3\times32$ lattice with $\beta=2.3$, $L_s=16$ and $m_f=0.02$.
}
\label{fig:effective_potential_3.45}
\end{figure}

\begin{figure}[htbp]
\centering
\includegraphics[width=\plotwidth,angle=-90]{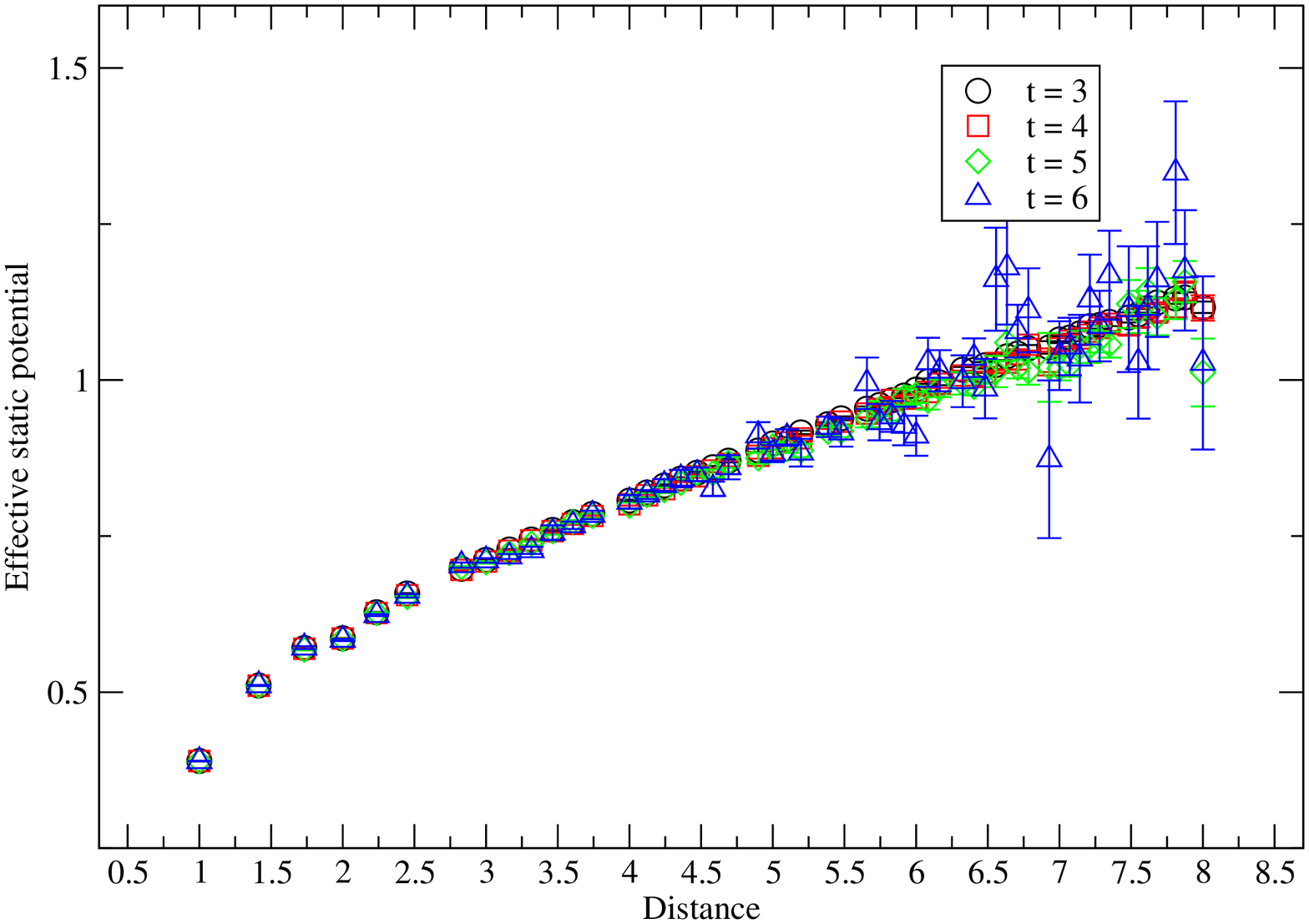}
\caption{%
Effective static potential $V_{eff}(\bfx,t)$ as a function of distance $|\bfx|$ at fixed time $t$ for a $16^3\times32$ lattice with $\beta=2.35\bar3$, $L_s=16$ and $m_f=0.02$.
}
\label{fig:effective_potential_3.53}
\end{figure}

\begin{figure}[htbp]
\centering
\includegraphics[width=\plotwidth,angle=-90]{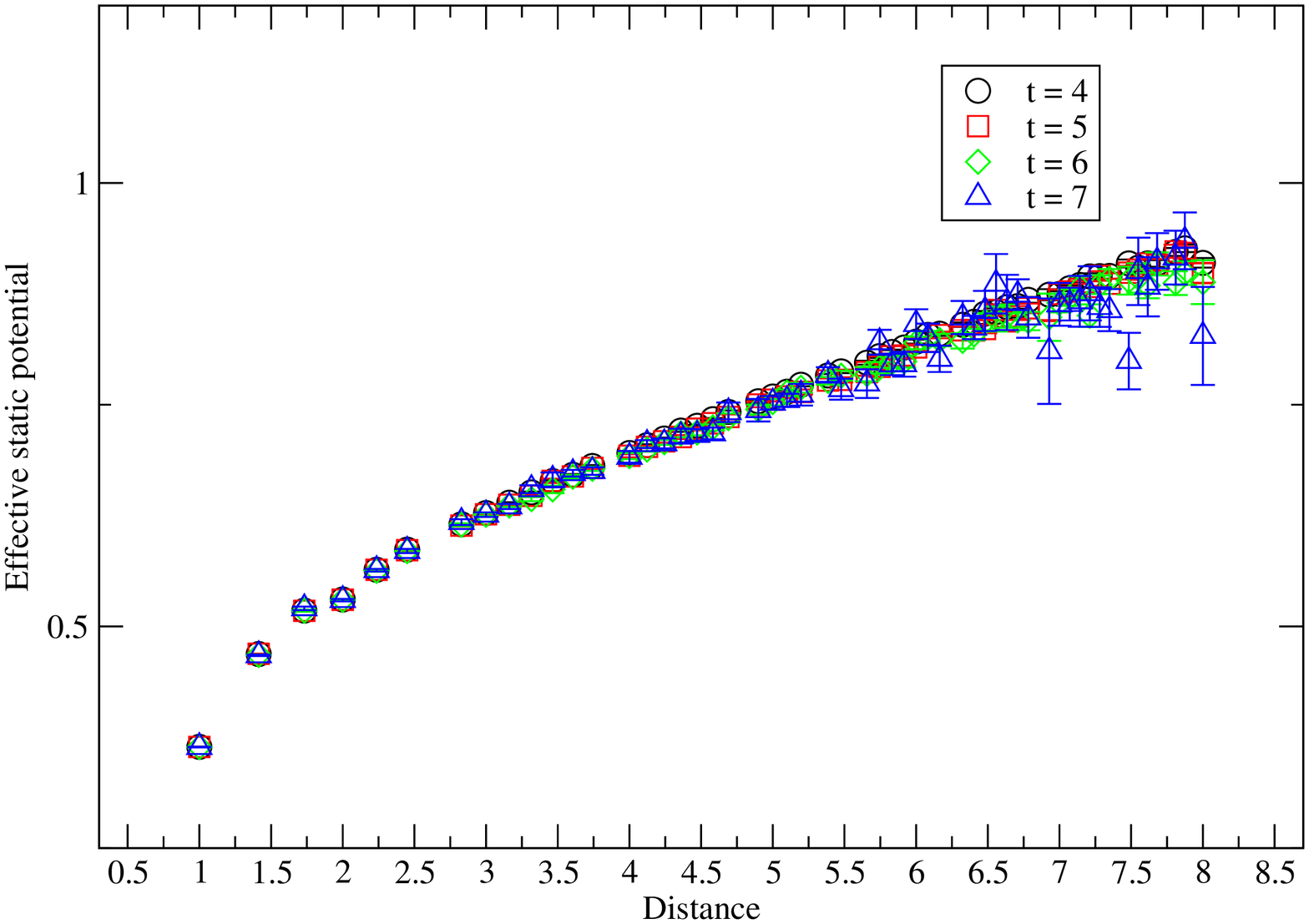}
\caption{%
Effective static potential $V_{eff}(\bfx,t)$ as a function of distance $|\bfx|$ at fixed time $t$ for a $16^3\times32$ lattice with $\beta=2.4$, $L_s=16$ and $m_f=0.02$.
}
\label{fig:effective_potential_3.60}
\end{figure}

Here we describe the analysis of our static potential measurements.
To begin, we first determined the plateau region of \Eq{wilson_loop} over which excited state contamination becomes negligible by investigating the effective potential \Eq{effective_potential} as a function of distance for a variety of time slice values.
\Figure{effective_potential_3.45}, \Fig{effective_potential_3.53} and \Fig{effective_potential_3.60} show the distance dependence of $V(\bfx,t)$ for each of the three values of $\beta$, and with $L_s=16$ and $m_f=0.02$.
For $\beta=2.3$, we take the plateau region to be the time range: 4-8, whereas for $\beta=2.35\bar3$ and $2.4$ we use the ranges: 5-9 or 5-10.
For fixed distances $|\bfx|$, the Wilson loops were then fit as a function of time to the formula \Eq{wilson_loop} within the previously determined time interval.
Finally the extracted potentials $V(\bfx)$ were fit as a function of distance to the Cornell potential given by \Eq{potential}.
\Table{hqpot} provides the fit results and respective errors which were determined by a jackknife analysis.
Systematic errors in this fitting procedure may be estimated by varying the upper and lower limits of the fit ranges.
These yield variations in the Sommer scale, however, which are comparable--if not smaller than--the statistical errors.

\begin{figure}[htbp]
\centering
\includegraphics[width=\plotwidth,angle=-90]{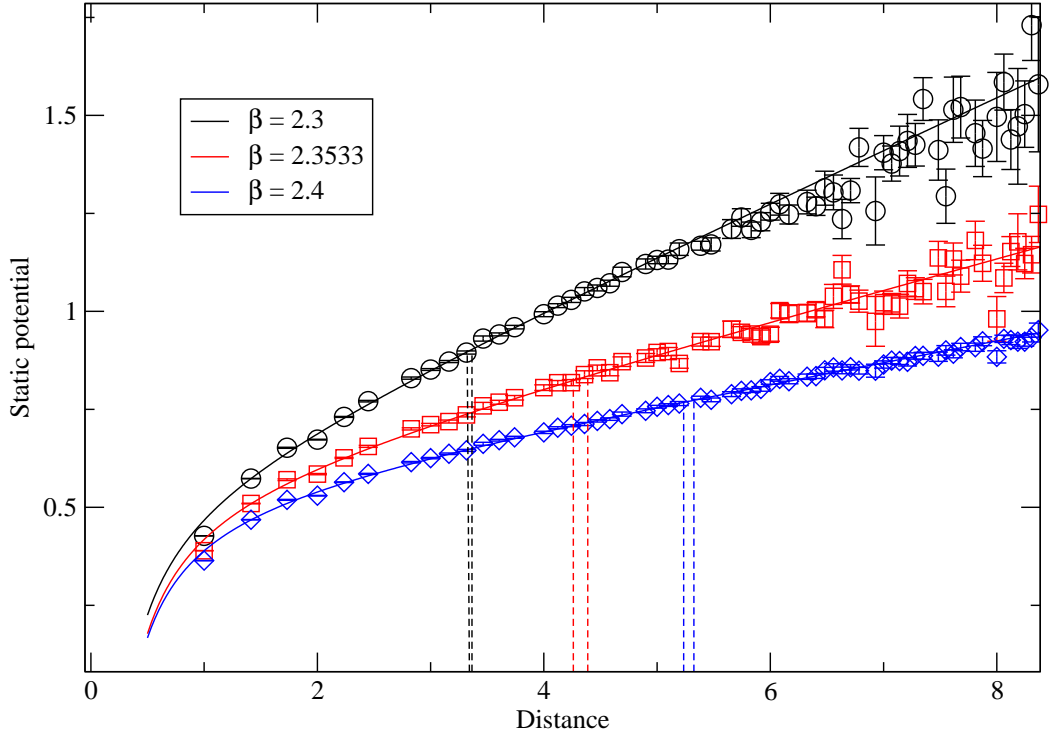}
\caption{%
Static potential as a function of distance for $16^3\times32$ lattices with $L_s=16$ and $m_f=0.02$.
The solid curves represent fits to the potential given in \Table{hqpot} and dashed lines indicate $1\sigma$ error bars for the corresponding values for the Sommer scale obtained from \Eq{sommer_scale}.
}
\label{fig:potential}
\end{figure}

\begin{figure}[htbp]
\centering
\includegraphics[width=\plotwidth,angle=-90]{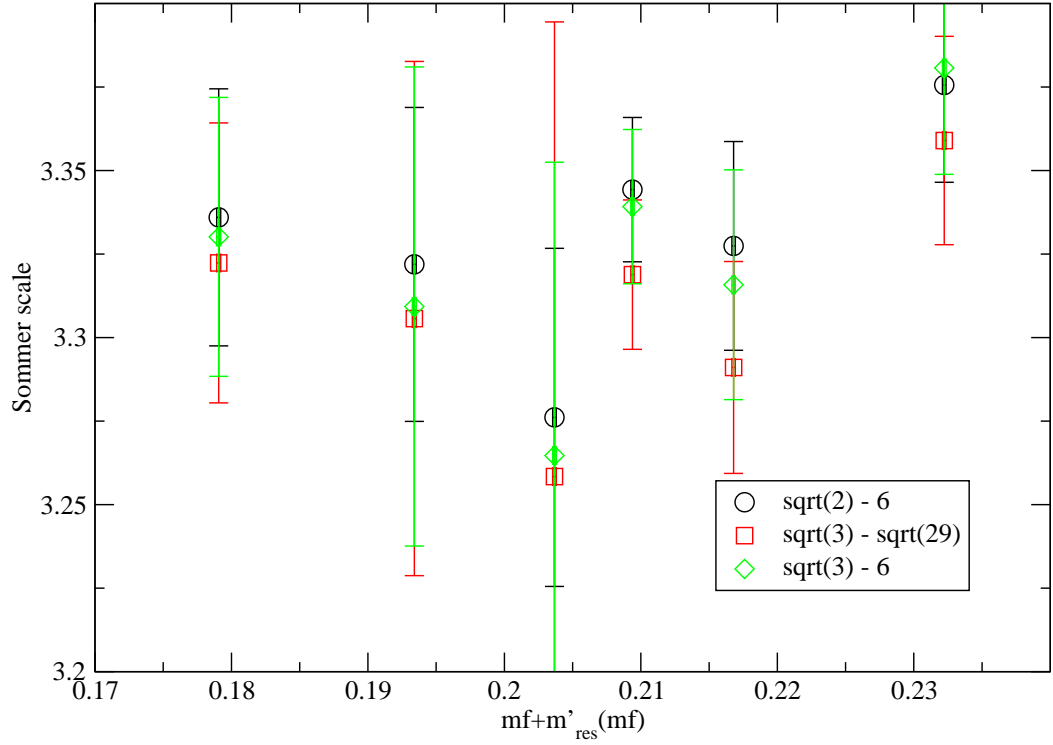}
\caption{%
Sommer scale as a function of $m_f+m_{res}^\prime(m_f) \approx m_g$ for $16^3\times32$ lattices with $\beta=2.3$.
}
\label{fig:r0_vs_mass}
\end{figure}

\Figure{potential} shows a plot of the static potential at the three different values of the coupling; the corresponding fit parameters may be found in \Table{hqpot}.
From $\sigma$ and $\alpha$ we determine the Sommer scale defined by \Eq{sommer_scale}.\footnote{Measurements of the Sommer scale were performed in collaboration with I. Mihailescu.}
As indicated in \Table{hqpot}, the Sommer scale ranges from between $r_0 \approx 3.3$ for $\beta=2.3$ and $r_0\approx 5.2$ for $\beta=2.4$.
Our static potential results provide evidence that the theory is confining, which is consistent with expectations.
\Figure{r0_vs_mass} shows a plot of $r_0$ as a function of $m_f+m_{res}^\prime(m_f)\approx m_g$ for $\beta=2.3$.
From this plot it appears that the Sommer scale has little discernible gluino mass dependence at the current level of statistics.

\subsection{Gluino condensate}
\label{sec:3_2}

In this section we describe the chiral limit extrapolation of the gluino condensate.
This extrapolation requires taking both the $L_s\to \infty$ and $m_f\to0$ limits in some fashion.
We begin by discussing the former.

For fixed $m_f$, we fit the gluino condensate defined in \Eq{condensate} as a function of $L_s$ using the best available, theoretically motivated fit formula:
\beq
b_0 + b_1 \frac{e^{-b_2 L_s}}{L_s}\ ,
\label{eq:pbp_vs_ls}
\eeq
which is based upon the functional dependence of $m_{res}$ on $L_s$.
Note, however, that the dislocation contribution to $m_{res}$ which appears in \Eq{mres} is absent in \Eq{pbp_vs_ls}.
The reason for this omission may be understood by observing that the chiral condensate is dominated by contributions from UV modes, whereas the dislocation term appearing in $m_{res}$ may predominantly be attributed to low energy phenomena \cite{Cheng:2008ge}.

\begin{figure}[htbp]
\centering
\includegraphics[width=\plotwidth,angle=-90]{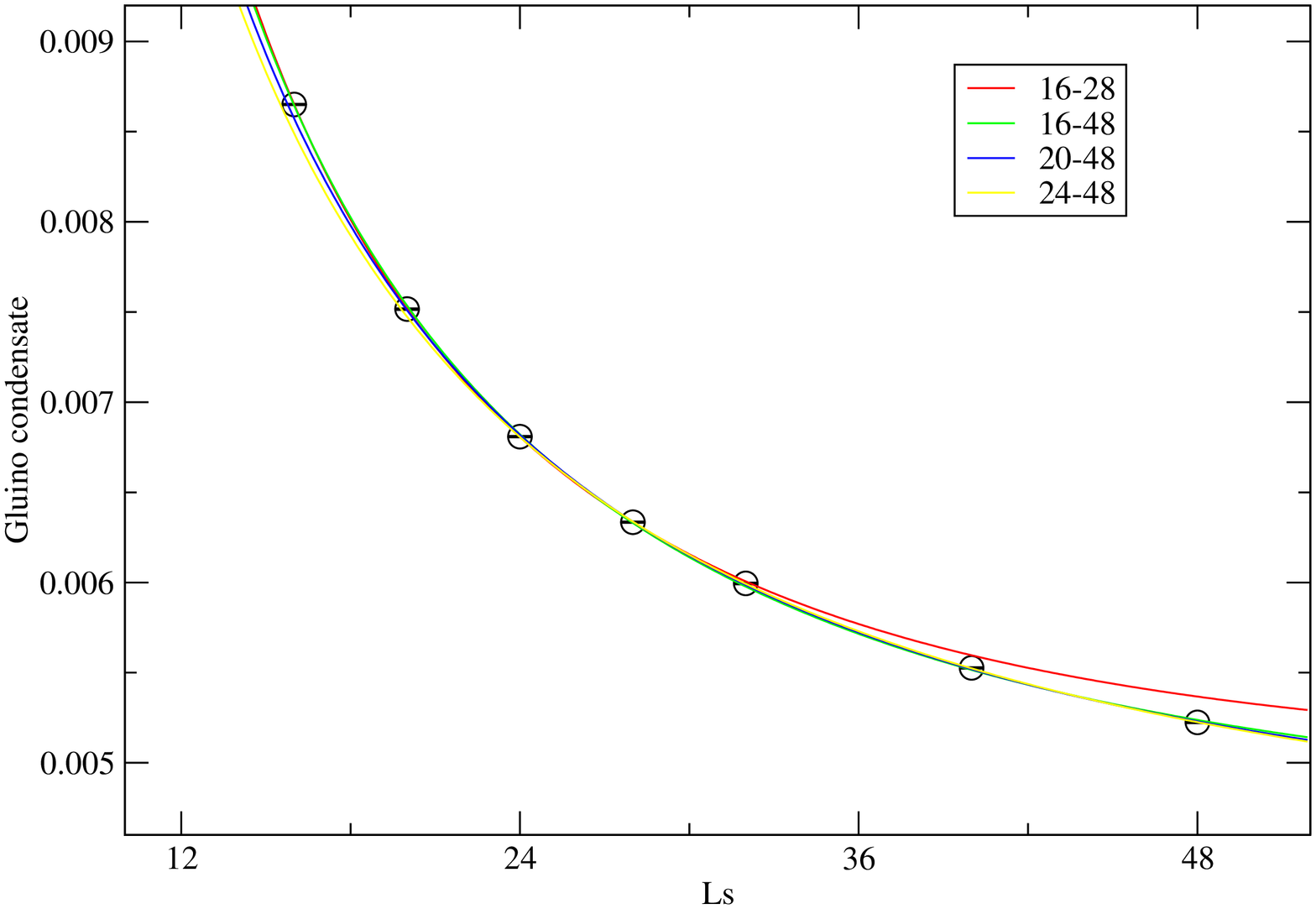}
\caption{%
Gluino condensate as a function of $L_s$ for $16^3\times32$ lattices with $\beta=2.3$ and $m_f=0.02$.
Solid curves represent fit results obtained for a variety of different $L_s$ ranges.
Details of the fits may be found in \Table{pbp_ls_fits}.
}
\label{fig:pbp_vs_ls}
\end{figure}

Before considering the chiral limit extrapolation of the condensate we investigate the $L_s$ dependence of the gluino condensate for $\beta=2.3$ and a fixed input gluino mass $m_f=0.02$.
The primary purpose of this investigation is two-fold.
First, we would like to test our assumptions leading to \Eq{pbp_vs_ls} by checking that the fit value for $b_2$ is consistent with that of $a_1$ obtained in \Eq{mres}.
Second, we wish to ascertain the systematic errors associated with an $L_s\to \infty$ extrapolation of the condensate, which come from the limited range of $L_s$ values used in the fit.
Results of our fits to \Eq{pbp_vs_ls} are provided in \Table{pbp_ls_fits} and \Fig{pbp_vs_ls}.
From these fit results we conclude that the fit over the $L_s$ range: 16-28 overestimates the $L_s\to \infty$ value of the chiral condensate at fixed $m_f$ by approximately 25\% compared to our most reliable estimates which are obtained from the fit ranges: 20-48 and 24-48.
We find large variations in the parameter $b_2$ with the $L_s$ fit range, however the results are within a factor of two of $a_1$ obtained from the residual mass extrapolations.

\begin{figure}[htbp]
\centering
\includegraphics[width=\plotwidth,angle=-90]{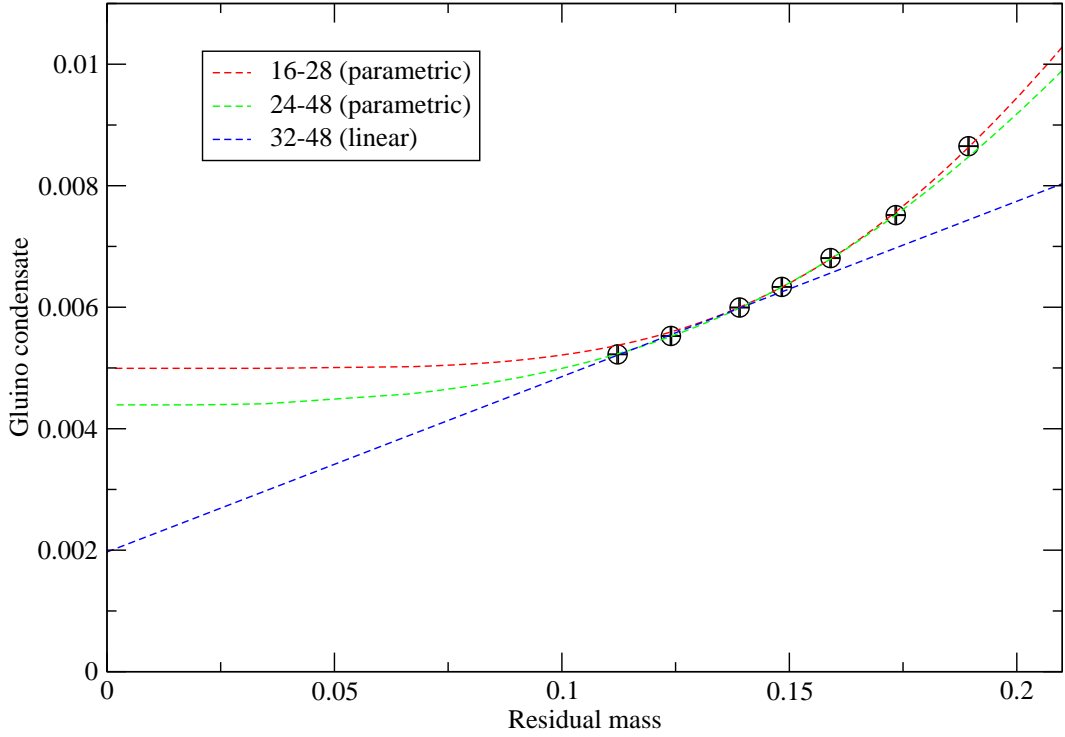}
\caption{%
Gluino condensate as a function of the residual mass for $16^3\times32$ lattices with $\beta=2.3$ and $m_f=0.02$.
The two curve labeled ``parametric'' represent curves obtained from the fit results obtained in \Table{mres_ls_fits} and \Table{pbp_ls_fits}.
The fit parameters used for the residual mass are the same, and given by the $L_s$ range: 24-48.
Parameters used for the condensate are indicated on the plot.
The curve labeled ``linear'' represents a linear extrapolation of the $L_s=32, 40$ and $48$ results.
}
\label{fig:pbp_vs_mres}
\end{figure}

As an alternative approach, one may consider a linear extrapolation of the gluino condensate as a function of the residual mass, as was performed in \cite{Giedt:2008xm}, in order to obtain an $L_s\to\infty$ value for the condensate at fixed $m_f$.
However such an approach assumes the same functional dependence on $L_s$ for both the condensate and the residual mass.
While indeed for sufficiently large $L_s$, each of these quantities will scale as $L_s^{-1}$ and such a scenario may be achieved, for the values of $L_s$ used in this study, it is believed that the chiral condensate is dominated by a UV divergent term which is proportional to the perturbative contribution to the residual mass.
Here, we compare the results of a linear extrapolation of the condensate as a function of the residual mass using the three data points which lie closest to the $L_s=\infty$ limit (i.e., $L_s = 32, 40$ and $48$).
Results from this linear extrapolation are plotted in \Fig{pbp_vs_mres} and yield a condensate value of $0.00197(4)$ at $m_f=0.02$.
Also plotted in \Fig{pbp_vs_mres} are parametric plots of the gluino condensate as a function of the residual mass using \Eq{mres}, \Eq{pbp_vs_ls}, and the fit parameters obtained in \Table{mres_ls_fits} and \Table{pbp_ls_fits}.
Specifically we use the $L_s$ range: 24-48 fit parameters for $m_{res}$ and the $L_s$ range: 24-48 and 28-48 fit parameters for the condensate.
A linear extrapolation of the condensate as a function of the residual mass yields a value which underestimates our $L_s$ extrapolation by approximately 100-150\%.

Finally we consider the chiral limit extrapolation of the condensate.
Such an extrapolation may be achieved by considering the individual $L_s$ and $m_f$ dependence of the condensate, as was considered in \cite{Fleming:2000fa}.
Here, we perform chiral limit extrapolations of the gluino condensate at a single lattice spacing ($\beta = 2.3$) using two different limit orders.
First, we perform a linear $m_f \to0$ extrapolation of the gluino condensate at fixed $L_s$ using the formula
\beq
c_0 + c_1 m_f\ ,
\label{eq:pbp_vs_mf}
\eeq
followed by an $L_s \to \infty$ extrapolation of the $m_f=0$ result using \Eq{pbp_vs_ls}.
We believe that this approach is more transparent than, for instance, a linear extrapolation of the condensate as a function of $m_{res}$ (followed by a linear $m_f\to0$ extrapolation), and most importantly, it correctly accounts for the functional form of \Eq{pbp_vs_ls}.

\begin{figure}[htbp]
\centering
\includegraphics[width=\plotwidth,angle=-90]{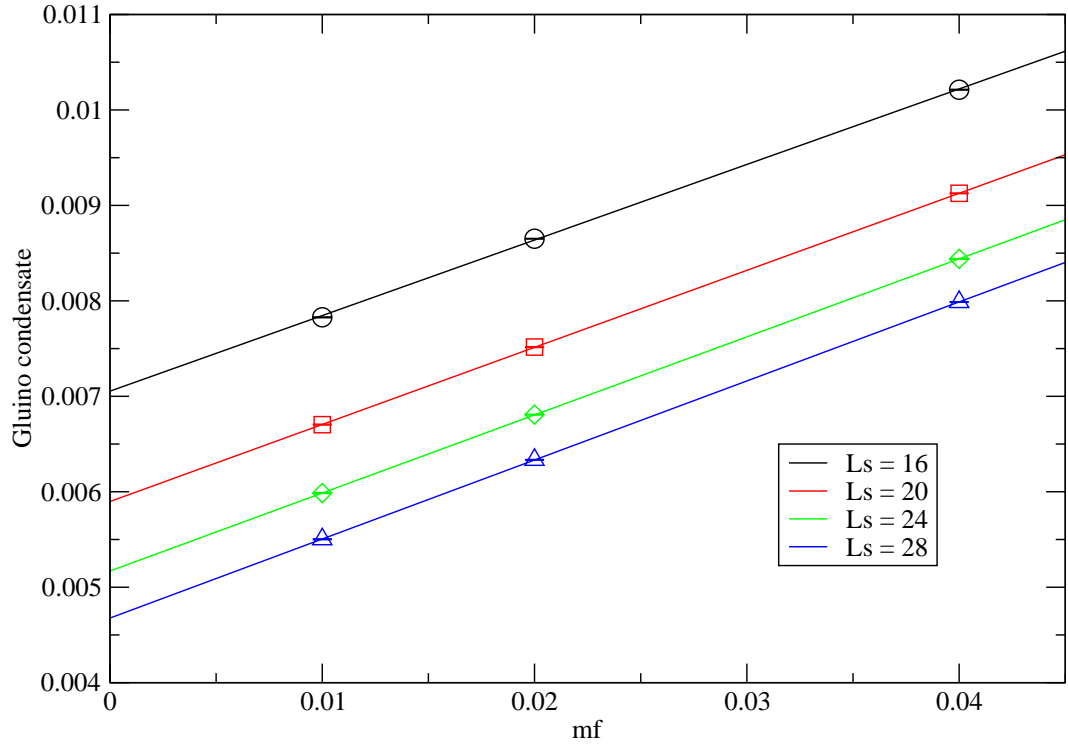}
\caption{%
Gluino condensate as a function of $m_f$ for $16^3\times32$ lattices with $\beta=2.3$.
Solid curves represent fit results which are obtained from \Table{pbp_mf_fits}.
}
\label{fig:pbp_vs_mf}
\end{figure}

\begin{figure}[htbp]
\centering
\includegraphics[width=\plotwidth,angle=-90]{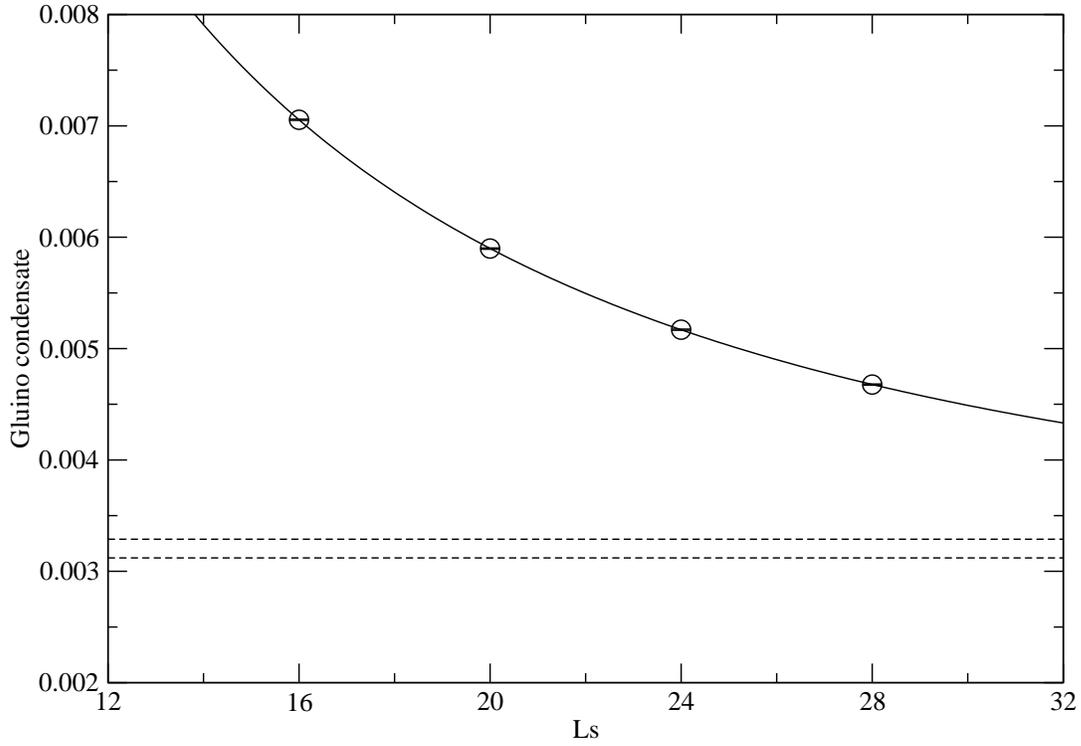}
\caption{%
$m_f =0$ extrapolated values of the gluino condensate as a function of $L_s$ for $16^3\times32$ lattices with $\beta=2.3$.
The solid curve represents the fit result obtained from \Table{pbp_chiral_limit_fit}.
The dashed lines indicate the $1\sigma$ statistical error bars associated with the chirally extrapolated value of the condensate.
}
\label{fig:pbp_chiral_limit}
\end{figure}

Following the double extrapolation procedure outlined above, we obtain an unrenormalized value for the gluino condensate in the chiral limit at finite lattice spacing.
Results from these fits are tabulated in \Table{pbp_mf_fits} and \Table{pbp_chiral_limit_fit} and plotted in \Fig{pbp_vs_mf} and \Fig{pbp_chiral_limit}, and yield the chiral limit value of 0.00320(9) for the condensate.
Chiral extrapolations have also been performed by reversing the order of limits and yield consistent results with comparable statistical error bars.
The fits of the condensate as a function of $L_s$ were performed over a fit range: 16-28 and, as such, we expect our chiral limit result to over estimate the true answer by approximately 25\%, as was found for our extended $L_s$ fits at $\beta=2.3$ and $m_f=0.02$, which are described above.
We have performed additional fits using other, phenomenologically motivated fit formulae to describe the $L_s$ dependence of the gluino condensate (e.g., \Eq{pbp_vs_ls}, without the $L_s^{-1}$ prefactor in the second term).
Such fits yield an approximate 20\% variation in the chirally extrapolated value of the gluino condensate as compared to that obtained by using \Eq{pbp_vs_ls}.

\subsection{Spectrum of the Hermitian DWF Dirac operator}

Here we study the eigenvalues and eigenvectors of the Hermitian DWF Dirac operator.
The primary purpose of this study is to establish to what degree the five dimensional theory reproduces the correct four-dimensional, continuum low energy physics and to obtain an independent measure of the residual mass and gluino condensate.
To achieve this aim, we measure the lowest 64 eigenvalues and eigenvectors of the Hermitian DWF Dirac operator on $8^3\times8$ lattices using the method of Kalkreuter-Simma (KS) \cite{Kalkreuter:1995mm}.
In this method, the Hermitian matrix $D_H^2$ is first diagonalized using a modified Rayleigh-Ritz diagonalization procedure, where we have exploited the relationship between degenerate eigenvectors $\Psi_{\Lambda_H}$ and $\Psi_{\Lambda_H}^c$ in order to eliminate any unnecessary minimizations of the Ritz functional.
Subsequently, the matrix $D_H$ is diagonalized using Jacobi's method within the subspace obtained from diagonalizing $D_H^2$.
These steps are then iterated until specified stopping criteria have been achieved.
A detailed description of the procedure used to diagonalize $D_H$ and the stopping criteria may be found in \cite{Blum:2001qg}.
For our purposes, the process was considered converged when the change in the eigenvalues between iterations was less than $1\times10^{-7}$.
We check this choice by increasing the stopping condition to $1\times10^{-10}$ on a test configuration and determined that eigenvalues changed by at most 0.02\% over the the range of eigenvalues considered.

The eigenvectors of $D_H$ obtained from the KS method are ordered such that the magnitude of $\Lambda_H$ are ascending with eigenvector number.
In all of the analysis that follows we use only the first 60 of the 64 eigenvalues obtained, since the last few may be unreliable \cite{Blum:2001qg}.
Since there is a two-fold degeneracy in the spectrum due to \Eq{hermitian_dirac_op_properties}, this yields a total of 30 non-degenerate eigenvectors and eigenvalues.
All measurements in this work were performed on $8^3\times8$ lattices with $\beta=2.3$ and a sea gluino mass: $m_f = 0.02$.
In physical units, the lowest 60 eigenvalues for a typical configuration range roughly between $0\leq r_0 |\Lambda_H| \lesssim 0.07$ for the ensembles and valence masses under consideration.

\begin{figure}[htbp]
\centering
\includegraphics[width=\plotwidth,angle=-90]{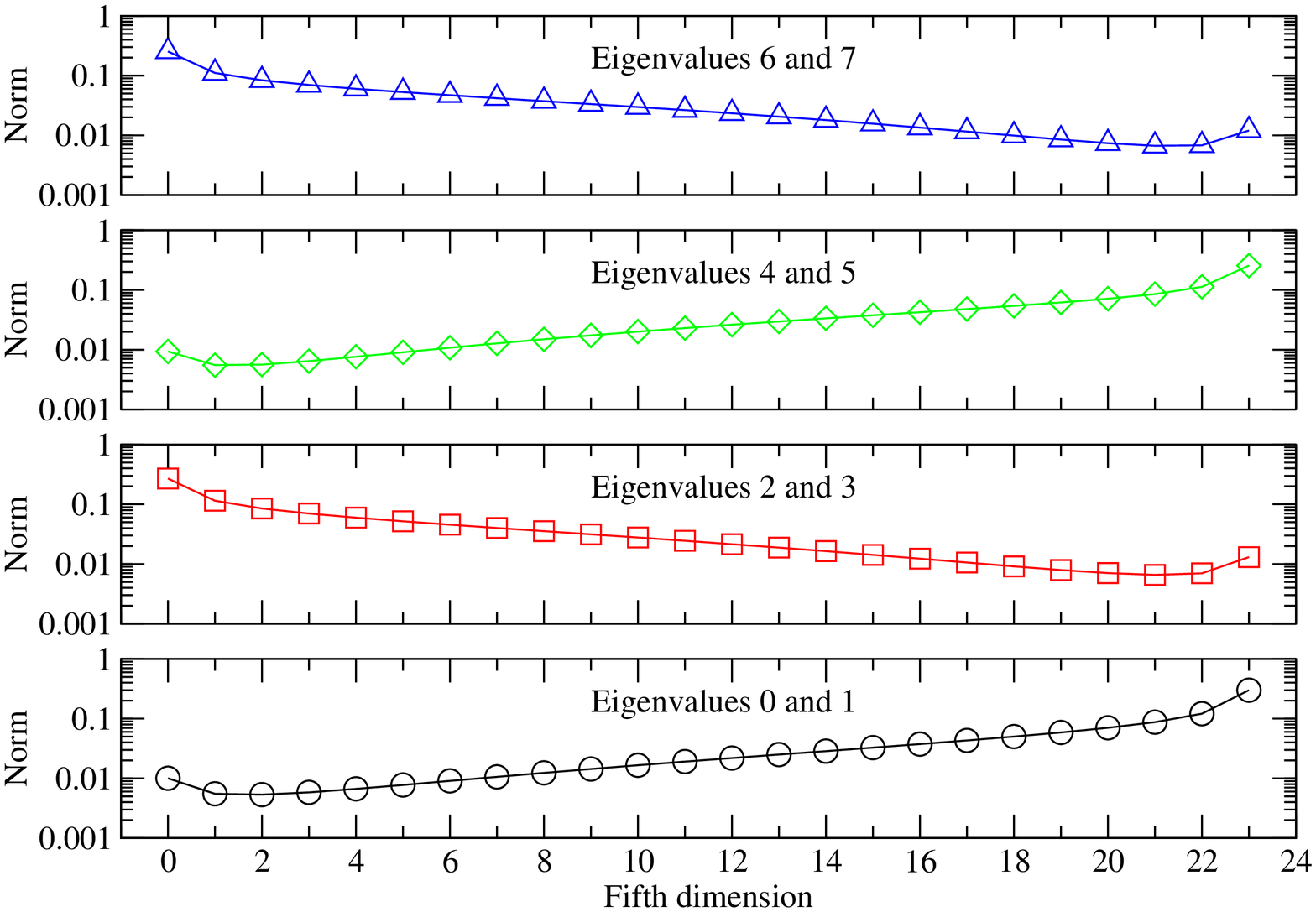}
\caption{%
Four dimensional norms $\calN(s)$ associated with the first eight eigenfunctions of $D_H$, measured on a typical background gauge field configuration generated on an $8^3\times8$ lattice with $\beta = 2.3$, $L_s=24$ and $m_f=m_v=0.02$.
}
\label{fig:four_dim_norms}
\end{figure}

\begin{figure}[htbp]
\centering
\includegraphics[width=\plotwidth,angle=-90]{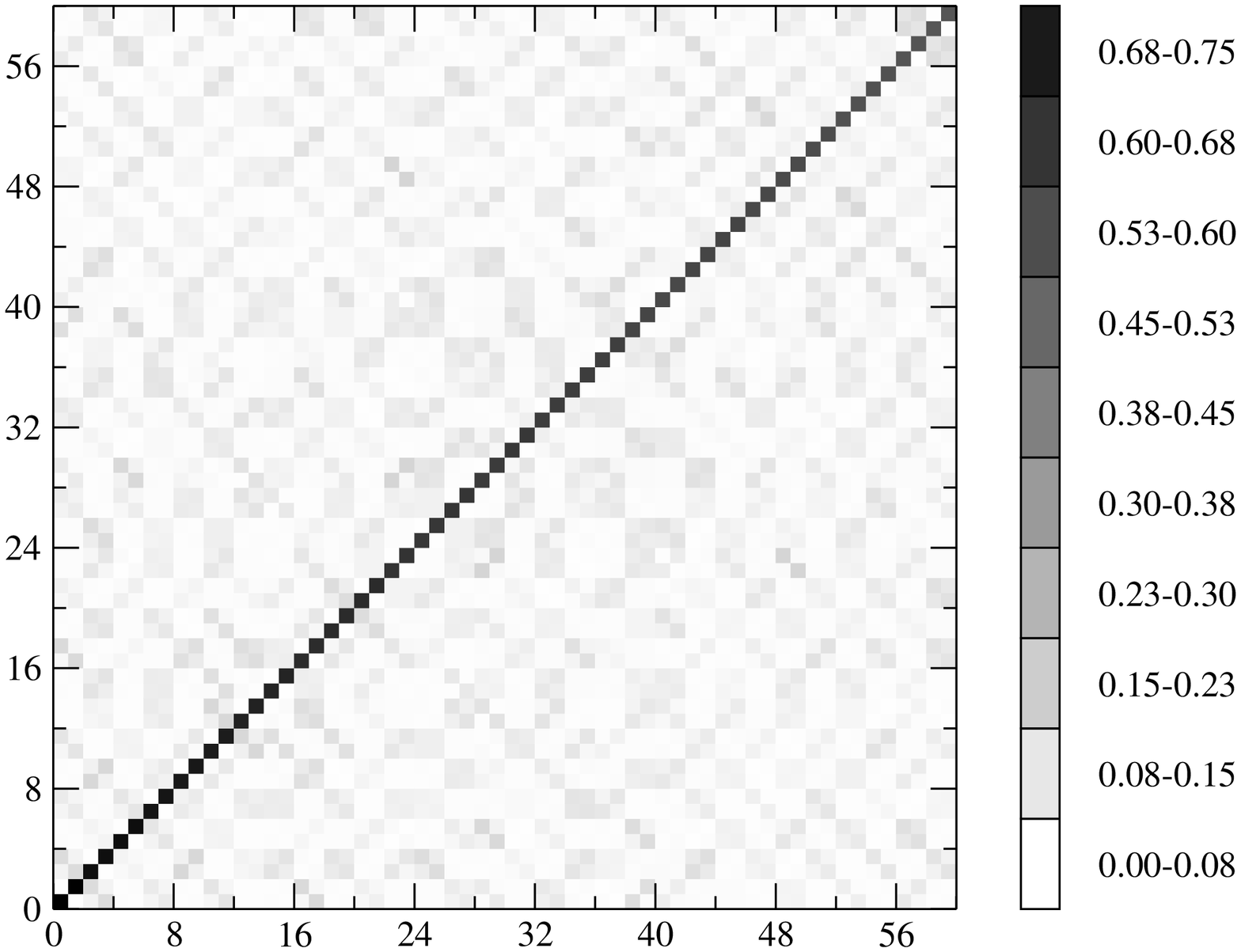}
\caption{%
Magnitude of the matrix elements $\langle \Lambda_H^\prime |\Gamma_s | \Lambda_H \rangle$, measured on a typical background gauge field configuration generated on a $8^3\times8$ lattice with $\beta = 2.3$, $L_s=24$ and $m_f=m_v=0.02$.
}
\label{fig:phys_gamma5}
\end{figure}

The lowest 60 eigenvectors were used to compute the matrix elements of the physical gamma matrix $\Gamma_s$ defined in \Eq{phys_gamma5}, as well as to determine the localization of wave functions in the fifth dimension, which may be characterized by $\calN_{\Lambda_H}(s)$ as defined in \Eq{four_dim_norm}.
\Figure{four_dim_norms} and \Fig{phys_gamma5} shows the results of these calculations for a representative background gauge field configuration in the ensemble: $\beta=2.3$, $L_s=24$ and $m_f=0.02$, with $m_v=0.02$.
As may be seen in these figures, the eigenvectors are exponentially localized on both boundaries of the fifth dimension and alternate with increasing $\Lambda_H$.
The near-diagonal nature of $\Gamma_s$ confirms that the low-lying eigenvectors of $D_H$ are near chiral modes, however these states are not necessarily near-zero modes of $D_H$.
We find that in every instance:
\beq
\sgn\left(\Lambda_H\right) =  \sgn\left(\left\langle \Lambda_H | \Gamma_s | \Lambda_H \right\rangle\right)\ ,
\label{eq:something}
\eeq
which is entirely consistent with continuum predictions in the regime where $\lambda < m_g$, according to \Eq{gamma5_matrix_elements}.

\begin{figure}[htbp]
\centering
\includegraphics[width=\plotwidth,angle=-90]{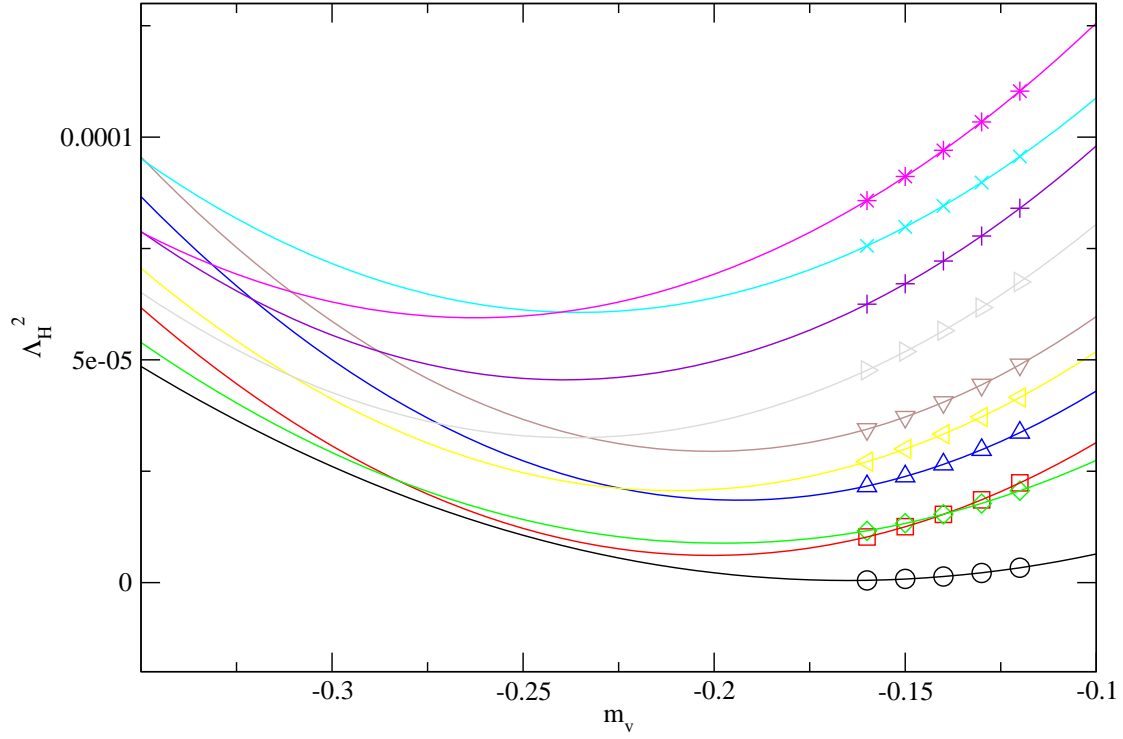}
\caption{%
Lowest 10 eigenvalues of $D_H^2$ on a typical background gauge field configuration generated on an $8^3\times8$ lattice with $\beta=2.3$, $L_s=24$ and $m_f=0.02$.
Solid lines represent results from a fit to \Eq{eig2_parameterization}.
}
\label{fig:eig2_vs_mf}
\end{figure}

Next we study the $m_v$ dependence of $\Lambda_H^2$ and attempt to extract the residual mass and chiral condensate for the $\beta=2.3$, $L_s=16$, $20$ and $24$, and $m_f=0.02$ ensembles.
Eigenvalues were computed for five values of $m_v$, ranging from $-0.16$ to $-0.12$, on a total of $150$ gauge configurations.
After appropriately reordering eigenvalues in order to account for possible level crossings, we fit the lowest 60 eigenvalues of $D_H^2$ as a function of $m_v$ to \Eq{eig2_parameterization} and extract the parameters $n_{5,i}$, $\lambda_i$ and $\delta m_i$ for each eigenvector.
\Figure{eig2_vs_mf} shows a plot of the fit results for the lowest 20 eigenvalues on a representative background field configuration for $L_s = 24$.
As may be seen in this plot, for a typical eigenvector, the splitting of $\pm\Lambda_H$ pairs due to the finite lattice spacing effects appear comparable to the spacings between adjacent eigenvectors.
As a qualitative measure, we conclude that the lattice spacing is not sufficiently small for the emergence of this continuum behavior of the Dirac spectrum.

\begin{figure}[htbp]
\centering
\includegraphics[width=\plotwidth,angle=-90]{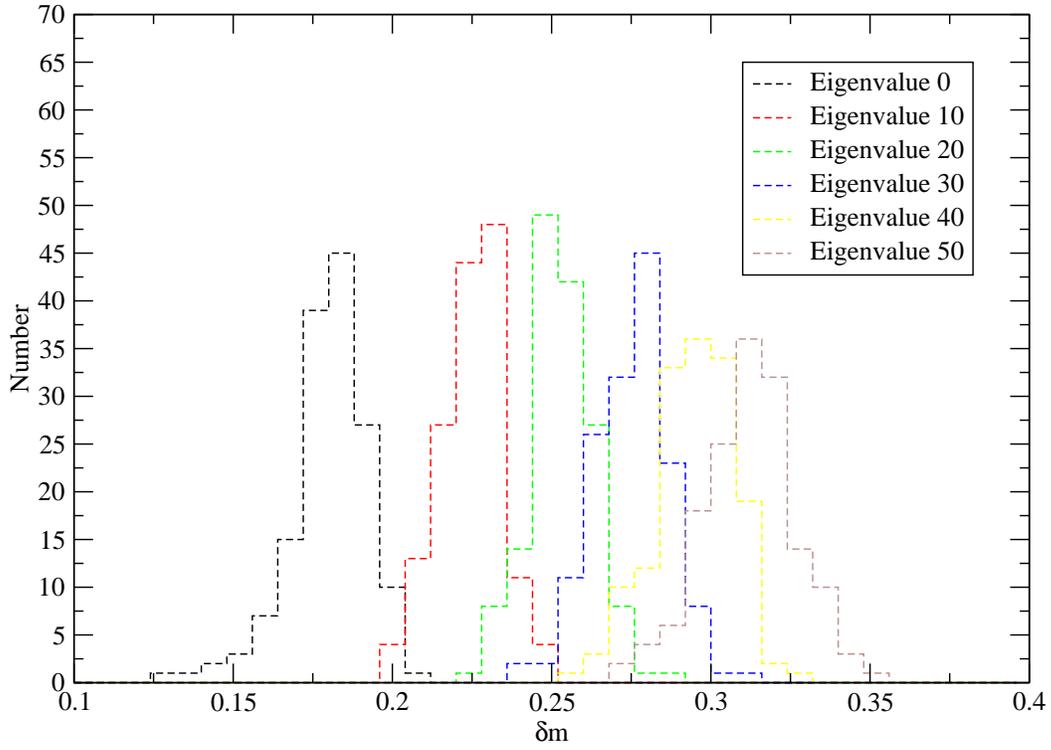}
\caption{%
Distribution of $\delta m$ values obtained from a fit to \Eq{eig2_parameterization} for an $8^3\times8$ lattice with $\beta=2.3$, $L_s=24$ and $m_f=0.02$.
}
\label{fig:dm_dist}
\end{figure}

In \Fig{dm_dist} we plot the distribution of $\delta m_i$ values at fixed eigenvalue number for the $L_s = 24$ ensemble.
Identifying the peak of the distribution of $\delta m_i$ values with the residual mass relies on an assumption that the spectrum of $D_H$ is in some sense continuum-like.
Unfortunately due to the lack of eigenvalue pairing, such interpretation is somewhat obscured.
For the lowest eigenvalue, the mean of the distribution appears consistent with the value of $m_{res}$ obtained from \Eq{mres_ratio}, however for larger eigenvalues, the mean increases in magnitude.

\begin{figure}[htbp]
\centering
\includegraphics[width=\plotwidth,angle=-90]{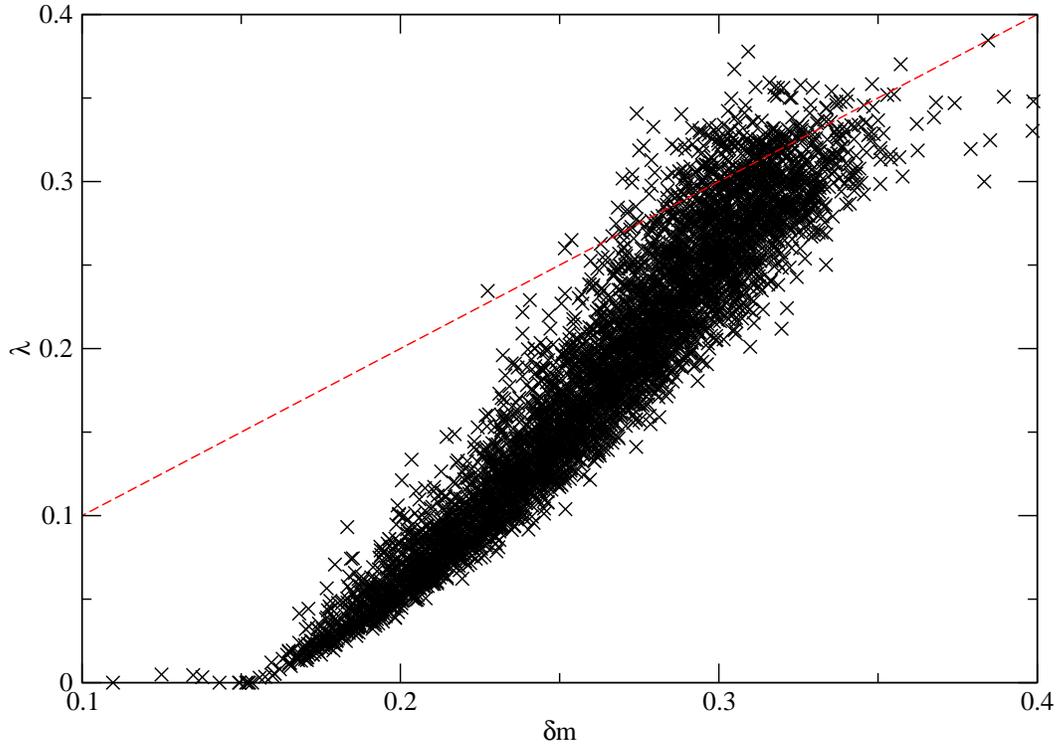}
\caption{%
Scatter plot of $|\lambda|$ verses $\delta m$ values obtained from the $m_v$ dependence of $\Lambda_H^2$, which have been fit to \Eq{eig2_parameterization}.
Eigenvalues where obtained for an $8^3\times8$ lattice with $\beta=2.3$, $L_s=24$ and $m_f=0.02$.
The dashed line represents the curve: $\lambda = \delta m$.
}
\label{fig:eig_vs_dm}
\end{figure}

\begin{figure}[htbp]
\centering
\includegraphics[width=\plotwidth,angle=-90]{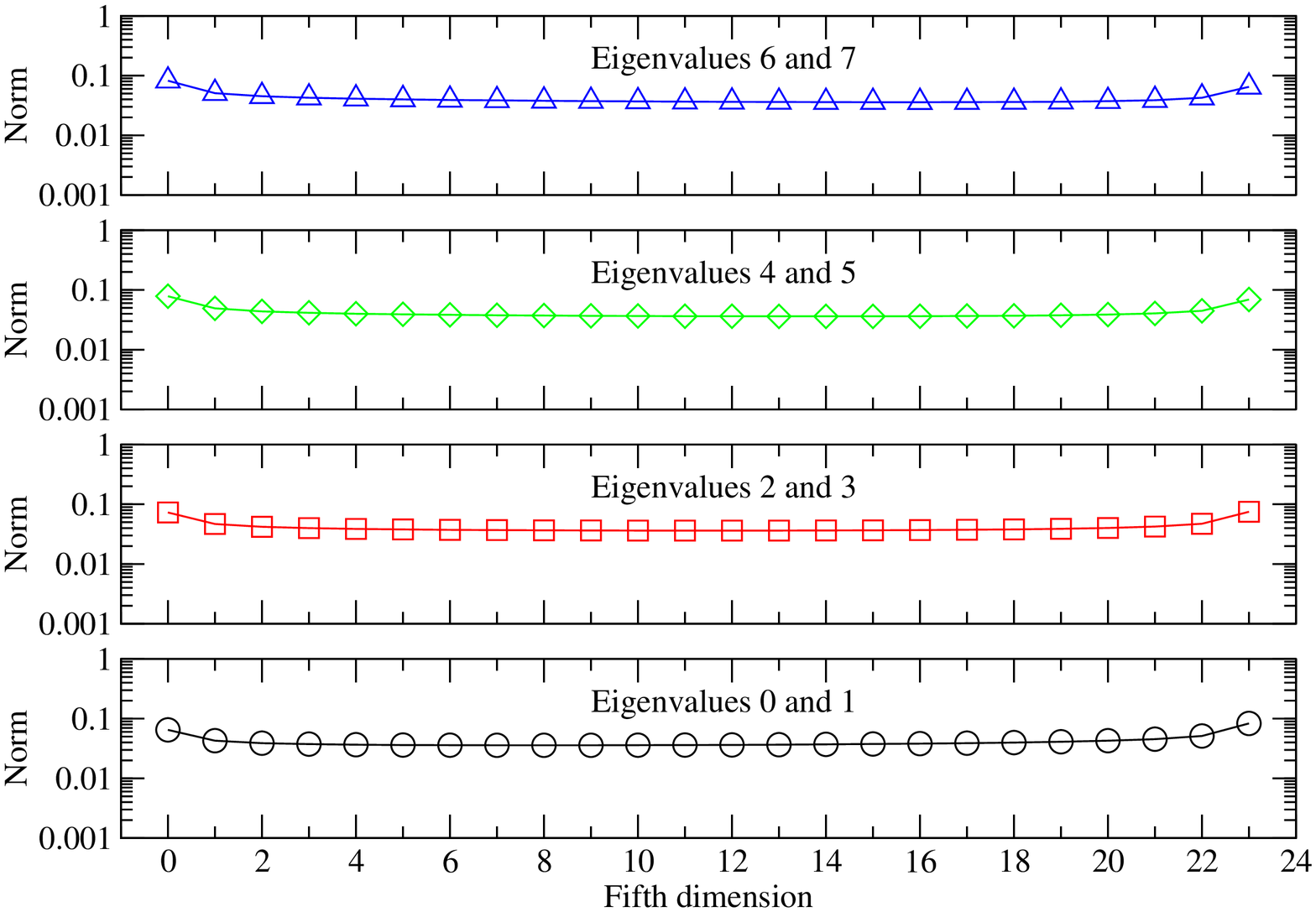}
\caption{%
Four dimensional norms $\calN(s)$ associated with the first eight eigenfunctions of $D_H$, measured on a typical background gauge field configuration generated on an $8^3\times8$ lattice with $\beta = 2.3$, $L_s=24$, $m_f=0.02$ and $m_v=-0.25$.
}
\label{fig:four_dim_norms_pq}
\end{figure}

\begin{figure}[htbp]
\centering
\includegraphics[width=\plotwidth,angle=-90]{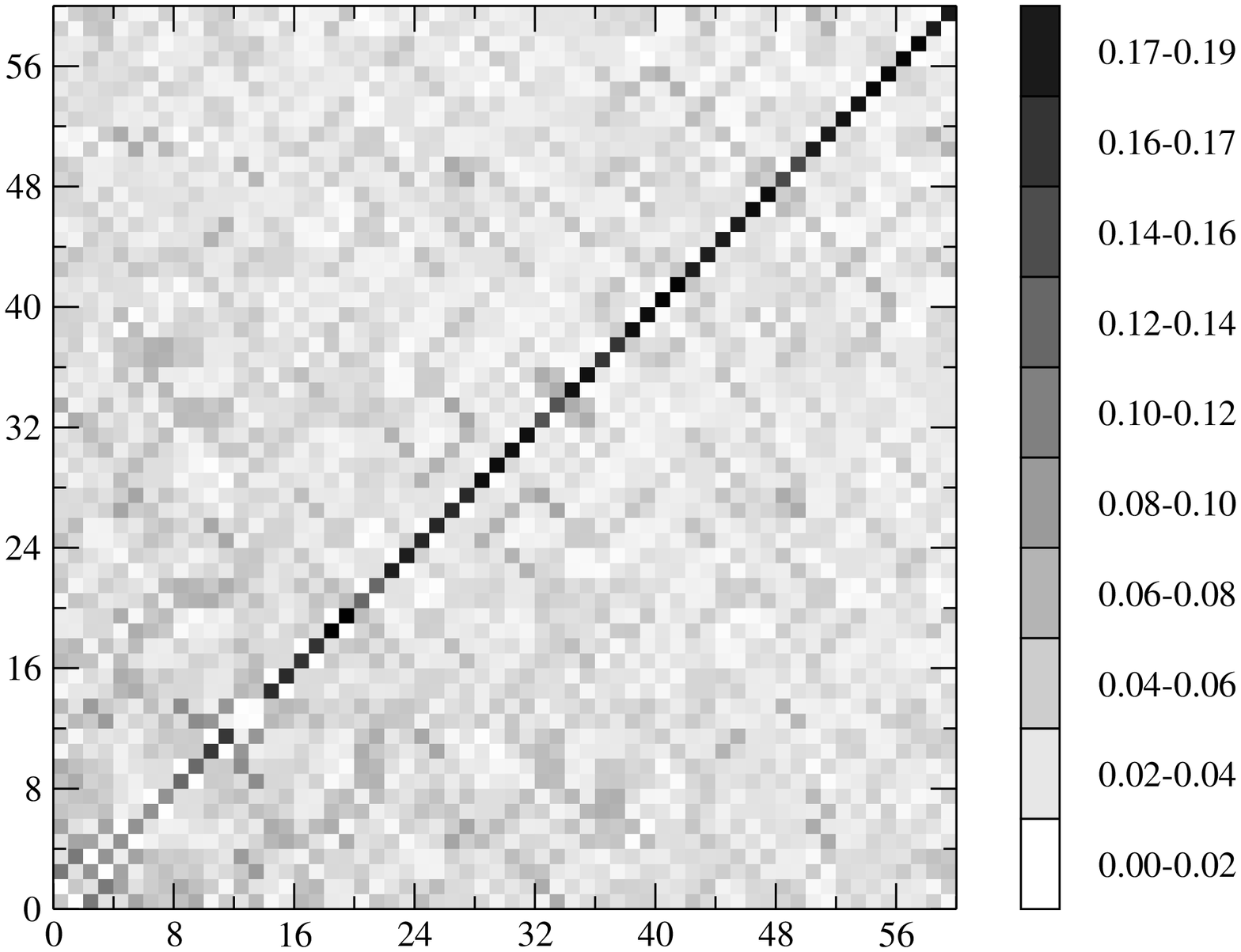}
\caption{%
Magnitude of the matrix elements $\langle \Lambda_H^\prime |\Gamma_s | \Lambda_H \rangle$, measured on a typical background gauge field configuration generated on a $8^3\times8$ lattice with $\beta = 2.3$, $L_s=24$, $m_f=0.02$, and $m_v=-0.25$.
}
\label{fig:phys_gamma5_pq}
\end{figure}

In \Fig{eig_vs_dm} we plot the values of $\lambda_i$ verses $\delta m_i$ for all $60\times150$ eigenvalues, which were extracted from fits to \Eq{eig2_parameterization} for the $L_s=24$ ensemble.
At low energies, it is evident that we are in a regime where $\lambda_i \lesssim \delta m_i$, which is consistent with our findings for the structure of $\Gamma_s$ in \Fig{phys_gamma5}.
If the distribution of $\delta m_i$ is highly peaked about $m_{res}$, one may in principle tune $m_v \approx -m_{res}$ such that $m_g < \lambda$ and expect off diagonal pairing to emerge in the matrix elements of $\Gamma_s$.
An attempt to produce off-diagonal pairing of $\Gamma_s$ by tuning $m_v\approx -m_{res}^\prime(m_f)$ was unsuccessful, but is likely due to the fact that off-diagonal pairing in $\Gamma_s$ may only occur when the $\pm$ eigenvalues of $D_H$ are paired.
In \Fig{four_dim_norms_pq} we plot $\calN_{\Lambda_H}(s)$ for the same configuration as in \Fig{four_dim_norms}, however with $m_v = -0.25$. 
\Figure{phys_gamma5_pq} displays the corresponding matrix elements of $\Gamma_s$ for this configuration.
From the first plot, we see that the lowest lying wave functions are no longer strongly localized on the fifth dimension boundaries.
As such, the diagonal structure of $\Gamma_s$ in \Fig{phys_gamma5_pq} disappears for the lowest few modes, although off-diagonal pairing in not very evident either.

\begin{figure}[htbp]
\centering
\includegraphics[width=\plotwidth,angle=-90]{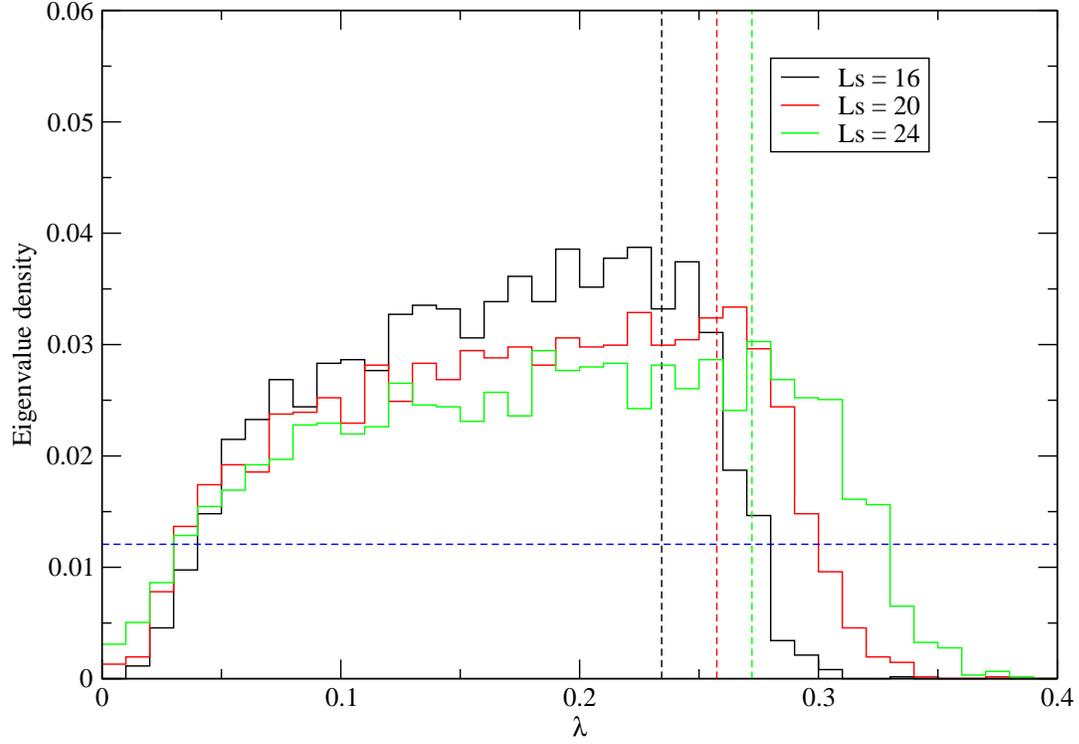}
\caption{%
Four dimensional eigenvalue density as a function of lambda for $8^3\times8$ lattices with $\beta=2.3$, two values of $L_s$ and $m_f=0.02$.
Results were obtained from the $m_f$ dependence of $\Lambda_H^2$, which have been fit to \Eq{eig2_parameterization}.
The dashed vertical lines represent the maximum value of $\lambda$ for which the distribution is reliably computed.
The dashed horizontal line corresponds to the chiral limit extrapolation value of the gluino condensate.
}
\label{fig:eig_dist}
\end{figure}

In \Fig{eig_dist} we plot the density of four dimensional eigenvalues $\rho^\prime(\lambda,m_g)$ for $L_s=16$, $20$ and $24$, which have been extracted from $60\times150$ eigenvalue measurements using \Eq{eig2_parameterization}.
The distribution exhibits a relatively constant region between $\lambda\approx 0.15$ and $0.25$, and a depleted region between $\lambda\approx 0$ and $0.15$ which may be attributed to finite volume effects.
Contrary to continuum expectations, there is no visible peak at $\lambda =0$, which one would normally attribute to near-zero modes.
If such a peak were to exist, it may be that its height is suppressed due to a finite bin size $\delta \lambda$, which is too large.
From \Fig{eig2_vs_mf}, however, it appears that the number of zero modes is few, suggesting that there is perhaps little topology change and that we are primarily within the zero topological charge sector of the theory.
This observation is consistent with the findings of \cite{Fleming:2000fa}, where fluctuations in the condensate as a function of MD time where studied as a possible indicator of topology change.
In that study, no large spikes in evolution of the condensate (which would correspond to the presence of zero-modes) were observed for the same choice of parameters.
Yet in light of the fact that the residual mass is so large in these simulations, it may also be that the fermions simply provide a poor measure of topology.
To make a more definitive statement regarding the topic, it would be interesting and beneficial to study the gauge fields directly instead.

\begin{figure}[htbp]
\centering
\includegraphics[width=\plotwidth,angle=-90]{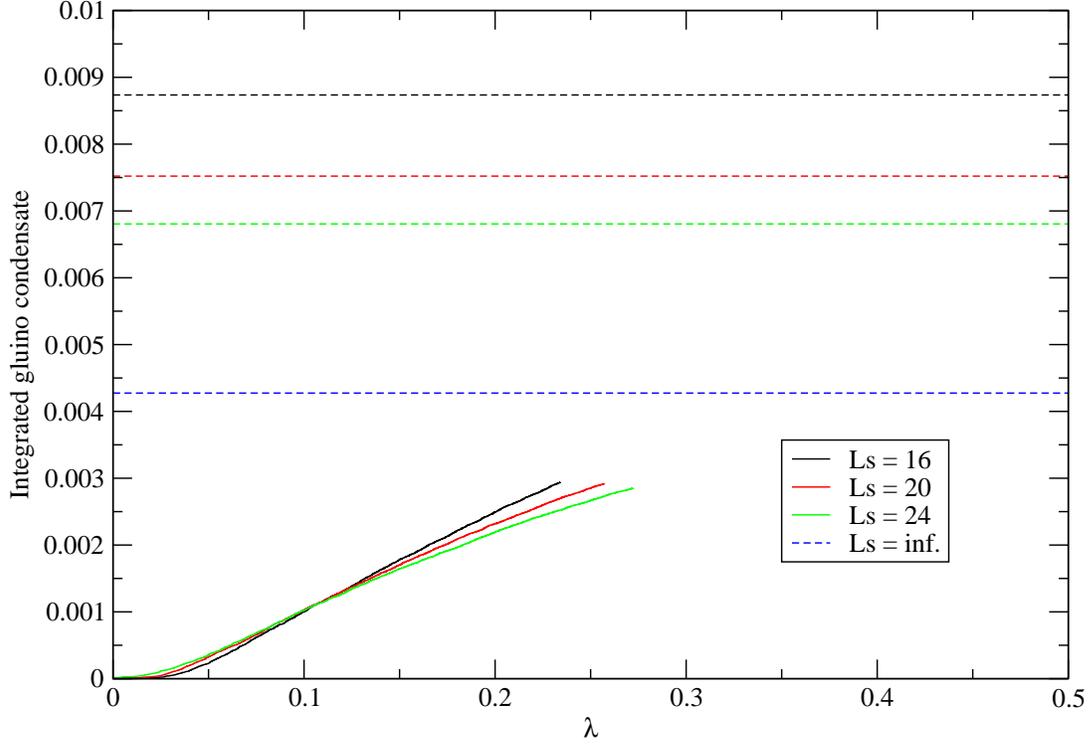}
\caption{%
Integrated condensate obtained from \Eq{dwf_condensate} with $m_v = m_f$ for eigenvalues in the range $0-\lambda$.
Solid curves indicates the integrated condensate as a function of $\lambda$ for $8^3\times8$ lattices with $\beta=2.3$, three values of $L_s$ and $m_f=0.02$
The corresponding horizontal dashed lines indicate the value of the gluino condensate obtained from \Eq{condensate}, as well as the $L_s = \infty$ extrapolated value of the condensate obtained in \Table{pbp_ls_fits} for the fit range $24-48$.
}
\label{fig:integrated_pbp}
\end{figure}

In principle one may extract an estimate of $\rho_{eff}^\prime(0,m_g)$ from the constant region of the eigenvalue density, compute the condensate and finally perform an $m_g\to0$ extrapolation.
But for the current choice of simulation parameters, the low lying spectrum appears to be too heavily distorted by the large residual mass for such an analysis to be reliable.
We may still extract some useful information from the eigenvalue density, however.
In our analysis of the chiral extrapolation of the gluino condensate described in \Sec{3_2}, we assumed that the $L_s$ dependent contribution to the condensate was dominated by UV modes.
We may check this assumption, by computing the contribution to \Eq{condensate} from modes in the interval: $0$-$\lambda$, using our fit values for $\lambda_i$ and $\delta m_i$ and \Eq{dwf_condensate} with $m_v = m_f$.
In \Fig{integrated_pbp}, we plot the integrated condensate as a function of the upper limit value $\lambda$ for $L_s = 16$, $20$ and $24$.
The contribution to the condensate from modes $|\lambda| r_0 \lesssim 0.8$ is approximately $1/3$ of the total value for each choice of $L_s$, which is consistent with our assumptions.

\begin{figure}[htbp]
\centering
\includegraphics[width=\plotwidth,angle=-90]{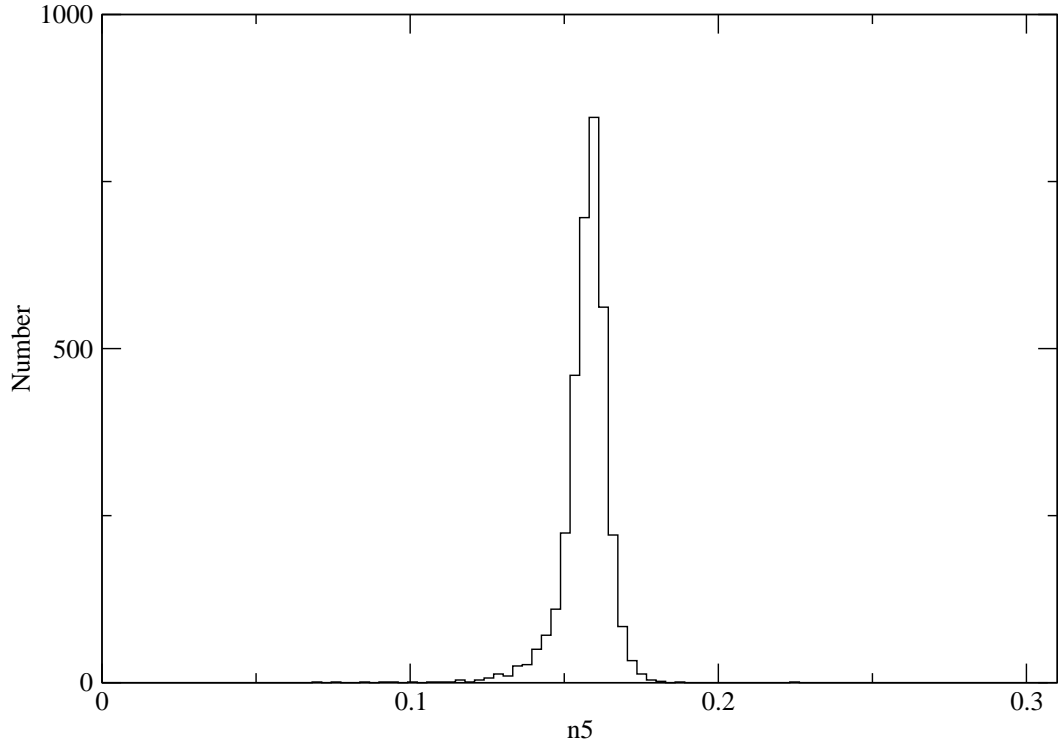}
\caption{%
Distribution of $n_5$ values for an $8^3\times8$ lattice with $\beta=2.3$, $L_s=24$ and $m_f=0.02$.
Results were obtained from the $m_v$ dependence of $\Lambda_H^2$, which have been fit to \Eq{eig2_parameterization}.
}
\label{fig:n_dist}
\end{figure}

Finally for completeness, \Fig{n_dist} provides a histogram of $n_5$ values extracted from the $\Lambda_H^2(m_v)$ fits for the $\beta=2.3$, $L_s=24$ and $m_f=0.02$ ensemble, where it is evident that $n_5$ is strongly peaked at approximately $n_5 \approx 0.16$.

\section{Conclusion}
\label{sec:4}

We have performed dynamical numerical simulations of N=1 SYM theory on $8^3 \times 8$ and $16^3\times32$ lattices using domain wall fermions with several goals in mind: to establish a lattice scale, assess the size of chiral symmetry breaking effects at finite $L_s$, and understand the degree to which DWFs are correctly reproducing the desired continuum physics at low energy.
In part, these efforts were motivated by the necessity to establish benchmarks for future studies.
As such, our work consists of a variety of basic measurements on $16^3\times32$ lattices, including measurements of the static potential, residual mass and gluino condensate.
We buttress these results by analyzing the eigenvalues and eigenvectors of the Hermitian DWF Dirac operator on $8^3\times8$ lattices, from which an independent estimate of the residual mass and chiral condensate may in principle be extracted.
We briefly summarize each of our measurements and results below.

We have presented results for the residual mass, which have been obtained using two different methods.
The first approach utilized a low energy identity relating the wall pseudo-scalar density to the midpoint pseudo-scalar density and assumptions about the decomposition of connected and disconnect parts of various pseudo-scalar correlators.
Using this approach we obtained residual masses which where approximately 5-20 times larger than the input gluino mass $m_f$.
A second approach for obtaining $m_{res}$, based upon the $m_v$ dependence of the eigenvalues of $D^2_H$, gave inconclusive yet qualitatively consistent results for the residual mass.
On an eigenvalue by eigenvalue basis, chiral symmetry breaking shifts in $m_f$ where comparable, if not larger than the value for $m_{res}$ obtained in the first approach.
The limited success of the second method in determining a precise value for the residual mass may be attributed to large finite lattice spacing artifacts.

The static potential for fundamental representation sources was determined using standard techniques for three different values of the coupling.
The potential exhibits characteristics indicative of confinement.
From the static potential we obtain estimates of the Sommer scale, which range from approximately $r_0 \approx 3.3$ at $\beta=2.3$ to $r_0 \approx 5.3$ at $\beta=2.4$.

We study properties of the low energy eigenvalues of the Hermitian DWF Dirac operator in an effort to understand to what degree the desired low energy physics is reproduced by the DWF formalism.
By studying the profile of the wave functions as a function of the fifth dimension coordinate, we have established that the low energy modes are indeed localized on the right and left boundaries of the fifth direction.
And, the matrix elements of the physical chirality operator $\Gamma_s$ are consistent with the presence of a large gluino mass compared to the typical low energy eigenvalues of $i \Dslash$.

We have performed a chiral extrapolation of the gluino condensate at a fixed value of the lattice spacing, $\beta=2.3$. 
The chiral extrapolation of the condensate was performed in a variety of different ways in order to elucidate some of the systematic errors inherent with the extrapolation.
For the parameters under consideration we find that various extrapolation procedures yield, at best, a value of the condensate which is reliable to approximately 25\%.
We attempt to provide an independent measurement of the gluino condensate from our studies of the Dirac spectrum.
However, given the current large values of the residual mass, a reliable estimate of the gluino condensate from the Banks-Casher relation was not possible.
 
Finally, we come to the general conclusion that there is, at present, no evidence to suggest that our simulations are not in the same universality class as SYM at non-zero gluino mass.
However, additional simulations will be required at far smaller residual masses and weaker couplings in order to achieve a more reliable measure of the quantities discussed in this paper.
Reducing the residual mass may be achieved with a variety of approaches; the simplest approach is to increase the size of the fifth dimension.
This method for reducing the residual mass becomes costly, however, once the dislocation term appearing in \Eq{mres} dominates over the perturbative contribution to the residual mass.
Alternatives to this approach, which may be more efficient, include going to weaker coupling, or using an improved action such as the DBW2 \cite{deForcrand:1999bi}, Iwasaki \cite{Iwasaki:1984cj,Iwasaki:1985we}, or auxiliary determinant \cite{Renfrew:2009aa} action.
The latter action combines the Gap DWF method of \cite{Vranas:1999rz,Vranas:2006zk} with the approach of \cite{Fukaya:2006vs} in order to suppress the residual mass while still maintaining adequate topological tunneling.
Such considerations will be left to a future study.
Finally we emphasis that scaling analysis at weaker coupling is essential in order to truly understanding the effects of finite lattice spacing on observables, and whether or not the gluino condensate survives the continuum limit.

\begin{acknowledgments}
M. G. E. would like to thank T. Blum, N. Christ, C. Dawson, C. Kim, R. Mawhinney and S. Takeda for helpful discussions, I. Mihailescu for fitting the static potential data presented in this work, and C. Jung for technical assistance with compiling and running the Columbia Physics System on BlueGene/L and QCDOC.
This research utilized resources at the New York Center for Computational Sciences at Stony Brook University/Brookhaven National Laboratory which is supported by the U.S. Department of Energy under Contract No. DE-AC02-98CH10886 and by the State of New York.
Early portions of this work were performed on QCDOC at Columbia University.
This work was supported by the U.S. Department of Energy under grant number DE-FG02-92ER40699.
\end{acknowledgments}

\appendix
\section{Correlators involving Majorana fermions}
\label{app:0}

In $\calN = 1$ SYM, the fundamental fermionic degrees of freedom are Majorana fermions rather than Dirac fermions.
As a result, correlation functions written in terms of propagator contractions may differ from that of conventional QCD.
Here we briefly review the construction of some of the basic observables studied in the paper, although for simplicity we limit the majority of the discussion to four dimensions.
Generalization of the results to the specific case of DWFs will be considered at the end of this section.

To begin with, consider the generic Majorana fermion path integral:
\beq
G^{(2N)} = \frac{1}{\Delta} \int [d\psi]\, e^{- \frac{1}{2} \psi^\Transpose M \psi} \calG_{i_1\ldots i_{2N}} \psi_{i_1}\ldots\psi_{i_{2N}} \ ,
\label{eq:fermion_path_integral}
\eeq
where the indices $i_k$ for $k=1,\ldots, 2N$ collectively represent space-time coordinate ($x$), spin ($\alpha$) and color ($c$), and summation over these indices is implied.
Due to the anti-commuting nature of the variables $\psi$, components of $\calG_{i_1\ldots i_{2N}}$ which are symmetric under the interchange of any two indices will not contribute to $G^{(2N)}$.
The Majorana matrix $M$ is related to the Dirac operator $D^4$ via the relation $M\equiv C D^4$, where $C$ is the charge conjugation matrix whose properties in four dimension are given in \Eq{charge_conjugation_matrix}.
We allow the Dirac operator and $\calG$ to depend on background field configurations and, as such, so may the normalization factor $\Delta$.
By definition we choose $\Delta$ such that $G^{(0)} \equiv 1$, and so it follows that $\Delta = \Pf (M)$.
Expectation values of operators involving the fermion fields $\psi$ in the full gauge theory are given by gauge averages over $G^{(2N)}$ with the probability measure $\Delta\, e^{-S_G} $, where $S_G$ is the gauge action.

The path integral given by \Eq{fermion_path_integral} may be evaluated in a standard fashion, and yields
\beq
G^{(2N)} = \calG_{i_1\ldots i_{2N}} \Delta_{i_1\ldots i_{2N}}\ ,
\eeq
where
\beq
\Delta_{i_1\ldots i_{2N}} =  \Pf \left(
\begin{array}{ccc}
M^{-1}_{i_1,i_1} & \ldots & M^{-1}_{i_1,i_N} \\
\vdots & \ddots       & \vdots \\
M^{-1}_{i_N,i_1} & \ldots &  M^{-1}_{i_N,i_N}
\end{array}
\right)
\eeq
is the Pfaffian constructed from Majorana propagators $M^{-1} = [D^4]^{-1} C^\dagger$, which start and end on the points ${i_1\ldots i_{2N}}$ in all possible combinations.
Note that since the matrix $M$ is antisymmetric, the diagonal matrix elements of the propagator vanish.

As an example of how this result may be applied, consider the two important cases $N=1$ and $N=2$, for which:
\beq
G^{(2)} &=&  \calG_{i_1 i_2} M^{-1}_{i_1 i_2}\ , \cr
G^{(4)} &=&  \calG_{i_1 i_2 i_3 i_4}(M^{-1}_{i_1 i_2} M^{-1}_{i_3 i_4} - M^{-1}_{i_1 i_3} M^{-1}_{i_2 i_4} + M^{-1}_{i_1 i_4} M^{-1}_{i_2 i_3}) \ .
\label{eq:contractions}
\eeq
In the case of the condensate $-\langle \mybar \psi \psi \rangle$ defined in \Eq{condensate}, we have:
\beq
\calG_{i_1 i_2} \sim \frac{1}{12 V} \delta_{x_1,x_2} C_{\alpha_1,\alpha_2} \delta_{c_1,c_2}\ ,
\eeq
which according to the expression for $G^{(2)}$ in \Eq{contractions} yields:
\beq
\frac{1}{12 V} \sum_{x}  \Tr S(x,x)
\label{eq:a1}
\eeq
prior to gauge averaging.
Here, $S = [D^4]^{-1}$ is the Dirac propagator and the trace is taken over both spin and color.
In the case of the pseudo-scalar correlator $\langle P(x) P(x^\prime) \rangle$, where $P(x) = \mybar \psi(x) \gamma_5 \psi(x)$, we have
\beq
\calG_{i_1 i_2 i_3 i_4} \sim \delta_{x_1,x} \delta_{x_2,x} \delta_{x_3,x^\prime} \delta_{x_4,x^\prime}  (C\gamma_5)_{\alpha_1,\alpha_2} (C\gamma_5)_{\alpha_3,\alpha_3}  \delta_{c_1,c_2} \delta_{a_3,a_3} \ ,
\eeq
which according to the expression for $G^{(4)}$ in \Eq{contractions} yields:
\beq
 2 \Tr \left[ S(x,x^\prime) \gamma_5 S(x^\prime,x) \gamma_5 \right] -   \Tr \left[ S(x,x^\prime) \gamma_5 \right] \Tr \left[ S(x^\prime,x) \gamma_5 \right] \  ,
\label{eq:a2}
\eeq
prior to gauge averaging.
Note that the connected and disconnected parts of this correlator, as defined in \Eq{pseudoscalar_correlator_parts}, correspond to the single and double trace terms respectively.

While the results presented here pertain to the four dimensional theory, they may easily be generalized to the case of DWFs, where the interpolating gluino fields $q(x)$  are expressed in terms of the fields $\Psi(x,s)$ at the fifth dimension boundary, according to \Eq{wall_gluino}.
Specifically the results for the gluino condensate and pseudo-scalar correlator given by \Eq{a1} and \Eq{a2} remain valid with the simple replacement $S \to S_q$, where
\beq
S_q(x,x^\prime) =&&   P_- D^{-1}_{x,0;x^\prime,L_s-1} P_-   +   P_- D^{-1}_{x,0;x^\prime,0} P_+    +  \cr
                 &&   P_+ D^{-1}_{x,L_s-1;x^\prime,L_s-1} P_-   +   P_+ D^{-1}_{x,L_s-1;x^\prime,0} P_+ \ ,
 \eeq
and $D$ is the DWF Dirac operator defined in \Eq{dwf_dirac_op}.


\bibliography{SUSY}
\bibliographystyle{apsrev}
\clearpage
\pagebreak



\clearpage
\pagebreak

\begin{table}[htbp]
\caption{%
Parameter values for the RHMC algorithm for $8^3\times8$ lattices.
$\lambda_{max}$ and $\lambda_{min}$ are the maximum and minimum eigenvalues of $\calD(M_5,m_f)$ required for the rational approximation, and $n_{MD}$ and $n_{MC}$ represent the degree of the rational approximation for the MD and MC accept/reject step, respectively.
}
\centering
\begin{tabular}{l c c|c c c c c c c c c}
\hline\hline
$\beta$ & $L_s$ & $m_f$ & $\lambda_{max}$ & $\lambda_{min}$ & $n_{MD}$ & $n_{MC}$ & $\lambda_{max}^{PV}$ & $\lambda_{min}^{PV}$ & $n_{MD}^{PV}$ & $n_{MC}^{PV}$ \\
\hline
2.3 & 12 & 0.02 & 4.0 & $4\times10^{-4}$ & 9 & 15 & 4.0 & $1\times10^{-3}$ & 6 & 9 \\
    & 16 & 0.02 & 4.0 & $4\times10^{-4}$ & 9 & 15 & 4.0 & $1\times10^{-3}$ & 6 & 9 \\
    & 20 & 0.02 & 4.0 & $4\times10^{-4}$ & 9 & 15 & 4.0 & $1\times10^{-3}$ & 6 & 9 \\
    & 24 & 0.02 & 4.0 & $4\times10^{-4}$ & 9 & 15 & 4.0 & $1\times10^{-3}$ & 6 & 9 \\
\hline
\hline
\end{tabular}
\label{tab:rhmc_parameters_small}
\end{table}

\clearpage
\pagebreak

\begin{table}[htbp]
\caption{%
Parameter values for the RHMC algorithm for $16^3\times32$ lattices.
An explanation of these parameter values may be found in \Table{rhmc_parameters_small}.
}
\centering
\begin{tabular}{l c c|c c c c c c c c c}
\hline\hline
$\beta$ & $L_s$ & $m_f$ & $\lambda_{max}$ & $\lambda_{min}$ & $n_{MD}$ & $n_{MC}$ & $\lambda_{max}^{PV}$ & $\lambda_{min}^{PV}$ & $n_{MD}^{PV}$ & $n_{MC}^{PV}$ \\
\hline
2.3           & 16 & 0.01 & 4.0 & $4\times10^{-4}$ & 9 & 15 & 4.0 & $1\times10^{-2}$ & 6 & 9 \\
              &    & 0.02 & 4.0 & $4\times10^{-4}$ & 9 & 15 & 4.0 & $1\times10^{-2}$ & 6 & 9 \\
              &    & 0.04 & 4.0 & $4\times10^{-4}$ & 9 & 15 & 4.0 & $1\times10^{-2}$ & 6 & 9 \\
              & 20 & 0.01 & 4.0 & $4\times10^{-4}$ & 9 & 15 & 4.0 & $1\times10^{-2}$ & 6 & 9 \\
              &    & 0.02 & 4.0 & $4\times10^{-4}$ & 9 & 15 & 4.0 & $1\times10^{-2}$ & 6 & 9 \\
              &    & 0.04 & 4.0 & $4\times10^{-4}$ & 9 & 15 & 4.0 & $1\times10^{-2}$ & 6 & 9 \\
              & 24 & 0.01 & 4.0 & $4\times10^{-4}$ & 9 & 15 & 4.0 & $5\times10^{-3}$ & 6 & 9 \\
              &    & 0.02 & 4.0 & $4\times10^{-4}$ & 9 & 15 & 4.0 & $5\times10^{-3}$ & 6 & 9 \\
              &    & 0.04 & 4.0 & $4\times10^{-4}$ & 9 & 15 & 4.0 & $5\times10^{-3}$ & 6 & 9 \\
              & 28 & 0.01 & 4.0 & $4\times10^{-4}$ & 9 & 15 & 4.0 & $5\times10^{-3}$ & 6 & 9 \\
              &    & 0.02 & 4.0 & $4\times10^{-4}$ & 9 & 15 & 4.0 & $5\times10^{-3}$ & 6 & 9 \\
              &    & 0.04 & 4.0 & $4\times10^{-4}$ & 9 & 15 & 4.0 & $5\times10^{-3}$ & 6 & 9 \\
              & 32 & 0.02 & 4.0 & $1\times10^{-4}$ & 9 & 15 & 4.0 & $1\times10^{-3}$ & 6 & 9 \\
              & 40 & 0.02 & 4.0 & $4\times10^{-5}$ & 9 & 15 & 4.0 & $5\times10^{-4}$ & 6 & 9 \\
              & 48 & 0.02 & 4.0 & $4\times10^{-5}$ & 9 & 15 & 4.0 & $5\times10^{-4}$ & 6 & 9 \\
\hline
$2.35\mybar3$ & 16 & 0.02 & 4.0 & $4\times10^{-4}$ & 9 & 15 & 4.0 & $1\times10^{-2}$ & 6 & 9 \\
              & 28 & 0.02 & 4.0 & $4\times10^{-4}$ & 9 & 15 & 4.0 & $5\times10^{-3}$ & 6 & 9 \\
\hline
2.4           & 16 & 0.02 & 4.0 & $4\times10^{-4}$ & 9 & 15 & 4.0 & $1\times10^{-2}$ & 6 & 9 \\
              & 28 & 0.02 & 4.0 & $4\times10^{-4}$ & 9 & 15 & 4.0 & $5\times10^{-3}$ & 6 & 9 \\
\hline
\hline
\end{tabular}
\label{tab:rhmc_parameters}
\end{table}

\clearpage
\pagebreak

\begin{table}[htbp]
\caption{%
MD evolution parameters and MC statistics for $8^3\times8$ lattices.
For each ensemble, the MD evolution trajectory length was $5\times\delta\tau$ MD time units.
A total number of trajectories for each ensemble is given by $N_{traj}$.
$N_{therm}$ is the number of initial trajectories before configurations in a given ensemble were deemed thermalized.
``Accept'' indicates the acceptance rate associated with the MC accept/reject step.
Ensemble averages of various functions of the Hamiltonian $H$ were performed using measurements on every trajectory.}
\centering
\begin{tabular}{l c c|c c c c c c c c c}
\hline\hline$\beta$ & $L_s$ & $m_f$ & $\delta_\tau$ & $N_{traj}$ & $N_{therm}$ & Accept & $\langle \Delta H\rangle$ & $\sqrt{\langle\Delta H^2\rangle}$ & $\langle e^{-\Delta H} \rangle$ 
\\
\hline
2.3 & 12 & 0.02 & 0.260 & 1000 & 300 & 0.626(18) & 0.498(28) & 1.095(27) & 0.956(39) \\
    & 16 & 0.02 & 0.220 &  750 & 300 & 0.749(20) & 0.146(26) & 0.577(19) & 1.004(19) \\
    & 20 & 0.02 & 0.240 & 1000 & 300 & 0.594(18) & 0.541(39) & 1.165(30) & 1.013(60) \\
    & 24 & 0.02 & 0.230 & 1000 & 300 & 0.627(18) & 0.470(35) & 1.965(43) & 0.966(42) \\
\hline
\hline
\end{tabular}
\label{tab:rhmc_statistics_small}
\end{table}

\clearpage
\pagebreak

\begin{table}[htbp]
\caption{%
MD evolution parameters and MC statistics for $16^3\times32$ lattices.
An explanation of these parameter values may be found in \Table{rhmc_statistics_small}}.
\centering
\begin{tabular}{l c c|c c c c c c c c c}
\hline\hline$\beta$ & $L_s$ & $m_f$ & $\delta_\tau$ & $N_{traj}$ & $N_{therm}$ & Accept & $\langle \Delta H\rangle$ & $\sqrt{\langle\Delta H^2\rangle}$ & $\langle e^{-\Delta H} \rangle$ 
\\
\hline
2.3           & 16 & 0.01 & 0.160 & 3125 & 500 & 0.758(19) & 0.193(28) & 0.673(9) & 1.016(14) \\
              &    & 0.02 & 0.160 & 3195 & 500 & 0.762(18) & 0.187(26) & 0.625(8) & 0.991(12) \\
              &    & 0.04 & 0.160 & 2790 & 500 & 0.776(19) & 0.152(26) & 0.577(8) & 1.003(12) \\
              & 20 & 0.01 & 0.155 & 2895 & 500 & 0.745(20) & 0.208(30) & 0.685(10) & 1.008(15) \\
              &    & 0.02 & 0.155 & 2655 & 500 & 0.753(21) & 0.212(31) & 0.674(10) & 0.993(16) \\
              &    & 0.04 & 0.160 & 2760 & 500 & 0.731(21) & 0.238(32) & 0.722(11) & 0.995(16) \\
              & 24 & 0.01 & 0.155 & 2855 & 500 & 0.775(19) & 0.142(26) & 0.578(8) & 1.019(13) \\
              &    & 0.02 & 0.130 & 2620 & 500 & 0.792(20) & 0.143(28) & 0.593(11) & 1.020(14) \\
              &    & 0.04 & 0.155 & 2610 & 500 & 0.760(21) & 0.175(32) & 0.690(10) & 1.050(17) \\
              & 28 & 0.01 & 0.135 & 2740 & 500 & 0.817(18) & 0.086(22) & 0.474(7) & 1.022(11) \\
              &    & 0.02 & 0.140 & 2855 & 500 & 0.796(19) & 0.158(24) & 0.536(7) & 0.974(11) \\
              &    & 0.04 & 0.155 & 2880 & 500 & 0.784(19) & 0.157(25) & 0.577(8) & 0.996(12) \\
              & 32 & 0.02 & 0.140 & 2730 & 500 & 0.757(20) & 0.180(29) & 0.654(9) & 1.016(15) \\
              & 40 & 0.02 & 0.140 & 2189 & 500 & 0.628(27) & 0.455(51) & 1.028(18) & 0.974(27) \\
              & 48 & 0.02 & 0.140 & 1555 & 500 & 0.608(34) & 0.586(76) & 1.247(25) & 1.031(46) \\
\hline                                          
$2.35\mybar3$ & 16 & 0.02 & 0.163 & 2575 & 600 & 0.782(20) & 0.160(27) & 0.583(9) & 0.996(13) \\
              & 28 & 0.02 & 0.140 & 2305 & 600 & 0.829(20) & 0.072(22) & 0.415(7) & 1.012(10) \\
\hline                                          
2.4           & 16 & 0.02 & 0.160 & 2710 & 750 & 0.824(18) & 0.093(20) & 0.432(7) & 0.996(10) \\
              & 28 & 0.02 & 0.140 & 2285 & 750 & 0.882(18) & 0.040(18) & 0.320(6) & 1.010(8) \\
\hline
\hline
\end{tabular}
\label{tab:rhmc_statistics}
\end{table}

\clearpage
\pagebreak

\begin{table}[htbp]
\caption{%
Gluino condensate $\langle \mybar q q\rangle$, $\langle \mybar q \gamma_5 q \rangle$ and average plaquette $\langle \mybar P \rangle$ for $8^3\times8$ lattices.}
\centering
\begin{tabular}{l c c|c c c}
\hline\hline
$\beta$ & $L_s$ & $m_f$ & $\langle \mybar q q\rangle$ & $\langle \mybar q \gamma_5 q \rangle\times10^{-5}$ & $\langle \mybar P \rangle$ \\
\hline
2.3 & 12 & 0.02 & 0.010631(40) & $  -1(6)$ & 0.73496(36) \\
    & 16 & 0.02 & 0.008736(42) & $  -5(9)$ & 0.73296(51) \\
    & 20 & 0.02 & 0.007522(37) & $ -10(6)$ & 0.73276(30) \\
    & 24 & 0.02 & 0.006806(34) & $  -1(7)$ & 0.73202(32) \\
\hline
\hline
\end{tabular}
\label{tab:meas_small}
\end{table}

\clearpage
\pagebreak

\begin{table}[htbp]
\caption{%
Gluino condensate $\langle \mybar q q\rangle$, $\langle \mybar q \gamma_5 q \rangle$ and average plaquette $\langle \mybar P \rangle$ for $16^3\times32$ lattices.}
\centering
\begin{tabular}{l c c|c c c}
\hline\hline
$\beta$ & $L_s$ & $m_f$ & $\langle \mybar q q\rangle$ & $\langle \mybar q \gamma_5 q \rangle\times10^{-6}$ & $\langle \mybar P \rangle$ \\
\hline
2.3           & 16 & 0.01 & 0.0078277(43) & $  11(6)$ & 0.733463(33) \\
              &    & 0.02 & 0.0086506(30) & $   3(5)$ & 0.733218(24) \\
              &    & 0.04 & 0.0102120(41) & $  -1(5)$ & 0.733203(36) \\
              & 20 & 0.01 & 0.0067032(46) & $  -1(6)$ & 0.732618(36) \\
              &    & 0.02 & 0.0075158(46) & $   5(6)$ & 0.732530(57)\\
              &    & 0.04 & 0.0091267(35) & $   1(6)$ & 0.732419(31) \\
              & 24 & 0.01 & 0.0059855(40) & $   2(7)$ & 0.732127(35) \\
              &    & 0.02 & 0.0068081(42) & $  -2(7)$ & 0.732110(41)\\
              &    & 0.04 & 0.0084396(36) & $   1(6)$ & 0.731983(37) \\
              & 28 & 0.01 & 0.0055027(46) & $  -8(7)$ & 0.731734(34) \\
              &    & 0.02 & 0.0063346(33) & $   6(6)$ & 0.731688(32) \\
              &    & 0.04 & 0.0079882(32) & $  -3(5)$ & 0.731628(38) \\
              & 32 & 0.02 & 0.0059947(47) & $  11(7)$ & 0.731418(45)\\
              & 40 & 0.02 & 0.0055266(43) & $  -2(10)$ & 0.730947(44)\\
              & 48 & 0.02 & 0.0052238(43) & $ -10(10)$ & 0.730727(50)\\
\hline
$2.35\mybar3$ & 16 & 0.02 & 0.0077917(72) & $   7(6)$ & 0.745243(40) \\
              & 28 & 0.02 & 0.0057106(61) & $   7(10)$ & 0.743853(38)\\
\hline
2.4           & 16 & 0.02 & 0.0068851(64) & $  -8(8)$ & 0.754149(30)\\
              & 28 & 0.02 & 0.0049179(82) & $ -16(9)$ & 0.752951(27)\\
\hline
\hline
\end{tabular}
\label{tab:meas}
\end{table}

\clearpage
\pagebreak

\begin{table}[htbp]
\caption{%
Fit results for the residual mass which was extracted from $R(t)$ for $16^3\times32$ lattices.
The column labeled ``time range'' indicates the plateau region over which the $R(t)$ was fit.}
\centering
\begin{tabular}{l c c|c c c}
\hline\hline
$\beta$ & $L_s$ & $m_f$ & time range & $\chi^2/\dof$ & $m_{res}$ \\
\hline
2.3           & 16 & 0.01 & 10-21 & 63.8/11 & 0.18682(13) \\
              &    & 0.02 & 10-21 & 15.0/11 & 0.18939(12) \\
              &    & 0.04 & 10-21 & 17.2/11 & 0.19223(13) \\
              & 20 & 0.01 & 10-21 & 10.8/11 & 0.17151(14) \\
              &    & 0.02 & 10-21 & 35.1/11 & 0.17340(14) \\
              &    & 0.04 & 10-21 & 49.4/11 & 0.17680(13) \\
              & 24 & 0.01 & 10-21 & 13.4/11 & 0.15807(13) \\
              &    & 0.02 & 10-21 & 18.4/11 & 0.15906(15) \\
              &    & 0.04 & 10-21 & 19.8/11 & 0.16367(13) \\
              & 28 & 0.01 & 10-21 & 28.6/11 & 0.14685(13) \\
              &    & 0.02 & 10-21 & 25.5/11 & 0.14834(13) \\
              &    & 0.04 & 10-21 & 111.7/11 & 0.15283(13) \\
              & 32 & 0.02 & 10-21 &  5.1/11 & 0.13905(14) \\
              & 40 & 0.02 & 10-21 & 23.2/11 & 0.12397(14) \\
              & 48 & 0.02 & 10-21 & 29.1/11 & 0.11225(16) \\
\hline                               
$2.35\mybar3$ & 16 & 0.02 & 10-21 & 45.5/11 & 0.14071(19) \\
              & 28 & 0.02 & 10-21 & 42.5/11 & 0.10269(18) \\
\hline                               
2.4           & 16 & 0.02 & 10-21 &  9.5/11 & 0.10082(15) \\
              & 28 & 0.02 & 10-21 & 24.6/11 & 0.06513(17) \\
\hline
\hline
\end{tabular}
\label{tab:mres_fits}
\end{table}

\clearpage
\pagebreak

\begin{table}[htbp]
\caption{Fit results for the residual mass as a function of $L_s$ for $16^3\times32$ lattices with $\beta=2.3$ and $m_f=0.02$.
The column labeled by ``$L_s$ range'' indicates the range of $L_s$ values used in the fit.
The parameters $a_0$, $a_1$ and $a_2$ are defined in \Eq{mres}.}
\centering
\begin{tabular}{l c c|c c c c c}
\hline\hline
$\beta$ & $L_s$ range & $m_f$ & $\chi^2/\dof$ & $a_0$ & $a_1$ & $a_2$ \\
\hline
2.3    & 16-48 & 0.02 & 53.5/4 & -6.20(3) & 0.0275(4)  & 7.03(5) \\
       & 20-48 &      &  5.9/3 & -6.37(7) & 0.0236(7)  & 7.43(10) \\
       & 24-48 &      &  2.2/2 & -6.30(5) & 0.0255(11) & 7.24(12) \\
\hline
\hline
\end{tabular}
\label{tab:mres_ls_fits}
\end{table}

\clearpage
\pagebreak

\begin{table}[htbp]
\caption{%
Static quark potential fit parameters for a subset of $16^3\times32$ lattices.
The column labeled by ``time range'' indicates the time window over which Wilson loops were fit at fixed distances in order to extract the potential.
The column labeled by ``distance range'' indicates the distance window over which the potential was fit.}
\centering
\begin{tabular}{l c c|c c c c c c}
\hline\hline
$\beta$ & $L_s$ & $m_f$ & time range & distance range & $V_0$ & $\alpha$ & $\sigma$ & $r_0$ \\
\hline
2.3           & 16 & 0.02 & 4-8   & $\sqrt{2}$-6           & 0.511(10) & 0.176(10) & 0.132(2)  & 3.344(21) \\
              &    & 0.02 & 4-8   & $\sqrt{3}$-$\sqrt{29}$ & 0.483(17) & 0.140(23) & 0.137(4)  & 3.319(22) \\
              &    & 0.02 & 4-8   & $\sqrt{3}$-6           & 0.501(18) & 0.161(23) & 0.134(4)  & 3.339(23) \\
              &    & 0.04 & 4-8   & $\sqrt{2}$-6           & 0.524(15) & 0.185(14) & 0.129(3)  & 3.376(29) \\
              &    & 0.04 & 4-8   & $\sqrt{3}$-$\sqrt{29}$ & 0.516(26) & 0.178(33) & 0.130(5)  & 3.381(32) \\
              &    & 0.04 & 4-8   & $\sqrt{3}$-6           & 0.534(26) & 0.200(32) & 0.127(5)  & 3.359(31) \\
              & 20 & 0.02 & 4-8   & $\sqrt{2}$-6           & 0.511(13) & 0.174(13) & 0.134(4)  & 3.322(47) \\
              &    & 0.02 & 4-8   & $\sqrt{3}$-$\sqrt{29}$ & 0.480(44) & 0.132(54) & 0.139(11) & 3.306(77) \\
              &    & 0.02 & 4-8   & $\sqrt{3}$-6           & 0.483(41) & 0.136(49) & 0.138(10) & 3.309(71) \\
              &    & 0.04 & 4-8   & $\sqrt{2}$-6           & 0.532(15) & 0.197(15) & 0.131(4)  & 3.327(31) \\
              &    & 0.04 & 4-8   & $\sqrt{3}$-$\sqrt{29}$ & 0.485(26) & 0.136(31) & 0.140(5)  & 3.291(31) \\
              &    & 0.04 & 4-8   & $\sqrt{3}$-6           & 0.507(27) & 0.163(32) & 0.135(5)  & 3.316(34) \\
              & 24 & 0.02 & 4-8   & $\sqrt{2}$-6           & 0.522(19) & 0.182(18) & 0.132(4)  & 3.340(38) \\
              &    & 0.02 & 4-8   & $\sqrt{3}$-$\sqrt{29}$ & 0.504(32) & 0.158(39) & 0.135(7)  & 3.330(41) \\
              &    & 0.02 & 4-8   & $\sqrt{3}$-6           & 0.511(31) & 0.166(38) & 0.134(6)  & 3.322(41) \\
              &    & 0.04 & 4-8   & $\sqrt{2}$-6           & 0.524(15) & 0.185(14) & 0.129(3)  & 3.376(29) \\
              &    & 0.04 & 4-8   & $\sqrt{3}$-$\sqrt{29}$ & 0.516(26) & 0.178(33) & 0.130(5)  & 3.359(31) \\
              &    & 0.04 & 4-8   & $\sqrt{3}$-6           & 0.534(26) & 0.200(33) & 0.127(5)  & 3.381(32) \\
\hline
$2.35\mybar3$ & 16 & 0.02 & 5-9   & $\sqrt{2}$-6           & 0.543(13) & 0.202(13) & 0.077(3)  & 4.324(64) \\
              &    & 0.02 & 5-9   & $\sqrt{3}$-$\sqrt{29}$ & 0.550(23) & 0.218(32) & 0.077(4)  & 4.313(76) \\
              &    & 0.02 & 5-9   & $\sqrt{3}$-6           & 0.569(23) & 0.240(23) & 0.074(4)  & 4.379(80) \\
\hline
2.4           & 16 & 0.02 & 5-9   & $\sqrt{2}$-6           & 0.528(7)  & 0.191(7)  & 0.053(2)  & 5.254(67) \\
              &    & 0.02 & 5-9   & $\sqrt{3}$-$\sqrt{29}$ & 0.534(16) & 0.199(22) & 0.052(3)  & 5.272(101) \\
              &    & 0.02 & 5-9   & $\sqrt{3}$-6           & 0.536(15) & 0.202(20) & 0.052(3)  & 5.288(94) \\
              &    & 0.02 & 5-10  & $\sqrt{2}$-6           & 0.533(5)  & 0.196(5)  & 0.052(1)  & 5.280(45) \\
              &    & 0.02 & 5-10  & $\sqrt{3}$-$\sqrt{29}$ & 0.532(12) & 0.196(16) & 0.053(2)  & 5.257(71) \\
              &    & 0.02 & 5-10  & $\sqrt{3}$-6           & 0.539(11) & 0.205(15) & 0.051(2)  & 5.306(68) \\
\end{tabular}
\label{tab:hqpot}
\end{table}

\clearpage
\pagebreak

\begin{table}[htbp]
\caption{%
Fit results for gluino condensate as a function of $L_s$ for $16^3\times32$ lattices with $\beta=2.3$ and $m_f=0.02$.
The column labeled by ``$L_s$ range'' indicates the range of $L_s$ values used in the fit.
The parameters $b_0$, $b_1$ and $b_2$ are defined in \Eq{pbp_vs_ls}.}
\centering
\begin{tabular}{l c c|c c c c c c}
\hline\hline
$\beta$ & $L_s$ range & $m_f$ & $\chi^2/\dof$ & $b_0$ & $b_1$ & $b_2$ \\
\hline
2.3    & 16-28 & 0.02 & 0.1/1    & 0.004994(53)  & 0.1057(18) & 0.0370(20) \\
       & 16-32 &      & 1.9/2    & 0.004939(39)  & 0.1040(14) & 0.0350(15) \\
       & 16-40 &      & 26.8/3   & 0.004758(20)  & 0.0989(7)  & 0.0289(7) \\
       & 16-48 &      & 62.9/4   & 0.004650(16)  & 0.0962(5)  & 0.0255(5)  \\
       & 20-48 &      & 14.8/3   & 0.004529(29)  & 0.0897(9)  & 0.0204(10) \\
       & 24-48 &      & 1.9/2    & 0.004392(62)  & 0.0839(18) & 0.0154(19) \\
\hline
\hline
\end{tabular}
\label{tab:pbp_ls_fits}
\end{table}

\clearpage
\pagebreak

\begin{table}[htbp]
\caption{%
Fit results for gluino condensate as a function of $m_f$ for $16^3\times32$ lattices with $\beta=2.3$.
The column labeled by ``$m_f$ range'' indicates the range of $m_f$ values used in the fit.
The parameters $c_0$ and $c_1$ are defined in \Eq{pbp_vs_mf}.}
\centering
\begin{tabular}{l c c|c c c c c}
\hline\hline
$\beta$ & $L_s$ & $m_f$ range & $\chi^2/\dof$ & $c_0$ & $c_1$ \\
\hline
2.3    & 16 & 0.01-0.04  & 40.6/1   & 0.0070544(51)  & 0.07915(19) \\
       & 20 &            &  0.7/1   & 0.0058979(55)  & 0.08073(19) \\
       & 24 &            &  0.8/1   & 0.0051697(49)  & 0.08177(18) \\
       & 28 &            &  0.5/1   & 0.0046770(51)  & 0.08279(18) \\
\hline
\hline
\end{tabular}
\label{tab:pbp_mf_fits}
\end{table}

\clearpage
\pagebreak

\begin{table}[htbp]
\caption{%
Fit results for the $m_f=0$ extrapolated value of the gluino condensate as a function of $L_s$ for $16^3\times32$ lattices with $\beta=2.3$.
The column labeled by ``$L_s$ range'' indicates the range of $L_s$ values used in the fit.
The parameters $b_0$, $b_1$ and $b_2$ are defined in \Eq{pbp_vs_ls}.}
\centering
\begin{tabular}{l c c|c c c c c c}
\hline\hline
$\beta$ & $L_s$ range & $m_f$ & $\chi^2/\dof$ & $b_0$ & $b_1$ & $b_2$ \\
\hline
2.3    & 16-28 & 0 & 0.07/1    & 0.00320(9)  & 0.1051(24) & 0.0334(27) \\
\hline
\hline
\end{tabular}
\label{tab:pbp_chiral_limit_fit}
\end{table}

\clearpage
\pagebreak

\clearpage
\pagebreak


\def\figdir{figures/}


\clearpage
\pagebreak

\clearpage
\pagebreak

\clearpage
\pagebreak

\clearpage
\pagebreak

\clearpage
\pagebreak

\clearpage
\pagebreak

\clearpage
\pagebreak

\clearpage
\pagebreak

\clearpage
\pagebreak

\clearpage
\pagebreak

\clearpage
\pagebreak

\clearpage
\pagebreak

\clearpage
\pagebreak

\clearpage
\pagebreak

\clearpage
\pagebreak

\clearpage
\pagebreak

\clearpage
\pagebreak

\clearpage
\pagebreak

\clearpage
\pagebreak

\clearpage
\pagebreak

\clearpage
\pagebreak

\clearpage
\pagebreak

\clearpage
\pagebreak

\clearpage
\pagebreak

\clearpage
\pagebreak

\clearpage
\pagebreak

\clearpage
\pagebreak

\clearpage
\pagebreak

\clearpage
\pagebreak

\clearpage
\pagebreak

\end{document}